\documentclass[12pt]{iopart}
\usepackage{iopams}
\usepackage{graphicx}
\usepackage{xcolor}
\usepackage[amssymb,Gray]{SIunits}
\newcommand{\mom}{Q}
\newcommand{\erf}{\mathrm{Erf}}
\newcommand{\abs}[1]{\ensuremath{\left\vert #1 \right\vert}}
\newcommand{\erw}[1]{\ensuremath{\langle #1 \rangle}}
\newcommand{\com}[2]{\left[#1,#2\right]}
\newcommand{\acom}[2]{\left\{#1,#2\right\}}
\newcommand{\order}[1]{\Or(#1)}
\newcommand{\nnl}{\nonumber\\}
\newcommand{\bra}[1]{\langle #1 \hspace{-2pt} \mid}
\newcommand{\ket}[1]{\mid \hspace{-1pt} #1 \rangle}
\renewcommand{\vec}[1]{\mathrm{\mathbf{#1}}}
\newcommand{\schr}{Schr{\"o}\-din\-ger}
\newcommand{\sne}{Schr{\"o}\-din\-ger--New\-ton equation}
\newcommand{\mplanck}{m_\text{Pl}}
\newcommand{\tplanck}{t_\text{Pl}}
\newcommand{\lplanck}{\ell_\text{Pl}}
\begin{document}

\topical{Gravitational Decoherence}

\author{Angelo Bassi$^{1,2}$, Andr\'e Gro{\ss}ardt$^{3,4}$ and Hendrik Ulbricht$^{5}$}
\address{$^1$ Department of Physics, University of Trieste, Strada Costiera 11, 34151 Miramare-Trieste, Italy}
\address{$^2$ Istituto Nazionale di Fisica Nucleare, Sezione di Trieste, Via Valerio 2, 34127 Trieste, Italy}
\address{$^3$ ZARM, University of Bremen, Am Fallturm 2, 28359 Bremen, Germany}
\address{$^4$ Centre for Theoretical Atomic, Molecular, and Optical Physics, School of
Mathematics and Physics, Queen's University Belfast, BT7 1NN, United Kingdom}
\address{$^5$ School of Physics and Astronomy, University of Southampton, SO17 1BJ, United Kingdom}
\ead{bassi@ts.infn.it}

\begin{abstract}
We discuss effects of loss of coherence in low energy quantum systems caused by or related to gravitation, referred to as gravitational decoherence. These effects, resulting from random metric fluctuations, for instance, promise to be accessible by relatively inexpensive table-top experiments, way before the scales where true quantum gravity effects become important. Therefore, they can provide a first experimental view on gravity in the quantum regime.
We will survey models of decoherence induced both by classical and quantum gravitational fluctuations; it will be manifest that a clear understanding of gravitational decoherence is still lacking. Next we will review models where quantum theory is modified, under the assumption that gravity causes the collapse of the wave functions, when systems are large enough. These models challenge the quantum-gravity interplay, and can be tested experimentally. In the last part we have a look at the state of the art of experimental research.  We will review efforts aiming at more and more accurate measurements of gravity ($G$ and $g$) and ideas for measuring conventional and unconventional gravity effects on nonrelativistic quantum systems.
\end{abstract}

%
%
\submitto{\CQG}
%
%
%

\section{Introduction}

The unification of quantum physics with general relativity is perhaps the most important and ambitious open problem in modern physics. Both theories work excellently when taken separately; when rigidly patched together to explain the evolution of the universe, they give impressive results,
the discovery of the cosmic microwave background and its properties being the most important one, 
although the resolution of the dark matter and dark energy puzzle might ultimately show that such patching was wrong in the first place~\cite{Will:1981}. Yet, a coherent unified theory does not exist.  

The general attitude is to quantize gravity in some way, which produced a variety of sophisticated formulations, string theory~\cite{green1987superstring} and loop quantum gravity~\cite{rovelli1998loop} arguably being the most prominent ones.
A minority of the scientific community favors the opposite approach, to ``gravitize'' quantum mechanics~\cite{Penrose:2014}, or to radically reconsider both theories~\cite{Sorkin:2009,Dowker:2011,Hooft:2016}.

From the experimental point of view, not much is really known. We know that matter is quantum and, when in a classical (i.\,e.~localized in {\it space}) state, it reacts to gravity as predicted by Newton or by Einstein. This has been verified for source masses ranging from $10^{-2}$~\kilo\gram~\cite{ritter1990experimental} to stars, galaxies, and beyond, and from distances ranging from $10^{-2}$~\meter~\cite{ritter1990experimental} to cosmological distances (but here again, the resolution of the dark matter/energy problem might reveal surprises). We also know that matter in spatial {\it quantum superpositions} responds to an external classical gravitational field as predicted by simply adding the Newtonian gravitational potential to the \schr\ equation~\cite{Colella:1975,nesvizhevsky2002quantum,Fixler:2007}.

However, the answers, both theoretical and experimental, to the big questions are still missing: Is gravity quantum? Do gravitons, the quantum counterpart of the recently discovered classical gravitational waves~\cite{Abbott:2016}, really exist? What is the gravitational field generated by a quantum superposition of matter? Does quantum mechanics still hold true in presence of sufficiently strong gravitational fields? The fact that after almost one century of intense theoretical research we do not have a solid clue about the above questions, is perhaps the indication that radically new thinking is required to approach the problem of unifying quantum mechanics and gravity.
And it certainly is not a viable option to simply wait until the Planck scale ($\lplanck = 10^{-35}~\meter$, $\tplanck = 10^{-44}~\second$, $\mplanck = 10^{-8}~\kilo\gram$) becomes accessible experimentally in order to have an indication on how to proceed theoretically, as this will require a long time. 

In recent years, however, pushed by the impressive technological developments,  the scientific community is exploring the possibility that the quantum-gravity interplay can be seen far away from the Planck scale, in nonrelativistic quantum systems and in table-top experiments~\cite{Amelino‐Camelia:2005}. This is the context of this review article.

Whatever the final outcome of the quantum gravity riddle, one prediction seems to be rather robust: spacetime as a dynamical variable fluctuates, and these fluctuations cause decoherence in quantum systems, which can be revealed, at least in principle,  in matter-wave interferometers, for example. These fluctuations can be classical, quantum, or both. Classical fluctuations can be a relic of primordial gravitational waves, generated during the early evolution of the universe, a relic of the big bang together with the cosmic microwave background, or the sum of the gravitational waves randomly produced by the large variety of physical sources scattered through the universe. Their spectrum is estimated theoretically, and constrained by observations~\cite{Abbott:2016}. Quantum fluctuations arise from the quantization procedure, as for any field. Not much is known, as a consistent quantum theory of gravity is missing.

Gravitational decoherence, as any noise, modifies the evolution of quantum systems with respect to what is predicted by the \schr\ equation applied to isolated systems. Under general assumptions, Gorini--Kossakowski--Sudarshan~\cite{Gorini:1976} and independently Lindblad~\cite{Lindblad:1976}, proved that the general structure of the master equation for the density matrix $\rho_t$ describing a quantum system interacting with an external environment is
\begin{equation}
\frac{\rmd}{\rmd t}\rho_t = -\frac{\rmi}{\hbar} \com{H}{\rho_t} + \sum_n \lambda_n \left( L_n  \rho_t L_n^{\dagger} -  \frac{1}{2} \rho_t L_n^{\dagger} L_n - \frac{1}{2} L_n^{\dagger} L_n \rho_t \right),
\end{equation}
where $H$ is the standard quantum Hamiltonian, possibly containing a corrective term arising from the interaction with the  environment, the Lindblad operators $L_n$ generate decoherence and dissipation, and the positive coefficients $\lambda_n$ set the strength of these effects. The goal of theoretical research on gravitational decoherence is to derive an explicit expression for the Lindblad operators and the coefficients, by assuming that quantum matter interacts with a stochastic gravitational background. We will review some of the models which have been presented in the literature.

A remark is due concerning the notion of ``gravitational decoherence'', which is used in quite different contexts with different meanings.
For the purpose of this review article, we use it in a rather broad sense, referring to any loss of coherence of quantum \emph{matter} 
fields, whose cause can be related to gravitational effects---be it due to a classical or quantum
gravitational background (as discussed in \sref{sec:decohgrav}) or due to modifications of 
quantum mechanics motivated from gravity (as in \sref{sec:oxfc}). As the main focus of this review article are experiments which are feasible in the laboratory (see \sref{sec:experiments}), our main concern are effects in the limit of nonrelativistic systems and weak gravitational fields.

The emergence of a classical spacetime structure from quantum gravity, as it is discussed, for instance, 
in minisuperspace models~\cite{Zeh:1986,Kiefer:1992} where the Friedmann universe acts as ``system''
and higher multipoles are treated as the ``environment'', is also often referred to as ``gravitational decoherence''. This, however, is an effect where the quantized spacetime itself decoheres, rather than quantum matter in the laboratory, and is beyond the scope of this review article in which we are interested in understanding the effects of gravity on such laboratory quantum systems.

In \sref{sec:decohgrav} we will consider the standard setup: (nonrelativistic) quantum mechanical matter interacting with a random gravitational field. It turns out that there exist a large number of quite different approaches towards this problem, many of them motivated by \emph{quantum} gravitational fluctuations but nonetheless applicable to perturbations of a classical spacetime background.
All the approaches we present make use of simplifications and approximations such as only considering the effect of conformal metric fluctuations. A general treatment of decoherence due to classical spacetime fluctuations is still lacking.

We will proceed to the discussion of quantum gravity effects as a possible source of spacetime fluctuations causing decoherence. Many such proposals exist, and we will review effects of a thermal graviton bath, of a minimal length scale, and of fluctuations of the time variable.

Finally, we have a look into a recently proposed decoherence effect due to gravitational time dilation which has been the starting point of an ongoing controversy. This effect is somewhat different from the previously discussed decoherence effects,
as decoherence occurs due to a coupling of the center-of-mass motion (acting as ``system'') to internal degrees of freedom (acting as ``bath'') of a complex quantum system, and gravity (or rather relativity) solely enters in the role of a transmitter which enables this coupling.

In \sref{sec:oxfc} we will go beyond the standard quantum formalism. Quantum theory has originated a lively debate about its meaning, as it gives a fundamental and distinctive role, in its very formulation, to measurements and observations, while these should represent only a subclass of all possible physical events, described at least in principle by the dynamical equations of the theory. This is not the case with quantum mechanics. Several solutions to  this conundrum have been proposed, among which is the idea that the \schr\ equation should be modified and supplemented  with nonlinear and stochastic terms, which end up in a collapse of the wave function in specific situations, like measurement processes. The phenomenology of models of spontaneous wave function collapse is well understood, and experiments are placing stronger and stronger bounds on the collapse parameters. However a relevant  question remains open: if the collapse is real as predicted by these models, what is its origin? Of course one can always claim that it is an intrinsic property of nature, but intuition tells that there should be an explanation. 

People, most notably Roger Penrose, proposed the idea that gravity might be the ultimate explanation for the collapse of the wave function. Gravity is universal, and its magnitude increases with the mass, which matches the requirements of the collapse, which is also universal, if quantum theory is, and becomes relevant when macroscopic systems are involved. In addition, there is no evidence thus far, nor a consensus among theoretical physicists, whether gravity should be quantized, therefore the possibility remains open for it to give the required nonlinear coupling to break quantum linearity and the associated measurement problem.

We will review four major attempts to link the collapse of the wave function to gravity: the Di\'osi--Penrose model, Adler's model, K{\'a}rolyh{\'a}zy's model, and the \sne. 

The Di\'osi--Penrose model falls within the class of models of spontaneous wave function collapse, and it is specified by setting the correlation function of the noise equal to the Newtonian gravitational potential. The model was first proposed by Di\'osi, and later revived by Penrose, based on independent arguments on the behavior of spacetime in presence of quantum superpositions.  

Adler's model also falls within the class of models of spontaneous wave function collapse, but now the assumption is that gravity has an irreducible, complex, rapidly fluctuating component. The  correlation function of this noise is left unspecified, and eventually has to be explained by a yet-to-be-formulated underlying theory.   

K{\'a}rolyh{\'a}zy's model is based on the assumption that spacetime is fuzzy below a specific scale, and this fuzziness causes a lack of quantum coherence in space (and time) for material systems, which is the stronger, the larger the systems. The quantum-gravity coupling is not modified, therefore it is linear and so technically this model falls under the class of decoherence models discussed in \sref{sec:decohgrav}, not under the class of collapse models. However, since K{\'a}rolyh{\'a}zy's  motivations spring more from the analysis of the quantum measurement problem, rather than from an attempt of arriving at a quantum theory of gravity, it is better discussed in context of the other models in \sref{sec:oxfc}.

Finally, the \sne\ is discussed, which is based on the assumption that matter is fundamentally quantum but gravity is fundamentally classical. Then the most natural way to couple matter and gravity leads to a nonlinearly modified \schr\ equation of the solitonic type, where gravity attracts gravitationally different parts of the wave function.  This equation does not fall within the class of models of spontaneous wave function collapse, although it also causes its ``collapse'' in a way which we will discuss. 

In \sref{sec:experiments} we move from theory to experiment, and we will review the state of the art in experimental research of the quantum-gravity interplay at low energy, typically in table-top experiments. We will summarize the most advanced and precise Newtonian and non-Newtonian gravity measurements by torsion pendulum type experiments, and also by matter-wave interferometry. Matter-wave interferometers, such as neutron, atom, and molecule interferometers, are sensitive to phase shifts triggered by gravitational effects, while the source mass, from which the gravitational field originates, is typically larger than 1~\kilogram\ and usually not closer than 1~\meter\ to the test mass. Interestingly, gravity experiments involving source and test bodies with the smallest gravitational force between each other are torsion pendulum experiments.
We will further summarize proposals for tests of gravity which aim to go beyond the present limitations in decreasing mass and distance in gravitational force sensing, as well as in increasing the mass in spatial quantum superpositions, and which aim to test general relativistic effects and even quantum gravity effects, amongst which are the tests of collapse type effects and gravitational decoherence.

\section{Classical and quantum spacetime fluctuations}\label{sec:decohgrav}
Any quantum state subject to some environment will become entangled to this environment, and
undergo decoherence when environmental degrees of freedom are traced out. 
The gravitational interaction with the spacetime background will lead to decoherence as well.
It is, however, different from the usual decoherence sources in at least two respects:
\begin{itemize}
\item The gravitational interaction cannot be omitted, contrary to the interaction with, for instance, gas particles and radiation which can---at least in principle---be evacuated or shielded completely.
\item As, to date, gravity is only understood classically, the theoretical description of gravitational
decoherence is necessarily incomplete, and predictions may depend on assumptions about the correct
behavior of gravity at the quantum scale.
\end{itemize}
These distinctive features of gravitational decoherence make it very interesting to understand the 
details of the possible decoherence mechanisms. Does gravitational decoherence put an ultimate bound
on the scales at which quantum features of nature can be observed? And are there predictions that
differ depending on whether the gravitational background is treated classically or quantum mechanically?
Could such predictions even lead to observable consequences which give insight into a quantum theory
of gravity?

On the other hand, due to the lack of a consistent and broadly accepted quantum theory of gravity
from which decoherence effects at low energies could be derived in a rigorous way,
there is no unanimous theoretical description for gravitational decoherence. Even the theoretical
tools within quantum field theory on curved spacetime~\cite{Wald:1994,Fredenhagen:2007} are not
far enough developed for definite predictions at the level of low energy laboratory physics.

Decoherence effects are, therefore, discussed in very different contexts, where quantum gravitational
features are introduced in a rather ad-hoc manner. Many of these ideas are based on either
some specific quantum gravitational model, and are then model dependent, or on general ideas about
gravity at the Planck scale, such as the existence of a minimal length and time scale, or spacetime
fluctuations which can be implemented in multiple ways.

The latter idea of a fluctuating spacetime is often introduced in terms of stochastic perturbations
of the \emph{classical} background metric. Such perturbations can be of fundamental origin, e.\,g.
as a consequence of quantum gravitational fluctuations. They can also stem from gravitational wave
noise which itself may be modeled classically or in terms of quanta of the gravitational field, i.\,e.
gravitons. However, as the theoretical description of laboratory quantum systems on
curved spacetime background itself still poses many questions, the respective models are based on highly
simplified assumptions, and the resulting predictions can differ considerably depending on the choice
of these assumptions.

We first review some of the predictions and models for effects resulting from fluctuations of the
classical spacetime background in \sref{sec:gd-class}. Then, in \sref{sec:gd-quant},
we discuss how these classical fluctuations can be linked to quantum gravitational effects,
and how decoherence is described in the formalism of perturbative quantum gravity.

Finally, we review the recent discussion about a decoherence effect which has been attributed
to gravitational time dilation~\cite{Pikovski:2015} in \sref{sec:gd-time}.
This effect is different in some regards from the previously mentioned ones, as it occurs in
a homogeneous gravitational field, and gravity acts as a transmitter of entanglement between
center of mass and internal degrees of freedom. Hence, the role of the environment which allows for
decoherence is not played by gravity itself in this case.

\subsection{Classical spacetime fluctuations}\label{sec:gd-class}
Decoherence can already occur in the straightforward approach of a quantum wave function that is evolving
on a curved spacetime metric, if this spacetime background contains stochastic fluctuations.
A common feature of all approaches to model such fluctuations is that they assume a weak 
gravitational field, such that the linear approximation to general relativity applies, i.\,e.
the metric tensor can be written as $g_{\mu\nu} = \eta_{\mu\nu} + \epsilon\,h_{\mu\nu}$ with a
small parameter $\epsilon$.

\subsubsection{Gravitational phase changes for matter-waves}

At the lowest significant order, gravitational effects on matter-waves reveal themselves in the
form of a phase shift of the wave function. An instructive derivation can be given in terms of
the Wentzel-Kramers-Brillouin (WKB) approximation~\cite{Wentzel:1926,Kramers:1926,Brillouin:1926}.
For this, we start from the Klein-Gordon equation for a scalar field $\phi$ in a curved spacetime:
\begin{eqnarray}\label{eqn:klein-gordon}
0 &= \left[ g^{\mu\nu} \, \nabla_\mu \nabla_\nu + \frac{m^2 c^2}{\hbar^2} \right] \phi \nnl
&= \left[ g^{\mu\nu} \, \left(\frac{\partial}{\partial x^\mu} \frac{\partial}{\partial x^\nu}
- \Gamma^\lambda{}_{\mu\nu} \frac{\partial}{\partial x^\lambda} \right)
 + \frac{m^2 c^2}{\hbar^2} \right] \phi \,,
\end{eqnarray}
where the covariant derivatives $\nabla_\mu$ in the first line are expressed in terms of the
Christoffel symbols and partial derivatives in the second line.
Throughout this review article we adopt the mostly minus metric signature $(+,-,-,-)$.
The WKB approximation is based on the ansatz
\begin{equation}
\phi = \rme^{\frac{\rmi}{\hbar} \varphi}
\end{equation}
for the field, followed by a series expansion in terms of $\hbar$.
Inserting this ansatz into \eref{eqn:klein-gordon} yields
\begin{equation}
0 = g^{\mu \nu} \, \left(
\frac{\rmi}{\hbar}\,\frac{\partial}{\partial x^\mu}\frac{\partial}{\partial x^\nu} \varphi
-\frac{\rmi}{\hbar} \,\Gamma^\lambda{}_{\mu\nu} \frac{\partial}{\partial x^\lambda} \varphi
-\frac{1}{\hbar^2}\,\frac{\partial}{\partial x^\mu} \varphi \, \frac{\partial}{\partial x^\nu} \varphi
\right) + \frac{m^2 c^2}{\hbar^2}\,,
\end{equation}
which at lowest order in $\hbar$ results in
\begin{equation}\label{eqn:eikonal}
g^{\mu \nu} \, \frac{\partial}{\partial x^\mu} \varphi \, \frac{\partial}{\partial x^\nu} \varphi
= m^2 c^2\,.
\end{equation}
\eref{eqn:eikonal} is the equivalent to the eikonal equation in the case of geometrical optics
(i.\,e.~the case $m=0$~\cite{Plebanski:1960}).
Now, in the linear approximation $g_{\mu\nu} = \eta_{\mu\nu} + \epsilon\,h_{\mu\nu}$,
the phase can be split as $\varphi = \varphi^{(0)} + \epsilon\,\varphi^{(1)}$ which results
in the following set of two equations at the zeroth and first order in $\epsilon$~\cite{Linet:1976a}:
\numparts\begin{eqnarray}
m^2 c^2 &= \eta^{\mu \nu} \, \frac{\partial}{\partial x^\mu} \varphi^{(0)} \,
\frac{\partial}{\partial x^\nu} \varphi^{(0)} \label{eqn:zero-order-eikonal} \\
0 &= 2\,\eta^{\mu \nu} \, \frac{\partial}{\partial x^\mu} \varphi^{(0)} \,
\frac{\partial}{\partial x^\nu} \varphi^{(1)}
+ h^{\mu \nu} \, \frac{\partial}{\partial x^\mu} \varphi^{(0)} \,
\frac{\partial}{\partial x^\nu} \varphi^{(0)} \,.
\end{eqnarray}\endnumparts
\eref{eqn:zero-order-eikonal} is simply the eikonal equation in flat Minkowski space.
Defining the constant vector $\xi^\mu = (E/c, p^1, p^2, p^3)$, where $E$ is the energy and
$p^i$ the momentum of the particle described by \eref{eqn:klein-gordon}, the relativistic
energy-momentum relation takes the form $\xi^\mu \xi_\mu = E^2/c^2 - p^2 = m^2 c^2$,
and it is evident that
\begin{equation}
\varphi^{(0)} = \eta_{\mu \nu} \,\xi^\mu \,x^\nu + \varphi^{(0)}_0 \,,
\end{equation}
with constant $\varphi^{(0)}_0$ solves \eref{eqn:zero-order-eikonal}.

It can then be shown~\cite{Linet:1976a} that the gravitational contribution to the phase
along the particle trajectory, parametrized by $\tau$ as $x^\mu = \tau\,\xi^\mu + x^\mu_0$,
satisfies the differential equation
\begin{equation}
\frac{\rmd \varphi^{(1)}}{\rmd \tau} = \frac{1}{2}\,h_{\mu \nu}\,\xi^\mu\,\xi^\nu \,.
\end{equation}

According to Linet and Tourrenc~\cite{Linet:1976a} (cf. also Refs.~\cite{Stodolsky:1979,Cai:1989}),
the total phase difference resulting after
integration along the particle trajectory can be categorized into three different types of
gravitational phase shift effects:
\begin{enumerate}
\item The term resulting from the $h_{00}$ metric element is the contribution of the Newtonian
gravitational potential,
\begin{equation}
\Delta \varphi = \frac{E}{2} \int_{t_0}^{t_1} h_{00} \,\rmd t \,.
\end{equation}
In the Earth's homogeneous potential this yields the experimentally confirmed~\cite{Colella:1975} phase factor $\Delta \varphi = m\, g\, \Delta z\,t$.
\item The contribution from the $h_{0i}$ metric elements is
\begin{equation}
\Delta \varphi = c \int_{t_0}^{t_1} h_{0i}\,p^i \,\rmd t \,.
\end{equation}
This effect can be understood as stemming from the rotation of the gravitational source
(i.\,e.~the Lense-Thirring effect~\cite{Misner:1973}), which is closely related to the Sagnac effect~\cite{Sagnac:1913}
(phase shift in a rotating interferometer).
\item Finally, the third contribution from the spatial $h_{ij}$ terms results from planar
gravitational waves. The phase shift is
\begin{equation}
\Delta \varphi = \frac{c^2}{E} \int_{t_0}^{t_1} h_{ij}\,p^i\,p^j \,\rmd t \,.
\end{equation}
\end{enumerate}

In the transverse traceless (TT) gauge, where only the spatial components $h_{ij}$ of the metric 
perturbation are nonzero, gravitational waves can be described by a mode 
decomposition~\cite{Reynaud:2004}:
\begin{equation}
h_{ij}(x) = \sum_{\sigma \in \pm} \varepsilon_{ij}^\sigma \,\int \frac{\rmd^4 k}{(2\pi)^4}\,
h^\sigma(k)\,\rme^{-\rmi k_\mu x^\mu} \,,
\end{equation}
where $\varepsilon$ is the polarization tensor for the two possible polarizations $\sigma \in \pm$
and $h^\sigma$ are the coefficients of the expansion for the respective polarizations.
Reynaud \etal~\cite{Reynaud:2004} consider the special case of unpolarized, isotropic waves, in which case the metric components can be characterized by a spectral distribution
$S(\omega)$, e.\,g.~for the $xy$-component:
\begin{equation}
\erw{h_{12}(t) h_{12}(0)} = \int \frac{\rmd \omega}{2\pi} S(\omega)\,\rme^{-\rmi \omega t} \,.
\end{equation}
The phase difference then becomes a stochastic variable, $\delta \varphi$. If this gravitational noise
is assumed to be Gaussian, the fringe visibility in an interferometer will be reduced as follows:
\begin{equation}
\mathcal{V} = \erw{\exp\left(\frac{\rmi}{\hbar} \delta \varphi\right)} = \exp\left(-\frac{\erw{(\delta \varphi)^2}}{2\hbar^2}\right)
 \equiv \exp\left(-\frac{\Delta \varphi^2}{2\hbar^2}\right) \,.
\end{equation}
The variance in the exponent depends not only on the spectrum of the gravitational wave background $S$,
but also on the geometry of the interferometer, encoded in an ``apparatus response function'' $A$.
It can then be written in the frequency domain as~\cite{Lamine:2006}
\begin{equation}
\Delta \varphi^2 = \int \frac{\rmd \omega}{2\pi} S(\omega)\,A(\omega)\,F(\omega) \,, 
\end{equation}
where $F$ is a filter function ensuring that the integral only runs over the relevant (unobserved)
frequencies.

The apparatus function $A(\omega)$ generally has a complicated form. In the idealized case of a
rhombic Mach-Zehnder interferometer of arm length $L$, with small aperture angle $\alpha$, and
particles with nonrelativistic velocities $v \ll c$, it is given by~\cite{Lamine:2006}
\begin{equation}
A_\text{MZ}(\omega) = \frac{4\,m^2\,v^4}{\hbar^2\,\omega^2}\,\sin^2(\alpha) \,
\left(1-\cos \left(\frac{\omega\,L}{v}\right)\right)^2
\end{equation}

Lamine \etal~\cite{Lamine:2006} argue that for a measurable decoherence effect, i.\,e.~a variance
$\Delta \varphi^2$ of the order of unity, even under ideal conditions in a wide angle (90\degree)
Mach-Zehnder interferometer one would need, for instance, a femtogram particle with a velocity of
1~\kilo\meter\per\second\ and 1~\meter\ arm length (see \sref{Gdeco}~C).

Nonetheless, such order of magnitude estimations can only provide a first impression, and the many 
approximations that need to be made in the course of arriving at this result (isotropy, Gaussianity,
specific interferometer design, etc.) leave room for unexpected effects.

\subsubsection{Conformal metric perturbations}\label{sec:conformal-perturbations}
In order to go beyond the discussion based on the semi-classical WKB approximation and the
phase change of a wave function on a gravitational background, one is looking for simple models
for quantum matter in a fluctuating spacetime.

One possible approach is to model fluctuations of spacetime in the form of conformal transformations
of the metric, in which case only
a single scalar field determines the metric structure. This has been done by
S{\'a}nchez-G{\'o}mez~\cite{SanchezGomez:1993}
for a non-propagating fluctuation and by Power and Percival~\cite{Power:2000} for conformal gravitational waves.

Conformal fluctuations of the flat Minkowski metric are such transformations which can be obtained
by a local rescaling (Weyl transformations):
\begin{equation}\label{eqn:conf-metric}
g_{\mu\nu} = \Omega(x)^2 \, \eta_{\mu\nu} \,,
\end{equation}
with some spacetime function $\Omega$. Generally, Einstein's field equations can be derived from the
Einstein-Hilbert action
\begin{equation}\label{eqn:einstein-hilbert}
S = \frac{1}{16\,\pi\,G} \int \rmd^4 x \, \sqrt{-g} R \,,
\end{equation}
where $R$ is the Ricci curvature scalar~\cite{Misner:1973,Wald:1984}.
In order to rewrite the Einstein-Hilbert action in a more suitable form, it is convenient to define $A(x) = \Omega(x) - 1$.
The action then takes the form of the Klein-Gordon action\footnote{We are only interested
in the physical case of $D=4$ dimensions, for $D\not=4$ the redefinition of $\Omega(x)$ must be chosen
differently.}
\begin{equation}\label{eqn:einstein-hilbert-conformal}
\frac{3}{8\,\pi\,G} \int \rmd^4 x \, \eta^{\mu\nu} \frac{\partial}{\partial x^\mu} A(x)
\frac{\partial}{\partial x^\nu} A(x) \,,
\end{equation}
and by variation of this action one finds that the ``conformal field'' $A(x)$ satisfies the massless,
flat space Klein-Gordon equation
$\eta^{\mu\nu}\partial_\mu\partial_\nu A(x) = 0$. The wave solutions of this equations can be interpreted 
as gravitational waves traveling at the speed of light $c$.

A matter-wave evolving in this curved spacetime will, at lowest order, be evolving in a Newtonian
gravitational potential $\Phi$, which is related to the metric in the usual way~\cite{Wald:1984} by
$g_{00} = 1+2\Phi/c^2$. The evolution is then described by a nonrelativistic potential
\begin{equation}\label{eqn:power-grav-potential}
V(x) = m \Phi = \frac{mc^2}{2}\,(g_{00} - 1) = mc^2\,A(x) + \frac{mc^2}{2}\,A(x)^2\,.
\end{equation}
Note that the gravitational constant $G$ does not appear anywhere in
\eref{eqn:power-grav-potential}. In general relativity, Einstein's field equations which follow
from the action~\eref{eqn:einstein-hilbert} ascertain that the metric components in the Newtonian
limit depend on $G$, e.\,g.~$g_{00} = 1 - 2 G m/(r c^2)$. In contrast, the Klein-Gordon equation which
follows from the variation of the conformal action~\eref{eqn:einstein-hilbert-conformal} does not
depend on $G$. This is not surprising, as the conformal approach \emph{violates} Einstein's field
equations. The easiest way to see this is by noting that the continuity equation
$\nabla^\nu T_{\mu\nu} = 0$ for the stress-energy tensor, which must be satisfied for a solution of
general relativity, is invariant under conformal transformations
$g_{\mu\nu} \to \Omega^2 g_{\mu\nu}$ only for trace free stress-energy tensors, i.\,e.~for
$T^\mu_\mu = 0$~\cite{Wald:1984}, which is usually not the case.
The conformal gravitational waves modeled by the field $A(x)$, therefore, should be considered as
fluctuations of the Minkowski vacuum rather than solutions of Einstein's equations.

If the conformal $A(x)$ is supposed to be only a stochastic fluctuation around the Minkowski metric,
then its average values must satisfy
$\erw{A(x)} = \erw{\partial_\mu A(x)} = \erw{A(x) \partial_\mu A(x)} = 0$.
The first relation is due to the fact that there should not be an overall imprint of the stochastic
fluctuations, and the last two relations are consequences of the conformal symmetry~\cite{Kar:1993}.

With the assumption of a Gaussian white noise in all four dimension, i.\,e.
\begin{equation}
\erw{A(x) A(x')} = A_0^2 \,
\exp\left(-\frac{1}{L^2} \sum_{\mu=0}^3 (x^\mu - {x^\mu}')^2\right) \,,
\end{equation}
S{\'a}nchez G{\'o}mez~\cite{SanchezGomez:1993} derives the (one-particle) master equation
\begin{equation}\label{eqn:sanchez-master}
\fl
\frac{\partial}{\partial t}\rho(\vec x,\vec x';t) = 
-\frac{\rmi}{\hbar} \com{H_0}{\rho(\vec x,\vec x';t)}
-\frac{8 m^2 c^3 A_0^2 L}{\hbar^2} \, \min\left(\frac{(\vec x-\vec x')^2}{L^2} ; 1\right) \, \rho(\vec x,\vec x';t) \,,
\end{equation}
where the first term is the standard unitary quantum evolution term for the Hamiltonian $H_0$ without
the stochastic fluctuations.

The gravitational constant enters again when the field amplitude $A_0$ is associated with the
Planck scale---thus, by bringing quantum gravity arguments into the game
(cf.~also \sref{sec:min-length}). The largest distance of two points which behave coherently
under a metric fluctuation
is the length scale $L$. Given the Gaussian distribution of the length rescalings, the uncertainty
of this length scale due to the metric fluctuation will be $A_0^2 L^2$. Requiring that this amount
equals the Planck length $\lplanck^2 = \hbar G / c^3$ yields $A_0^2 \approx \hbar G / (L^2 c^3)$.
The pre-factor in \eref{eqn:sanchez-master} then is of the order of $G m^2 / (\hbar L)$.

For a single particle, effects are minuscule (e.\,g.~for $L=1~\micro\meter$ one finds a coherence
over 1~\centi\meter~for more than $10^{30}~\second$). For a macroscopic system, however, the pre-factor
$G m^2 / L^3$ in \eref{eqn:sanchez-master} becomes $G m \rho$, and can provide a significant amount
of decoherence. Fluctuations that are caused by astrophysical gravitational waves rather than quantum
gravity can of course become arbitrarily large, in principle, depending on their source, and there is
no a-priori reason for them to be related to the Planck length in any way.

The fluctuations considered above are isotropic and can be thought of as non-propagating.
Contrary to these non-propagating fluctuations, Power and Percival~\cite{Power:2000}
consider conformal waves that are directed in positive and negative $x$-direction, traveling
with velocity $c$.
For this purpose, they split the wave $A(x)$ in the parts propagating in those two directions:
\begin{equation}
A(t,x) = A_0 \left( \xi^+(ct-x) + \xi^-(ct+x) \right) \,,
\end{equation}
where $A_0$ is the amplitude and $\xi^\pm$ are the fluctuations for the two propagation directions.

The average values must vanish as before, specifically $\erw{\xi^\pm} = 0$. The first order correlation
is $\erw{\xi^+(x)\xi^-(x')} = 0$ and $\erw{\xi^+(x)\xi^+(x')} = \erw{\xi^-(x)\xi^-(x')} = g^{(1)}(x-x')$.
Power and Percival~\cite{Power:2000} use a Dyson expansion in order to determine the time evolution of
the stochastic operator:
\begin{eqnarray}\label{eqn:dyson-expansion}
\fl
\rho(t) = \rho(0) - \frac{\rmi}{\hbar} \int_0^t \rmd t' \, \left(\erw{H(t')}\rho(0) + \rho(0)\erw{H^\dagger(t')}\right) \nnl
-\frac{1}{\hbar^2} \int_0^t \rmd t' \int_0^{t'} \rmd t'' \, \Big(\erw{H(t')H(t'')}\rho(0) 
+ \erw{H(t')\rho(0)H^\dagger(t'')} \nnl
+ \rho(0)\erw{H^\dagger(t')H^\dagger(t'')}\Big) + \dots \,,
\end{eqnarray}
where $H$ is the Hamiltonian describing the dynamics of the conformal field $A$.
It turns out that the first order terms cancel completely, as one would expect, as they only yield
an overall phase for the wave function.

Without further specification of the function $g^{(1)}$, the evolution of the density matrix is found to
follow the equation
\begin{eqnarray}\label{eqn:power-second-order-master}
\fl
\rho(x,x';t) = \rho(x,x';0) +\frac{m^2 A_0^4}{\hbar^2} \nnl
\times \Bigg(
\int_0^t \rmd t' \int_0^t \rmd t'' g^{(1)}(ct'-ct''-x+x') g^{(1)}(ct'-ct''+x-x') \rho(x,x';0) \nnl
-2 \int_0^t \rmd t' \int_0^{t'} \rmd t'' \, \left( g^{(1)}(ct'-ct'') \right)^2 \rho(x,x';0) \Bigg)
+ \dots
\end{eqnarray}

Note that although the second order terms in \eref{eqn:dyson-expansion} in principle depend on
the second order correlations $g^{(2)}$ as well, only the first order correlation function $g^{(1)}$
appears in \eref{eqn:power-second-order-master}.
For a Gaussian correlation, $g^{(1)}(x) = \exp(-x^2/L^2)$, the second order master
equation~\eref{eqn:power-second-order-master} takes the much simpler form
\begin{equation}\label{eqn:power-master}
\fl
\frac{\partial}{\partial t}\rho(\vec x,\vec x';t) = 
-\frac{\rmi}{\hbar} \com{H_0}{\rho(\vec x,\vec x';t)}
-\sqrt{\frac{\pi}{2}}\frac{m^2 c^3 A_0^4 L}{\hbar^2} \, \left(1-\rme^{-2\frac{(x-x')^2}{L^2}}\right) \, \rho(\vec x,\vec x';t) \,.
\end{equation}
The main differences compared to the non-propagating fluctuations~\eref{eqn:sanchez-master}
are in the exponential dependence on spatial separation rather than the sharp cut-off
in~\eref{eqn:sanchez-master}, as well as in that the field amplitude enters only with the fourth
power instead of second, due to the fact that all first order terms cancel.

Both~\eref{eqn:sanchez-master} and~\eref{eqn:power-master} exhibit the typical structure known
also from collisional decoherence~\cite{Breuer:2002}: a quadratic decay for short distances which goes to zero
(exponentially or incontinuously through the cut-off) for longer distances.
This structure is also found in collapse models, and is discussed in \sref{sec:dp} in context of the Di\'osi--Penrose model.

\subsubsection{Effective \schr\ equation in a fluctuating spacetime}

Going beyond the conformal fluctuations in the previous section and considering generic fluctuations
of the metric, G\"okl\"u and L\"ammerzahl~\cite{Goeklue:2008} derive an effective \schr\ equation by
considering a Klein-Gordon field on a fluctuating spacetime. The scalar field is coupled minimally
to gravity, yielding the usual curved spacetime Klein-Gordon equation
\begin{equation}
g^{\mu\nu} \nabla_\mu \nabla_\nu \phi + \frac{m^2 c^2}{\hbar^2} \phi = 0 \,,
\end{equation}
where $\nabla_\mu$ denotes the covariant derivative. Using the method of Kiefer and
Singh~\cite{Kiefer:1991}, the Klein-Gordon equation is expanded in powers of $1/c^2$,
yielding
\begin{eqnarray}
\fl
\rmi \hbar \partial_t \psi = -\frac{\hbar^2}{2m} \nabla^2 \psi
+\frac{m}{2} \left(\delta_{lm} h_{(0)}^{il} h_{(0)}^{jm} - h_{(0)}^{00} \right) \psi
+\frac{1}{2} \acom{\rmi \hbar \partial_i}{h_{(1)}^{i0}+\delta_{lm}h_{(0)}^{il}h_{(1)}^{m0}} \psi \nnl
-\frac{\rmi \hbar}{4} \partial_t \left(\delta_{ij}h_{(0)}^{ij}
-\frac{1}{2}\delta_{ij}\delta_{lm}h_{(0)}^{il}h_{(0)}^{jm}\right)\psi
+\frac{\rmi \hbar}{4} h_{(1)}^{i0} \partial_i \delta_{ij}h_{(0)}^{ij} \psi \,,
\end{eqnarray}
where $\nabla^2 = \partial_i (\sqrt{g^{(3)}} g^{ij} \partial_j \psi) / \sqrt{g^{(3)}}$ is the Laplace--Beltrami operator in the three-dimensional spatial hypersurface, and $g^{(3)}$ is the determinant of the spatial part of the metric. A simultaneous transformation of both the wave function and the Hamiltonian
according to
\begin{equation}
\psi \to (g^{(3)})^{1/4} \psi \,, \quad\quad
H \to (g^{(3)})^{1/4} H (g^{(3)})^{-1/4} + \frac{\rmi \hbar}{4} \partial_t \ln g^{(3)}
\end{equation}
turns this into a Hermitian ``flat-space form''.
An effective \schr\ equation of the form $\rmi \hbar \partial_t \psi = (H_0 + H_p(t))\psi$ is obtained
after averaging over the spatial fluctuations, where $H_0$
corresponds to the usual, unperturbed evolution, and the term $H_p(t)$ encodes the effect of the time-dependent
metric fluctuations.

Under the assumption that these fluctuations are isotropic, Breuer \etal~\cite{Breuer:2009}
arrive at the master equation
\begin{equation}\label{eqn:breuer-master}
\frac{\partial}{\partial t}\rho(t) = 
-\frac{\rmi}{\hbar} \com{H_0}{\rho(t)}
-\frac{\tau_c}{8 m^2\,\hbar^2} \com{\vec p^2}{\com{\vec p^2}{\rho(t)}} \,,
\end{equation}
where $\tau_c$ is a free parameter corresponding to the magnitude of the random metric fluctuations.
This master equation, which has been previously found by Haba~\cite{Haba:2000},
is quite different from those discussed in the previous section.
Although also being in Lindblad form, it describes localization in energy
rather than position, implying energy conservation: $\frac{\rmd}{\rmd t}\erw{H_0} = 0$.

The origin of these differences remains unexplained, and the large variety of approaches with
seemingly different outcomes makes a further investigation of the question how classical spacetime
fluctuations should be modeled desirable.

\subsection{Quantum spacetime fluctuations}\label{sec:gd-quant}

Although there is no unanimously accepted high-energy quantum theory of gravity, it is often
assumed that the perturbative quantization of general relativity is applicable at low energies.
It is then treated as an effective field theory, the limiting case of the yet unknown full quantum
gravity theory. In this regime, the non-renormalizability of said theory does not pose an immediate
problem. One must, however, keep in mind that this treatment of gravity in analogy to the known
quantum field theories for matter fields is a hypothesis which is, so far, lacking experimental
evidence.

Even in a quantum theory of gravity, general relativity must emerge as a classical limit. In this
regime, the semi-classical Einstein equations
\begin{equation}\label{eqn:sce-in-decoherence-part}
R_{\mu \nu} - \frac{1}{2} g_{\mu \nu} R = \frac{8 \pi G}{c^4} \,
\bra{\Psi} \hat{T}_{\mu \nu} \ket{\Psi}
\end{equation}
are valid, i.\,e.~quantum matter enters as a source of spacetime curvature as an expectation value
of the stress-energy operator. However, since this stress-energy operator is derived from quantized
fields, one must account for quantum fluctuations which lead to corrections
to~\eref{eqn:sce-in-decoherence-part}. A rigorous approach towards modeling these fluctuations and
their consequences is provided by stochastic gravity, a review of which is given in~\cite{Hu:2008}.

\subsubsection{Thermal graviton background}\label{sec:thermalgravitons}

If perturbative quantum gravity is assumed to be correct, decoherence effects arise due to background
gravitons, in a very similar fashion in which photons are a source for decoherence.
Such a decoherence effect due to a thermal spectrum of gravitons has been studied by Blencowe~\cite{Blencowe:2013}.

A scalar field $\phi$ of mass $m$ on a spacetime with the metric $g_{\mu\nu}$ can be described
by the action
\begin{equation}\label{eqn:scalar-field-grav-action}
S = \int \rmd^4 x \, \sqrt{-g} \left(\frac{R}{16\pi\,G}
+ \frac{1}{2} g^{\mu\nu} \partial_\mu \phi \partial_\nu \phi
- \frac{m^2 c^2}{2\hbar^2} \phi^2 \right) \,,
\end{equation}
where $g$ denotes the metric determinant.
The first term is the Einstein-Hilbert action, describing the spacetime dynamics, and the second
and third term describe the dynamics of the field evolving on the curved spacetime.\footnote{Note
that for a scalar field the covariant derivative is simply the partial derivative and, hence,
there is no need for writing a covariant derivative in \eref{eqn:scalar-field-grav-action}.}

In the linear approximation, where $g_{\mu\nu} = \eta_{\mu\nu} + \epsilon\,h_{\mu\nu}$,
the action can then be
decomposed~\cite{Arteaga:2004} into a sum $S=S_\text{matter} + S_\text{grav} + S_\text{int}$
with the flat space Klein-Gordon action
\begin{equation}
S_\text{matter} = \int \rmd^4 x \left(\frac{1}{2} \eta^{\mu\nu} \partial_\mu \phi \partial_\nu \phi
- \frac{m^2 c^2}{2\hbar^2} \phi^2 \right) \,,
\end{equation}
and the graviton action
\begin{eqnarray}
\fl
S_\text{grav} = \frac{\epsilon^2}{32\pi\,G} \, \int \rmd^4 x \Bigg(
- \frac{1}{2} \eta^{\alpha\beta} \partial_\alpha h^{\mu\nu} \partial_\beta h_{\mu\nu}
+ \eta^{\alpha\beta} \partial_\nu h^{\mu\nu} \partial_\alpha h_{\mu\beta}
- \eta^{\alpha\beta} \partial_\mu h \partial_\nu h^{\mu\nu}\nnl
+ \frac{1}{2} \eta^{\mu\nu} \partial_\mu h \partial_\nu h \Bigg) + \order{\epsilon^3} \,,
\end{eqnarray}
where $h^{\mu\nu} = \eta^{\mu\alpha}\eta^{\nu\beta} h_{\alpha\beta}$ and
$h = \eta^{\mu\nu} h_{\mu\nu}$. The interaction term is
\begin{equation}
S_\text{int} = \int \rmd^4 x \left(
\frac{\epsilon}{2} T^{\mu\nu} h_{\mu\nu} + \frac{\epsilon^2}{4} U^{\mu\nu\alpha\beta} h_{\mu\nu}
h_{\alpha\beta} \right) + \order{\epsilon^3} \,.
\end{equation}
The stress energy tensor of the scalar field is
\begin{equation}
T_{\mu\nu} = \partial_\mu \phi \partial_\nu \phi - \frac{1}{2} \eta_{\mu\nu} \eta^{\alpha\beta}
\partial_\alpha \phi \partial_\beta \phi - \frac{m^2 c^2}{2 \hbar^2} \eta_{\mu\nu} \phi^2 \,,
\end{equation}
and the tensor $U$ is given by
\begin{eqnarray}
\fl
U_{\mu\nu\alpha\beta} = -2 \eta_{\nu\alpha}\partial_\mu \phi \partial_\beta \phi
+ \eta_{\mu\nu} \partial_\alpha \phi \partial_\beta \phi \nnl
+\left(\frac{1}{2}\eta_{\mu\alpha}\eta_{\nu\beta} -\frac{1}{4} \eta_{\mu\nu}\eta_{\alpha\beta}\right)
\left(\eta^{\rho\sigma} \partial_\rho \phi \partial_\sigma \phi + \frac{m^2c^2}{\hbar^2} \phi^2 \right)
\,.\end{eqnarray}

Blencowe~\cite{Blencowe:2013} now uses the closed time path integral approach~\cite{Calzetta:2008},
from which the evolution equation for the density matrix follows immediately:
\begin{eqnarray}\label{eqn:blencowe-master}
\fl
\rho_\text{matter}[\phi,\phi';t] = \int \rmd\phi_0 \, \rmd\phi_0'
\int_{\phi_0}^{\phi} \rmd \phi^+ \int_{\phi_0'}^{\phi'} \rmd \phi^-
\rho_\text{matter}[\phi_0,\phi_0';0] \nnl
\times \exp\left(\frac{\rmi}{\hbar} \left(S_\text{matter}[\phi^+]-S_\text{matter}[\phi^-]
+S_\text{FV}[\phi^+,\phi^-]\right)\right) \,,
\end{eqnarray}
where $S_\text{FV}[\phi^+,\phi^-]$ is the Feynman--Vernon influence 
action~\cite{Feynman:1963,Calzetta:2008}.

In the case where the environment action $S_\text{grav}$ is quadratic in the fields $h_{\mu\nu}$,
the initial density matrix of the environment is Gaussian, and the interaction term
$S_\text{int}$ is bilinear, i.\,e.~both $h_{\mu\nu}$ and $\phi$ appear linearly,
the influence action takes a simpler form~\cite{Calzetta:2008,Caldeira:1983}. The explicit
calculation is straightforward but tedious, following chapter 3 of~\cite{Calzetta:2008}.

Blencowe~\cite{Blencowe:2013} uses the Born approximation, i.\,e.~the assumption that
the environment is approximately constant over the time intervals of interest, and obtains the
master equation
\newcommand{\wtt}{\widetilde{T}}
\begin{eqnarray}\label{eqn:blencowe-master-pre}
\fl
\frac{\partial}{\partial t}\rho(t) =
-\frac{\rmi}{\hbar}\, \com{H_\text{matter}}{\rho(t)} \nnl
- \int_0^t \rmd \tau \int \rmd^3 r \, \rmd^3 r' \,\Bigg(
N(\tau,\vec r-\vec r') \, \left(2\,\com{T_{\mu\nu}}{\com{\wtt^{\mu\nu}}{\rho(t)}}
- \com{T_\mu^\mu}{\com{\wtt_\nu^\nu}{\rho(t)}}\right) \nnl
-\rmi \, D(\tau,\vec r - \vec r') \,
\left( 2\,\com{T_{\mu\nu}}{\acom{\wtt^{\mu\nu}}{\rho(t)}}
- \com{T_\mu^\mu}{\acom{\wtt_\nu^\nu}{\rho(t)}} \right)\Bigg) \,,
\end{eqnarray}
where we simply write $\rho \equiv \rho_\text{matter}$ now, and define
$T_{\mu\nu} = T_{\mu\nu}(\tau,\vec r)$, $\widetilde{T}_{\mu\nu} = T_{\mu\nu}(-\tau,\vec r')$.
The expressions $N$ and $D$ are the noise and dissipation kernels~\cite{Blencowe:2013},
respectively
\numparts\begin{eqnarray}
N(t,\vec r) &= \frac{G}{4\,\pi^2} \, \int \rmd^3 k \, \frac{\exp(\rmi\,\vec k \cdot \vec r)}{k}\,
(1+2\,n(k))\,\cos k t\,  \\
D(t,\vec r) &= \frac{G}{4\,\pi^2} \, \int \rmd^3 k \, \frac{\exp(\rmi\,\vec k \cdot \vec r)}{k}\,
\sin k t \,,
\end{eqnarray}\endnumparts
where $n(k)$ is the Bose-Einstein occupation number at a given temperature $T$.
These functions are a consequence of the assumption that the background field is described by
a thermal distribution of gravitons.

In the nonrelativistic limit, the dominant contribution to the stress-energy tensor comes from
the component $T_{00} \approx \frac{m^2 c^2}{2 \hbar^2} \phi^2$, and the master
equation~\eref{eqn:blencowe-master-pre} simplifies further. For decoherence effects,
only the noise term proportional to $N$ needs to be considered, while the dissipation term can be neglected.

Blencowe~\cite{Blencowe:2013} proceeds with the introduction of a basis of coherent states
\begin{equation}
\ket{\alpha} = \exp\left[-\frac{1}{2} \int \rmd^3 k \, \abs{\alpha(\vec k)}^2
+ \int \rmd^3 k \, \alpha(\vec k) \, \hat{a}^\dagger(\vec k) \right] \ket{0} \,,
\end{equation}
with
\begin{equation}
\alpha(\vec k) = \varphi_0 \, R^3 \, \left(\frac{m^2 c^2 + \hbar^2 k^2}{4 \hbar^2}\right)^{1/4} \,
\rme^{-\rmi\,\vec k \cdot \vec r_0 - (kR)^2/2} \,,
\end{equation}
which satisfy
\begin{equation}
\bra{\alpha} \phi(\vec r) \ket{\alpha} \sim
\exp\left(-\frac{(\vec r - \vec r_0)^2}{2R^2}\right) \quad\quad \text{and} \quad\quad
\bra{\alpha} \dot{\phi}(\vec r) \ket{\alpha} = 0 \,,
\end{equation}
and can, therefore, be understood as models for macroscopic material objects in the shape of
``Gaussian matter balls''. 
The radius is supposed to be much larger than the Compton wavelength, $R \gg \lambda_c = h/(mc)$.
One further assumes that these balls have negligible free spreading.
For two such Gaussian matter distributions in a superposition of distinct rest mass energies
$E_1 - E_2 = \Delta E$ one then finds a decoherence rate
\begin{equation}
\Gamma_\text{dec} = \frac{k_B T}{\hbar} \, \left(\frac{\Delta E}{E_P}\right)^2 \,,
\end{equation}
$E_P = \sqrt{\hbar c^5 / G}$ being the Planck energy.

\subsubsection{Generic perturbations}

A more generic approach is taken by Anastopoulos and Hu~\cite{Anastopoulos:2013} who derive a
master equation for gravitational decoherence in a more generic way. 
Starting from the action~\eref{eqn:scalar-field-grav-action} for a scalar field interacting with
the gravitational field, the usual Arnowitt--Deser--Misner (ADM) formalism~\cite{Arnowitt:1959}
is employed in order to perform a 3+1 decomposition
of the action and the metric. An expansion to first order in $\kappa = 16\pi\,G$ yields the perturbations
$h_{ij} = \delta_{ij} + \kappa \gamma_{ij}$ for the spatial metric, $N = 1 + \kappa n$ for the lapse,
and $N^i = \kappa n^i$ for the shift function. A Legendre transformation then results in the Hamiltonian
(in natural units)
\begin{eqnarray} \label{eqn:ana-hu-hamiltonian}
\fl
H
= \int \rmd^3 x \,\Bigg[
\frac{1}{\kappa}\,\left(\Pi^{ij}\Pi_{ij} - \frac{1}{2}\Pi^2 + \kappa^2 V_g \right)
+ \frac{1}{2} \left( \pi^2 + \partial_i \phi \partial^i \phi + m^2 \phi^2 \right) \nnl
-\frac{\kappa}{2} \gamma^{ij} \partial_i \phi \partial_j \phi
+ \frac{\kappa \gamma}{4} (\partial_i \phi \partial^i \phi + m^2 \phi^2 - \pi^2)
+ n \mathcal{H} + n_i \mathcal{P}^i
\Bigg] \,,
\end{eqnarray}
with $\gamma = \delta^{ij} \gamma_{ij}$. $\Pi^{ij}$ are the conjugate momenta of the spatial perturbation
$\gamma^{ij}$, $\pi$ is the conjugate momentum of the scalar field $\phi$, and the self-gravitational 
potential $V_g$ (containing the second order terms in $\gamma$) is given by
\begin{equation}
V_g = -\frac{1}{2}\partial_k \gamma_{ij} \partial^i \gamma^{kj}
-\frac{1}{2}\partial_k \gamma \partial^k \gamma
+ \partial_i \gamma \partial_k \gamma^{ik}
+ \frac{1}{4} \partial_k \gamma_{ij} \partial^k \gamma^{ij} \,.
\end{equation}
The Hamiltonian and momentum constraints $\mathcal{H}$ and $\mathcal{P}^i$ generate gauge transformations
which correspond to temporal and spatial reparametrizations of the free fields.
It is important to note that such reparametrizations in general relativity are pure gauge.
As pointed out by Anastopoulos and Hu:
\emph{``Any postulate of dynamical or stochastic fluctuations that correspond to space and time
reparametrizations conflicts with the fundamental symmetries of GR''}~\cite{Anastopoulos:2013}.

For the quantization of the Hamiltonian~\eref{eqn:ana-hu-hamiltonian}, Anastopoulos and 
Hu~\cite{Anastopoulos:2013} fix a gauge in which both the longitudinal part of the metric fluctuation
$\gamma^{ij}$ and the transverse trace of the gravitational conjugate momentum $\Pi$ vanish.
This choice ensures that the Lorentz frame introduced by the 3+1 decomposition remains invariant
under reparametrizations of space and time.

Writing the Hamiltonian operator as $\hat{H} = \hat{H}_0 + \kappa \hat{V}_g + \hat{H}_\text{int}$,
the interaction Hamiltonian $\hat{H}_\text{int}$ is expressed in terms of creation and annihilation
operators $\hat{b}^\dagger_\pm(\vec k)$ and $\hat{b}_\pm(\vec k)$ corresponding to the transverse
traceless metric perturbations and their conjugate momenta. The index $\pm$ denotes the two
polarizations, and with the operator $\hat{J}_\pm(\vec k)$, derived from the normal ordered
stress operator, the interaction Hamiltonian can be written as\footnote{See~\cite{Anastopoulos:2013} for the precise definition of $\hat{J}$ and $\hat{b}$.}
\begin{equation}\label{eqn:ana-hu-interaction-hamiltonian}
\fl
\hat{H}_\text{int} = \sum_{\sigma \in \pm} \int \frac{\rmd^3 k}{(2\pi)^3} \left[
\abs{\vec{k}} \hat{b}^\dagger_\sigma(\vec k) \hat{b}_\sigma(\vec k)
- \frac{1}{2}\sqrt{\frac{\kappa}{\abs{\vec{k}}}} \left(
\hat{b}_\sigma(\vec k) \hat{J}^\dagger_\sigma(\vec k)
+ \hat{b}^\dagger_\sigma(\vec k) \hat{J}_\sigma(\vec k) \right) \right] \,.
\end{equation}
This Hamiltonian models the gravitational environment which consists of harmonic oscillators
coupled to the matter degrees of freedom encoded in $\hat{J}$.

The next step is to derive a master equation for the matter degrees of freedom.
For this purpose, one starts with the assumption that the initial state factorizes in matter
and gravitational degrees of freedom, i.\,e.~$\rho = \rho_\text{mat} \otimes \rho_\text{grav}$,
and traces out the gravitational part. The only issue is the choice of the initial state
$\rho_\text{grav}$ of the gravitational field. Assuming that it is Gaussian at a ``temperature''
$\Theta$, the Hamiltonian~\eref{eqn:ana-hu-interaction-hamiltonian} straightforwardly yields
the master equation\footnote{Note that the derivation of this master equation to second order in 
$\sqrt{\kappa}$ does not make use of the Born--Markov approximation.}
\numparts\label{eqn:ana-hu-master}\begin{eqnarray}
\fl
\frac{\partial}{\partial t} \rho_\text{mat}(t)
= -\rmi \com{\hat{H}_0+\frac{\kappa}{2}\hat{V}}{\rho_\text{mat}} \nnl
- \sum_{\sigma \in \pm} \int \frac{\rmd^3 k}{(2\pi)^3} \frac{\kappa}{4 \abs{\vec{k}}} \Bigg[
\left( \coth\left(\frac{\abs{\vec k}}{2\Theta}\right) + 1 \right)
\com{\hat{J}^\dagger_\sigma(\vec k)}{\com{\hat{K}_\sigma(\vec k)}{\rho(t)}} \nnl
+ \left( \coth\left(\frac{\abs{\vec k}}{2\Theta}\right) - 1 \right)
\com{\hat{J}_\sigma(\vec k)}{\com{\hat{K}^\dagger_\sigma(\vec k)}{\rho(t)}} \Bigg]
\end{eqnarray}
with
\begin{equation}
\hat{K}_\sigma(\vec k) = \int_0^\infty \rmd s \, \rme^{-\rmi \abs{\vec k} s}
\rme^{-\rmi \hat{H}_0 s} \hat{J}_\sigma(\vec k) \rme^{\rmi \hat{H}_0 s} \,.
\end{equation}\endnumparts
In the limit of a nonrelativistic particle, restricted to one dimension, the momentum space
master equation simplifies considerably, yielding
\begin{equation}\label{eqn:ana-hu-master-nonrel}
\fl
\frac{\partial}{\partial t} \rho(\vec p,\vec p';t) = -\frac{\rmi}{2\,\hbar \,m}
\com{p^2}{\rho(\vec p,\vec p';t)}
- \frac{4 \pi\, G\, \Theta}{9 \,\hbar^2\,m^2\,c^5} \com{p^2}{\com{p^2}{\rho(\vec p,\vec p';t)}} \,,
\end{equation}
and the decoherence rate
\begin{equation}
\Gamma_\text{dec} = \frac{G\, \Theta}{\hbar^2\,m^2\,c^5} \, \erw{p}^2 \, \Delta p^2
\end{equation}
if $\erw{p}$ is the momentum expectation value and $\Delta p$ the variance of a momentum superposition state.
As the decoherence described by \eref{eqn:ana-hu-master-nonrel} is due to a difference in
\emph{energy} rather than in momentum, a state peaked at $+p$ and $-p$ will not decohere.
This is a feature this model has in common with the previously discussed decoherence effect
due to thermal gravitons~\cite{Blencowe:2013} discussed above in \sref{sec:thermalgravitons}.

A remark is necessary concerning the interpretation of the ``temperature'' variable $\Theta$.
As Anastopoulos and Hu~\cite{Anastopoulos:2013} point out, an initial state $\rho_\text{grav}$
as a thermal state at a certain temperature makes sense in the case where metric perturbations are of
a \emph{classical} nature. For fluctuations that are \emph{inherently quantum}, the natural choice for
the initial state is the quantum field theoretical vacuum state of the graviton field. As gravitons interact only weakly, their
thermalization cannot be taken for granted. In this sense, the parameter $\Theta$ is to be understood
as interpolating between classical and quantum gravitational fluctuations.
The previously discussed model of thermal gravitons~\cite{Blencowe:2013} should also be seen in this
light.

A very similar approach, without the restriction to a scalar field, has been followed
earlier by Haba and Kleinert~\cite{Haba:2002} and was re-derived by Oniga and Wang~\cite{Oniga:2016,Oniga:2016a}. They arrive at a rather generic master equation for 
gravitational decoherence under minimal assumptions which has a structure similar to~\eref{eqn:ana-hu-master}.

\subsubsection{Minimal length}\label{sec:min-length}

The previously described decoherence effect due to a thermal graviton background only
requires the assumption that gravity is quantized at low energies in a similar fashion as,
e.\,g., electrodynamics. Another feature which is usually associated with quantum theories of
gravity at high energies is the existence of a minimal length scale, which is often assumed to
be of the order of the Planck length $\lplanck$.

Power and Percival~\cite{Power:2000} make use of this idea in order to arrive at a master equation
which depends only on this minimal length scale as a free parameter. For the discussion of this idea,
we recall the master equation~\eref{eqn:power-master} in \sref{sec:conformal-perturbations}, which resulted from considering conformal
fluctuations of the (classical) spacetime metric. There, the master equation contained two free parameters: the amplitude $A_0$ of the conformal fluctuation, as well as the coherence length scale $L$.
Relating this to quantized gravity,
Power and Percival~\cite{Power:2000} estimate the amplitude of conformal quantum fluctuations to
be of the order of $A_0 \approx (\lplanck/l_\text{cut-off})^2$, where 
the proposed minimal length scale is used as a cut-off length $l_\text{cut-off}$.
This result can be directly inserted into the master equation~\eref{eqn:power-master} obtained for the classical conformal fluctuation. The coherence length scale is assumed to coincide with the cut-off length, i.\,e.~$L=l_\text{cut-off}$, resulting in a prediction for the
evolution of the density matrix which has only the cut-off scale as a free parameter:
\begin{equation}\label{eqn:power-master-2}
\fl
\frac{\partial}{\partial t}\rho(\vec x,\vec x';t) = 
-\frac{\rmi}{\hbar} \com{H_0}{\rho(\vec x,\vec x';t)}
-\sqrt{\frac{\pi}{2}}\frac{m^2 \hbar^2 G^4}{c^9 l_\text{cut-off}^7} \, \left(1-\rme^{-2\frac{(x-x')^2}{l_\text{cut-off}^2}}\right) \, \rho(\vec x,\vec x';t) \,.
\end{equation}
Note that, as in \sref{sec:conformal-perturbations}, it is assumed that fluctuations travel at the speed of light.

Wang \etal~\cite{Wang:2006} provide a discussion which is based on the model by
Power and Percival~\cite{Power:2000}. They add further degrees of freedom to the gravitational
field allowing also for shearing actions. For this purpose, the metric is rewritten as
$g_{\mu\nu} = (1+A)^2\,\gamma_{\mu\nu}$, where $A$ is the conformal field known from the previous
discussion, but the rescaled metric $\gamma_{\mu\nu}$ takes the place of the Minkowski metric.
Its time components are the same as in the Minkowski case, $\gamma_{00} = 1$ and $\gamma_{0i} = 0$,
while the spatial components $\gamma_{ij}$ vary and are normalized such that $\det(\gamma_{ij})=-1$.
In contrast to the conformal approach, this ansatz does not violate Einstein's equations.

Instead of the free Klein-Gordon equation $\eta^{\mu\nu}\partial_\mu\partial_\nu A(x) = 0$,
the dynamics of the field $A$ then are given by the conformal constraint Hamiltonian~\cite{Wang:2006}
\begin{equation}\label{eqn:wang-conformal-hamiltonian}
H^{(\text{CF})} = -\frac{3c^2}{8\pi\,G} \left( \dot{A}^2 + c^2\,\gamma^{ij} \frac{\partial}{\partial x_i} A
\frac{\partial}{\partial x_j} A \right) \,.
\end{equation}
This is almost the Hamiltonian for a massless, scalar Klein-Gordon field, however belonging to
negative energy due to the positive sign between temporal and spatial derivatives.
The total Hamiltonian can be re-written as a sum of the conformal Hamiltonian and a second
term describing gravitational waves: $H = H^{(\text{CF})} + H^{(\text{GW})}$.

Assuming that quantum fluctuations lead to a gravitational background for times shorter than
$\tau_0 = \lambda \tplanck$, i.\,e.~a multiple of the Planck time defined by the factor $\lambda$,
and modeling these fluctuations as standing waves in a box of unit volume
with a frequency cut-off at $\omega_0 = 2\pi/\tau_0$, one can calculate the zero point energy
\begin{equation}
H^{(\text{GW})}_0 \approx \frac{2\pi^2 \, c^2\, \mplanck}{\lambda^4\,\lplanck^3} \,,
\end{equation}
which would be of the order of $\lambda^{-4} \times 10^{98}~\kilo\gram\per\meter\cubed$.
However, Wang \etal~\cite{Wang:2006} point out that the conformal field gives a contribution
of equal magnitude and opposite sign: $H^{(\text{CF})} = -H^{(\text{GW})}_0$. The interpretation they provide is
that every pair of zero point energy gravitons of opposite helicities carries the energy
$\hbar \omega/2$ but is accompanied by a quantum of the conformal field with \emph{negative}
energy $-\hbar\omega$. The gravitational wave terms, being of first post-Newtonian order,
give no contribution in the Newtonian limit. The conformal field, on the other hand, leads to a
Newtonian gravitational potential and decoherence effects as discussed in
\sref{sec:conformal-perturbations}. From~\eref{eqn:wang-conformal-hamiltonian}
together with the condition $H^{(\text{CF})} = -H^{(\text{GW})}_0$ one can show that the amplitude of the conformal
field is $A_0^2 = 4\pi/(3\lambda^2)$. This can be inserted into~\eref{eqn:power-master}
and yields a lower bound on $\lambda$ by comparison with experimental observations, where the
dependence on experimental parameters is $\lambda \sim m^{2/3} \, (\delta \rho/\rho_0)^{-1}$.
From atom interferometric data~\cite{Peters:1997} one concludes that $\lambda \geq 7600$.
Under the assumption that this model is an accurate description of reality, this implies that
the scale where quantum gravitational effects become relevant cannot be lower than about four
orders of magnitude above the Planck scale, and refined experiments would further increase this
value.

Garay~\cite{Garay:1998,Garay:1999} provides a model for decoherence based on the assumption of a
spacetime foam~\cite{Carlip:1997}.
The spacetime foam leaves an imprint also at low energies as a consequence of
assumed nonlocal interactions. Specifically, given a basis $\{h_i[t]\}_i$ of the local,
gauge-invariant interactions, the dominant contribution in the Euclidean action comes from bilocal
interactions
\begin{equation}\label{eqn:garay-interaction}
I_\text{int} = \frac{1}{2} \int \rmd t \, \rmd t' \, c^{ij}(t-t')\,h_i[t]\,h_j[t'] \,.
\end{equation}
Garay considers three separate scales which are assumed to separate:
the Planck scale $\lplanck$, the scale $l$ of low energy physics, and the scale $r$ beyond which gravitational
fluctuations are of relevance.
The parameter $\epsilon \sim \rme^{-(r/\lplanck)^2} (r/\lplanck)^4 (l/\lplanck)^{-2}$ is introduced, which suppresses
both gravitational fluctuations ($r \gg \lplanck$) and low energy scales ($l \gg \lplanck$) far from the Planck
scale. It can be shown that the interactions for a scalar field $\phi$ are of the
order $h_i[t] \sim \phi^2/\lplanck^2$ and the bilocal interaction term is of the order
$I_\text{int} \sim \rme^{-(r/\lplanck)^2} (r/\lplanck)^4$, while further terms are of higher order in 
both $\epsilon$ and $r/l$.

Assuming that the correlation function $c^{ij}(t-t')$ is Gaussian, after a Wick rotation back into
the Lorentz frame and making use of the influence functional formalism
(cf.~\eref{eqn:blencowe-master}) the resulting master equation at lowest order is
\begin{equation}\label{eqn:garay-master}
\frac{\partial}{\partial t}\rho(t) = -\frac{\rmi}{\hbar} \com{H_0}{\rho}
- \int_0^\infty \rmd \tau \, c^{ij}(\tau) \, \com{h_i}{\com{h_j}{\rho}} \,
\end{equation}
where the second term describes the effect of the fluctuations while the first term is the usual
Hamiltonian evolution. Dissipation terms are absent from~\eref{eqn:garay-master}, as they
are of higher order in $r/l$.
The correlation functions $c^{ij}$ are not predicted by this model but are associated with a thermal
bath, quite similar to the model by Blencowe~\cite{Blencowe:2013}.
The reasoning behind this approach comes from the comparison of the master
equation~\eref{eqn:garay-master} with the equations for quantum Brownian motion~\cite{Erdos:2010}.

A critical view at this result~\cite{Anastopoulos:2007} raises the question whether the performed
separation of scales is physically reasonable, given that in general relativity the local structure
of spacetime is strongly linked with the locality of interactions. It is also pointed out that
the nonlocal interaction~\eref{eqn:garay-interaction} is, at the level of the resulting master
equation, indistinguishable from the effect of local additional phenomenological fields.
The Born--Markov approximation may not be a good assumption for the dynamics induced by a spacetime
foam. If the approximation does not hold, the resulting master equation could be of a vastly different
shape.
Similar criticisms~\cite{Anastopoulos:2007} apply to a related derivation~\cite{Mavromatos:2006}
of decoherence in the flavor sector for neutrinos due to a spacetime foam motivated by string theory.

\subsubsection{Time fluctuations}

A different approach to introduce quantum fluctuations of spacetime in quantum mechanics is through
the effect of such fluctuations on the time parameter in the \schr\ equation.

Intuitively speaking, a free quantum system evolves along a geodesic with the proper time $\tau$
given by the metric, for instance
\begin{equation}
\ket{\psi} \to \rme^{\rmi \omega \tau} \ket{\psi} \,,
\end{equation}
with some frequency $\omega$.
If the metric is considered a quantum object itself, which is in a quantum state
$\ket{g_{\mu\nu}}$, then one expects the quantum state of the system to become entangled with the state
of the metric: $\ket{\psi}\otimes\ket{g_{\mu\nu}}$. Specifically, if one allows for linear
superposition states of metric fluctuations, i.\,e.~$\frac{1}{2}(\ket{g_{\mu\nu}}+\ket{g'_{\mu\nu}})$,
then the system state $\ket{\psi}$ becomes entangled to the metric state and evolves according to
\begin{equation}
\ket{\psi}\otimes\frac{1}{2}\left(\ket{g_{\mu\nu}}+\ket{g'_{\mu\nu}}\right)
\to \frac{1}{2}\left(\rme^{\rmi \omega \tau}\ket{\psi}\otimes\ket{g_{\mu\nu}}
+\rme^{\rmi \omega \tau'}\ket{\psi}\otimes\ket{g'_{\mu\nu}}\right) \,,
\end{equation}
where $\tau$ and $\tau'$ are the proper times corresponding to the metrics $g_{\mu\nu}$ and
$g'_{\mu\nu}$, respectively. The fluctuations of the metric propagate to the system state and
lead to a loss of coherence. Kok and Yortsever discuss decoherence on the grounds of these
considerations~\cite{Kok:2003}.

A slightly more sophisticated model by Milburn~\cite{Milburn:1991} treats the dynamics of quantum
states as probabilistic, finite jumps rather than a continuous unitary evolution. This is put
on more rigorous mathematical grounds, motivated from time as a statistical variable, by
Bonifacio~\cite{Bonifacio:1999}, however with essentially identical physical implications.

Instead of the continuous evolution of the density matrix,
\begin{equation}\label{eqn:milburn-usual-master-eq}
\rho(t+\tau) = \exp\left(-\frac{\rmi}{\hbar} \, H \, \tau\right) \rho(t)
\exp\left(\frac{\rmi}{\hbar} \, H \, \tau\right) \,,
\end{equation}
one assumes that a change of the system occurs with a certain probability, specifically that
there is a probability $p_n(\tau_0,\tau)$ that $n$ random time shifts by $\tau_0$ lead to a total
time shift of $\tau$. The density matrix then satisfies an equation
\begin{equation}\label{eqn:milburn-final-master}
\rho(t+\tau) = \sum_{n=0}^\infty p_n(\tau_0,\tau) \,
\exp\left(-\frac{\rmi}{\hbar} \, H \, n\,\tau_0\right) \rho(t)
\exp\left(\frac{\rmi}{\hbar} \, H \, n\,\tau_0\right) \,.
\end{equation}
In the limit $\tau_0 \to 0$ this yields the continuous equation~\eref{eqn:milburn-usual-master-eq}.
For the specific choice of a Poisson distribution,
\begin{equation}
p_n(\tau_0,\tau) = \frac{\tau^n}{\tau_0^n\,n!}\,\rme^{-\tau/\tau_0} \,,
\end{equation}
the master equation takes the form
\begin{equation}
\frac{\partial}{\partial t} \rho(t) =
\frac{1}{\tau_0}\,\left(\exp\left(-\frac{\rmi}{\hbar} \, H \, \tau_0\right) \rho(t)
\exp\left(\frac{\rmi}{\hbar} \, H \, \tau_0\right) - \rho(t)\right) \,,
\end{equation}
or after expansion in the, presumably small, parameter $\tau_0$
which indicates the time scale where deviations from continuous evolution become apparent:
\begin{equation}
\frac{\partial}{\partial t} \rho(t) =
-\frac{\rmi}{\hbar} \com{H}{\rho(t)} - \frac{\tau_0}{2\hbar^2} \com{H}{\com{H}{\rho(t)}} \,.
\end{equation}

As pointed out by Anastopoulos and Hu~\cite{Anastopoulos:2008}, these considerations are in fact
more general, and the semigroup equation~\eref{eqn:milburn-final-master} follows quite generally
for any physical quantity and its corresponding generator. For example, the above results remain
valid if time is replaced by position and the time translation generator $H$ by the momentum $p$.
They criticize that in Milburn's formalism there is no requirement to relate the incremental
random time shifts with gravity or the Planck scale. This association is a mere intuitive hypothesis,
and considering that many quantum effects are only poorly modeled by classical statistics, not
necessarily a very compelling one.

Note also that, as Milburn points out himself~\cite{Milburn:1991}, Lorentz invariance requires the
random time shifts to be accompanied by a similar spatial structure, because the temporal jumps
will appear as position fluctuations in another reference frame.

\subsection{Decoherence due to gravitational time dilation}\label{sec:gd-time}
Decoherence can occur even in a classical and completely static spacetime, where the gravitational
field itself exhibits no quantum or stochastic fluctuations.
Instead, this type of decoherence effect stems from a coupling of internal and external degrees of
freedom in a complex quantum system, which can be triggered by gravity.

This has been discussed as a consequence of time dilation in a quite intuitive fashion by
Pikovski \etal~\cite{Pikovski:2015}, based on earlier work by the same authors~\cite{Zych:2011}
on the interference of clocks; their results have since been the starting point of an ongoing controversy~\cite{Bonder:2016,Bonder:2015a,Diosi:2015,Adler:2016,Zeh:2015,Carlesso:2016a,Pang:2016,Pikovski:2017,Toros:2017,Pikovski:2017a,Toros:2017a}.

A ``clock'' in this context refers to any system which can be modeled as point-like (i.\,e.~with a
single center-of-mass coordinate) and contains some oscillatory internal degrees of freedom.
Quantum mechanically, such a ``clock'' is typically described by a product state
\begin{equation}
\ket{\Psi} = \ket{\psi_\text{cm}} \otimes \ket{\Psi_\text{clock}} \,,
\end{equation}
with $\psi_\text{cm}$ the center-of-mass position wave function and $\Psi_\text{clock}$ the internal clock
state. In nonrelativistic settings, the clock undergoes a dynamical evolution described by a Hamiltonian 
$H = H_\text{cm} + H_\text{clock}$. Hence, the evolution strictly separates into the center-of-mass motion
and the internal dynamics (``ticking'') of the clock:
\begin{equation}
\ket{\Psi(t)} = \rme^{-\frac{\rmi}{\hbar}\,H\,t} \ket{\Psi}_0
= \rme^{-\frac{\rmi}{\hbar}\,H_\text{cm}\,t} \ket{\psi_\text{cm}}_0 \otimes
\rme^{-\frac{\rmi}{\hbar}\,H_\text{clock}\,t} \ket{\Psi_\text{clock}}_0 \,.
\end{equation}
In relativistic contexts, however, this strict separation does no longer hold.

The proper time measured by a clock between two spacetime events $A$ and $B$ depends on the
spacetime trajectory the clock takes from $A$ to $B$. In special relativity, this is of course known
as the ``twin paradox'', the reason for which is special relativistic time dilation, i.\,e.~the fact
that moving clocks tick at a slower rate. On the other hand, general relativity predicts that
clocks in a stronger gravitational field tick at a slower rate (and, therefore, clocks on Earth tick
the faster the higher their altitude).

Two clocks moving from $A$ to $B$ on \emph{different} spacetime trajectories will, consequently,
show a proper time difference $\Delta \tau$. For a quantum superposition of two clocks---assuming
that the linear superposition principle for wave functions remains valid in this relativistic
context---one expects constructive interference if $\Delta \tau$ is an integer multiple of both the 
period $E/\hbar$ of the clocks center-of-mass matter-wave and the oscillation period of the clock.
The first condition is simply the equivalent to the nonrelativistic requirement for the optical
path length to be an integer multiple of the wavelength. The second condition poses an additional coherence criterion which is absent
in the nonrelativistic limit.

If the considered system is not a ``clock'', containing one single internal degree of freedom,
but rather a complex system (such as a large molecule) with many degrees of freedom of different
proper frequencies, coherent superpositions become increasingly unlikely for proper time differences
$\Delta \tau \not= 0$, leading to an effective decoherence of wave functions due to the time
dilation effect~\cite{Pikovski:2015}.

Attempting to put this intuitive expectation on rigorous mathematical grounds, one faces several
difficulties. Quantum mechanics, strictly speaking, makes no sense in a relativistic context as
any interacting quantum theory is facing the problem that it must account for particle creation
and annihilation~\cite{Currie:1963}. In principle, relativistic effects should be described in the
language of quantum field theory---and in the gravitational case in terms of quantum field theory
on curved spacetime~\cite{Wald:1994,Fredenhagen:2007}.

In situations where the particle number is approximately conserved, the single particle \schr\ equation can, nonetheless,
be obtained from a relativistic Klein-Gordon field as the $c\to\infty$ limit, either by making use
of the Foldy-Wouthuysen transformation~\cite{Foldy:1950}, or by a WKB like
expansion~\cite{Kiefer:1991,Laemmerzahl:1995}. Extending this derivation to the next order in $1/c^2$ one obtains an $\order{1/c^2}$-Hamiltonian for a single particle which includes relativistic
corrections. In the special relativistic case, these corrections are well known, for instance from
the theory of the hydrogen atom.
In a homogeneous gravitation potential the same method yields the Hamiltonian\footnote{The
Hamiltonian given here is the one following 
for a Rindler spacetime, while L\"ammerzahl~\cite{Laemmerzahl:1995} derives the Hamiltonian in
a Schwarzschild spacetime. The differences are discussed in~\cite{Toros:2017}.}
\begin{equation}
H_\text{one-particle} = \frac{p^2}{2m} + \frac{p^4}{8m^3c^2}
- g \, m \,  z + \frac{g}{4 m c^2} \acom{z}{p^2} \,.
\end{equation}
In nonrelativistic physics one would assume the same single particle Hamiltonian to also describe
the center of mass of a system with internal degrees of freedom. In the context of relativity,
however, the internal energy contributes to the (inertial and gravitational) mass of the system.
The rest mass energy $m c^2$ must then be replaced by the full relativistic energy
$m c^2 + H_\text{intern}$ of the system, which contains the Hamiltonian for the internal
degrees of freedom. The full Hamiltonian to $\order{1/c^2}$ reads:
\begin{eqnarray}
\fl
H = \frac{p^2 c^2}{2(mc^2+H_\text{intern})} + \frac{p^4 c^4}{8 (mc^2+H_\text{intern})^3} \nnl
-\frac{g}{c^2}(mc^2+H_\text{intern})  + \frac{g}{4 m c^2} \acom{z}{p^2} + H_\text{intern} \nnl
\fl \phantom{H}
= \frac{p^2}{2m} - g \, m \, z + H_\text{intern} + \frac{p^4}{8 m^3 c^2} + \frac{g}{4 m c^2} \acom{z}{p^2}
- \frac{H_\text{intern}}{mc^2}\, \left( \frac{p^2}{2 m} + g \, m \, z \right) \,.
\label{eqn:cm-int-coup-full-hamiltonian}
\end{eqnarray}
The nonrelativistic part of this Hamiltonian strictly separates into center of mass and internal
motion, as expected. The relativistic $\order{1/c^2}$ corrections, on the other hand, contain
a special relativistic coupling of $H_\text{intern}$ to the center-of-mass energy, as well as
a gravitational coupling term $-g\,z\,H_\text{intern}/c^2$. This coupling is
the starting point for the derivation of the gravitationally induced decoherence
effect~\cite{Pikovski:2015}.

A remark is at order concerning the meaning of the Hamiltonian~\eref{eqn:cm-int-coup-full-hamiltonian}:
It is well known that the notion of center-of-mass coordinates is ambiguous
in relativistic contexts~\cite{Pryce:1948}. Specifically, a definition of center-of-mass coordinates
which satisfy canonical commutation relations, and at the same time transform covariantly under
Lorentz transformations does not exist. One consequence is that any choice of center-of-mass
coordinates that allows for a straightforward canonical quantization is frame dependent.
Such a definition of center-of-mass coordinates based on the form of the Poincar\'e group
generators has been introduced by Krajcik and Foldy~\cite{Krajcik:1974} for Minkowski space,
and recently has been extended to the situation of an accelerated observer (or a homogeneous
gravitational field)~\cite{Toros:2017}. The result is indeed the
Hamiltonian~\eref{eqn:cm-int-coup-full-hamiltonian}.

Once the existence of the coupling term $H_\text{coup} = -g\,z\,H_\text{intern}/c^2$ between center of
mass and the internal degrees of freedom is accepted, the quantum mechanical description is quite
straightforward, and follows the usual prescription of open quantum systems.
The full Hamiltonian can then be written as $H = H_\text{cm} + H_\text{intern} + H_\text{coup}$,
where the center of mass plays the role of the system and the internal degrees of freedom act as a
bath. In the Born--Markov approximation, after tracing out the bath degrees of freedom, the master equation
for the system is obtained as~\cite{Pikovski:2015}
\begin{eqnarray}\label{eqn:pikovski-full-master}
\fl
\frac{\partial}{\partial t} \rho_\text{cm}(t) = -\frac{\rmi}{\hbar}\,
\com{H_\text{cm} + \frac{g}{c^2}\,z\,\erw{H_\text{coup}}}{\rho_\text{cm}(t)}
-\frac{\Delta E^2}{\hbar^2 c^4}\nnl \times
\int_0^t \rmd t' \, \com{\left(g\,z+\frac{p^2}{2m^2}\right)}{\com{\rme^{-\frac{\rmi}{\hbar}\,H_\text{cm}\,t'}\,\left(g\,z+\frac{p^2}{2m^2}\right)\,
\rme^{\frac{\rmi}{\hbar}\,H_\text{cm}\,t'}}{\rho_\text{cm}(t)}} \,,
\end{eqnarray}
with $\Delta E^2 = \erw{H_\text{coup}^2}-\erw{H_\text{coup}}^2$.
Assuming that the center-of-mass motion is slow compared to decoherence, momentum terms can be
neglected, and only the linear term in $t$ is considered, which leads to
\begin{equation}
\fl
\frac{\partial}{\partial t} \rho_\text{cm}(t) = -\frac{\rmi}{\hbar}\,
\com{H_\text{cm} + \frac{g}{c^2}\,z\,\erw{H_\text{coup}}}{\rho_\text{cm}(t)}
-\frac{\Delta E^2\,g^2}{\hbar^2 c^4}\,t\, \com{z}{\com{z}{\rho_\text{cm}(t)}} \,,
\end{equation}
and a decoherence rate
\begin{equation}\label{eqn:pikovski-dec-rate}
\Gamma_\text{dec} = \frac{\Delta E \, g \, \Delta z}{\sqrt{2}\,\hbar\,c^2} \,,
\end{equation}
which depends on both the variance of the internal energy and the variance in the width of the
center-of-mass wave-function.

It is also interesting to assess to which extent this gravitationally induced decoherence effect
can be detected in the lab. \Eref{eqn:pikovski-dec-rate} tells that one needs to prepare a system
with a large energy spread, and largely delocalized in space. A vacuum chamber is needed in order to
decrease decoherence from gas particles, and a refrigerator in order to decrease thermal decoherence.
An analysis performed in~\cite{Carlesso:2016a} shows that current technology cannot sufficiently 
suppress competing decoherence effects in order to allow for a detection of the time dilation induced
decoherence; new schemes must be designed.

In addition to the question whether or not this decoherence effect is observable in practice,
its compatibility with the equivalence principle has been questioned~\cite{Bonder:2015a,Bonder:2016,Diosi:2015}. In the case of a free falling system,
there is no contradiction, as long as one accepts that an accelerated observer can see effects
which are absent in inertial frames---a fact that is well known, for instance, in the case of
the Unruh effect~\cite{Fulling:1973,Davies:1975,Unruh:1976}. The equivalence principle simply
states that for a free falling particle both an accelerated observer (in empty space) and an
observer at constant altitude on Earth (``earthbound'') agree on their observations. It does
not require these observations to be identical with the expectation in an inertial frame.

The requirement of a stationary system (in order to disregard the momentum terms
in~\eref{eqn:pikovski-full-master}), however, implies that not only the observer is earthbound
but the system is as well. Then the full Hamiltonian must also include some external potential
which keeps the system from falling and which, for a consistent description, must \emph{include}
relativistic corrections up to $\order{1/c^2}$. A-priori, there is no reason to exclude the
possibility that these correction terms contribute to the coupling of internal degrees of freedom
and the center of mass, and thus can overshadow or even cancel the gravitational
coupling. Whether or not the decoherence effect also exists in a system that is held at constant height by an external potential is still subject of a debate~\cite{Toros:2017,Pikovski:2017a,Toros:2017a}.

\section{Spontaneous wave function collapse and the role of gravity}
\label{sec:oxfc}

In the previous section we discussed decoherence effects induced by gravity, within the standard
framework of quantum physics. In this section, we will review proposals which attempt to modify
quantum theory, motivated from gravity.

The reason for such modifications is that, despite the success of Quantum Mechanics in explaining the structure of matter, the debate about its meaning and limits of validity never ceased. The main reason of concern lies in a straightforward consequence of the linearity of the \schr\ equation, which is the superposition principle. It implies that superpositions of matter states are possible, and this is what one actually observes in experiments with atoms and molecules~\cite{hornberger2012colloquium, arndt2014testing, cronin2009optics}. More than this, multi-particle superpositions, namely entangled states, show how different quantum physics can be from classical physics, in the form of non-locality~\cite{Bell:1964,Bell:2004, Aspect:1982}. They are also the basis for the new quantum technologies~\cite{Dowling:2003}. 

Quantum Mechanics contains no inner limit of validity, and it seems natural, until experiments prove the contrary, to extend it also to the description of the macroscopic domain of ordinary objects. In the end, macroscopic objects are made of atoms, which are quantum. But this poses a problem, first  addressed by Schr\"odinger himself~\cite{Schroedinger:1935}: situations can occur, according to the \schr\ equation, where {\it macroscopic} quantum superpositions of matter can be created, in plain contradiction with our daily experience.  Therefore, a straightforward application of the \schr\ equation to macroscopic object seems inconsistent.  

The standard formulation of Quantum Mechanics provides a tentative resolution of the conflict by assuming the collapse of the wave function at the end of a measurement~\cite{Neumann:1955}. But clearly this is a phenomenological recipe, which cannot be accepted at the fundamental level.  Weinberg among many, wrote: \textit{``The Copenhagen interpretation assumes a mysterious division between the microscopic world governed by quantum mechanics and a macroscopic world of apparatus and observers that obeys classical physics. During measurement the state vector of the microscopic system collapses in a probabilistic way to one of a number of classical states, in a way that is unexplained, and cannot be described by the time-dependent \schr\ equation.''}~\cite{Weinberg:2012}

To cope with this problem, the scientific community provided alternative formulations and theories: Bohmian mechanics~\cite{Broglie:1927, Bohm:1952,Bohm:1952a, Duerr:1992, Holland:1993, Duerr:2009}, many worlds~\cite{Everett:1957,Witt:1970,Wallace:2002} and models of spontaneous wave function collapse (collapse models)~\cite{Ghirardi:1986,Ghirardi:1990a,Bassi:2003,Bassi:2013} are among the most popular. Here we will review the third option, because of its potential relation to gravity.

The fundamental idea behind collapse models is that the \schr\ equation is only approximately valid. It should be supplemented with  additional {\it stochastic} and {\it nonlinear} terms, which account in a dynamical way for the collapse of the wave function in measurement-like situations, instead of assuming it phenomenologically as the Copenhagen interpretation does. Here a first problem arises: how can one consistently modify such a fundamental equation as the \schr\ equation, without running into theoretical and experimental inconsistencies?

After decades of research~\cite{Gisin:1989,Polchinski:1991,Wiseman:2001,Adler:2004,Bassi:2013b}, the situation seems now rather clear~\cite{Adler:2004}. If one asks for i) norm conservation and ii) no superluminal signaling, then  the structure of the modified \schr\ equation is fixed, as we will now show. But first a comment about these two requirements. Norm conservation is needed if one wants to interpret the wave function in a reasonable way as objectively describing the state of a quantum systems. One consequence of its violation would be that mass or number of particles spontaneously decays in time.
The second requirement, that no signal should be able to be sent faster than the speed of light, is needed for consistency with special relativity. A collapse that does not meet this requirement would allow for {\it instantaneous} communication at {\it arbitrarily large} distances~\cite{Gisin:1989},
which is in contradiction to relativistic causality. Note that this instantaneous signaling has nothing to do with the fact that the \schr\ equation is a \emph{nonrelativistic} equation; it is only due to the nonlocality of the collapse.

The  requirement of no superluminal signaling has a precise formulation~\cite{Gisin:1989}: whatever the evolution for the state vector, the dynamics for the density matrix must be {\it linear}. This basically implies that the master equation is of the Lindblad type~\cite{Lindblad:1976,Gorini:1976,Breuer:2002}, with possibly negative coefficients~\cite{Bassi:2013b}. By adding complete positivity, one has the true Lindblad structure with positive coefficients. 

This also implies that when adding nonlinear terms to the \schr\ equation, one has to include  appropriate stochastic terms, which wash away any nonlinearity at the density matrix level. Nonlinear and deterministic evolutions for the state vector unavoidably imply a nonlinear evolution for the density matrix. In other words, nonlinearity and stochasticity have to go hand in hand to have a consistent dynamics for the wave function. This is the basic reason why the structure of consistent nonlinear modifications of the \schr\ equation is uniquely identified.

Let us then consider the modified \schr\ equation, written in the It\^{o} language~\cite{Arnold:1974}: 
\begin{equation}\label{eq:a3}
\rmd\phi_{t}=\left[-\rmi\hat{H}_0\,\rmd t+\sqrt{\lambda}\,\hat{A}\,\rmd W_{t}+\hat{O}\,\rmd t\right]\phi_{t},
\end{equation}
with $\hbar=1$. The term $\sqrt{\lambda}\,\hat{A}\,\rmd W_{t}$ encodes the coupling of the quantum system (represented by the wave function $\phi_{t}$) with the noise $W_t$ (a Wiener process being the first and easiest choice), via the self-adjoint operator $\hat{A}$, on whose eigenstates the state will eventually collapse. The coupling constant $\lambda$ sets the strength of the collapse mechanism. Note the absence of the imaginary unit in front of this term, which makes the dynamics non standard. This is necessary, in order to arrive at a nonlinear equation of the collapse type. The operator $\hat{O}$ will be fixed by the requirement of no superluminal signaling.  

The norm of $\phi_{t}$ is not conserved. Using It\^o calculus, one can easily write down the equation for the normalized vector $\psi_{t}=\phi_{t}/\|\phi_{t}\|$: 
\begin{eqnarray}
\fl
\rmd\psi_{t} = \left[-\rmi\hat{H}_{0} \rmd t+\sqrt{\lambda}(\hat{A}-\langle \hat{A}\rangle_{t})\rmd W_{t}+\lambda\left(\frac{3}{2}\langle \hat{A}\rangle_{t}^{2}-\frac{1}{2}\langle \hat{A}^{2}\rangle_{t}-\hat{A}\langle \hat{A}\rangle_{t}\right)\rmd t\right.\nonumber \\
+\left.\left(\hat{O}-\frac{1}{2}\langle(\hat{O}^{\dagger}+\hat{O})\rangle_{t}\right)\rmd t\right]\psi_{t},
\end{eqnarray}
with $\langle \hat{A}\rangle_{t} = \langle \psi_t|\hat{A}|\psi_t\rangle$. As expected, the normalized vector evolves according to a nonlinear
stochastic dynamics. The stochastic ensemble of pure states $\rho_{t}^{W}=|\psi_{t}\rangle\langle\psi_{t}|$
obeys the following dynamics: 
\begin{eqnarray}
\fl
\rmd\rho_{t}^{W} = -\rmi[H,\rho_{t}^{W}] \nonumber \\
+ \lambda\left(4\langle \hat{A}\rangle_{t}^{2}\rho_{t}^{W}-\langle \hat{A}^{2}\rangle_{t}\rho_{t}^{W} -2\hat{A}\langle \hat{A}\rangle_{t}\rho_{t}^{W}-2\rho_{t}^{W}\hat{A}\langle \hat{A}\rangle_{t}-\hat{A}\rho_{t}^{W}\hat{A}\right)\rmd t \nonumber \\
 +\left(\hat{O}^{\dagger}\rho_{t}^{W}+\rho_{t}^{W}\hat{O}-\langle(\hat{O}^{\dagger}+\hat{O})\rangle_{t}\rho_{t}^{W}\right)\rmd t \nonumber \\
+(\text{extra terms})\,\rmd W_{t}.
\end{eqnarray}
When taking the expectation value to compute the dynamics for the
density matrix $\rho_{t}=\mathbb{E}[\rho_{t}^{W}]$, the `extra terms'
average to 0, while the remaining terms generate a nonlinear evolution
for the ensemble, as already stated. This leads to superluminal signaling, which can be avoided by choosing $\hat{O}=-(\lambda/2)\hat{A}^{2}+2\lambda(\hat{A}-\langle \hat{A}\rangle_{t})\langle \hat{A}\rangle_{t}$,
in which case all nonlinear terms cancel, and the equation for $\rho_{t}$
becomes of the Lindblad type: 
\begin{equation} \label{eq:li}
\frac{\rmd}{\rmd t}\rho_{t}=-\rmi[\hat{H}_{0},\rho_{t}]-\frac{\lambda}{2}[\hat{A},[\hat{A},\rho_{t}]].
\end{equation}
In turn, the equation for the normalized state vector $\psi_{t} $ reads:
\begin{equation}
\label{eq:fghr}
\rmd\psi_{t}=\left[-\rmi\hat{H}_{0}\rmd t+\sqrt{\lambda}(\hat{A}-\langle \hat{A}\rangle_{t}) \rmd W_t-\frac{\lambda}{2}(\hat{A}-\langle \hat{A}\rangle_{t})^{2}\rmd t\right]\psi_{t}\,.\label{eq:1}
\end{equation}
This is the structure of all collapse models. One can also proceed backwards~\cite{Bassi:2013b} and show how, given~\eref{eq:li}, one unravels~\eref{eq:1}. It is not difficult to show that the new terms stochastically drive the state vector toward one of the eigenstates of the operator $\hat{A}$, with a probability equal to the Born rule~\cite{Ghirardi:1990a,Adler:2001}.  The above equation can be generalized to include more than one operator, therefore more than one noise~\cite{Ghirardi:1990a}; non-self-adjoint operators~\cite{Ghirardi:1990a}; more general noises, either complex~\cite{Adler:2001a}, non white~\cite{Bassi:2002,Adler:2007b,Adler:2008}, or both~\cite{Gasbarri:2017}. 

So far, we presented the abstract formulation of collapse models.
A \emph{physical} model of spontaneous wave function collapse must
justify the classical world emerging from an underlying quantum world: wave functions of macro-objects should be well-localized in {\it space}. The collapse should be negligible for microscopic system, and grow in strength with the {\it mass/size} of the system. This suggest taking for the collapse operator the \textit{local mass density} $\hat{m}({\bf x})=\sum_{i}m_{i}\delta^{(3)}({\bf x}-\hat{{\bf x}}_{i})$, coupled to a white noise $w({\bf x},t)$ spread through space. Then~\eref{eq:1} becomes
\begin{eqnarray}
\label{eq:sch}
\fl
\rmd\psi_t = \Bigg[-\frac{\rmi}{\hbar}\hat{H}\,\rmd t +\sqrt{\lambda} \int \! \rmd^3 {x}\, (\hat{m}({\bf x})-\langle \hat{m}({\bf x})\rangle_t)\,\rmd W_t(\mathbf{x}) \nnl
\,- \frac{\lambda}{2}
\int\!\rmd^3 x\!\int\! \rmd^3 y\,{\cal G}({\bf x}-{\bf y})
(\hat{m}({\bf x})-\langle \hat{m}({\bf x}) \rangle_t )
(\hat{m}({\bf y})-\langle \hat{m}({\bf y}) \rangle_t )\,\rmd t
\Bigg]\psi_t,
\end{eqnarray}
(now we have re-introduced $\hbar$) where $W_t(\mathbf{x})$ is a family of Wiener processes, with spatial correlation function equal to ${\cal G}({\bf x}-{\bf y})$. We assume that ${\cal G}({\bf x}-{\bf y}) = {\cal G}(|{\bf x}-{\bf y}|)$, to respect translational and rotational symmetries. This equation was first proposed by Ghirardi, Pearle and Rimini and is the mass-proportional version~\cite{Ghirardi:1990,Ghirardi:1995} of the Continuous Spontaneous Localization (CSL) model~\cite{Ghirardi:1990a}, if one takes a Gaussian correlation function: ${\cal G}({\bf x}-{\bf y}) = \exp(-({\bf x}-{\bf y})^2/4r_C^2)$, where the correlation length $r_C$ (with dimension [L]) is a phenomenological parameter, and $\lambda = \lambda_{\text{\tiny CSL}}/m_0^2$, with $m_0$ a reference mass---equal to the mass of a nucleon, which is added to the definition of lambda to make the model mass-dependent---and $\lambda_{\text{\tiny CSL}}$ (with dimension [T$^{-1}$]) is the collapse rate, the second phenomenological parameter of the model. 

This model has been extensively studied in the literature~\cite{Bassi:2003,Bassi:2013}. Three properties are particularly important: i) The collapse  drives the wave function towards states, which are localized in {\it space}, thus suppressing spatial superpositions. ii) If one considers composite systems such as rigid objects, then the center-of-mass dynamics satisfies a single-particle collapse equation, with an effective collapse rate $\lambda_{\text{\tiny CSL}}^{\text{\tiny CM}}$ which grows with the mass/size of the system ($\lambda_{\text{\tiny CSL}}^{\text{\tiny CM}} = f(N) \lambda_{\text{\tiny CSL}}$, where $f(N)$ is a monotonically increasing function of the total number $N$ of nucleons and of the geometry of the system). Therefore, with a suitable choice of the parameters (the original choice was $\lambda_{\text{\tiny CSL}} = 10^{-16}$ s$^{-1}$ and $r_C = 10^{-7}$ m~\cite{Ghirardi:1986}) one can show that the collapse effects are negligible for microscopic systems, while for composite systems the amplification mechanism makes sure that the center-of-mass wave-function of macroscopic objects is always well-localized in space. iii) Quantum measurements  are consistently described, and their entire phenomenology (definite outcomes, Born probability rule, von Neumann projection postulate, role of the operators as observables) emerges from the dynamics~\cite{Bassi:2007,Bassi:2007a}.

For future reference, we write the master equation for the statistical operator $\rho_{t}=\mathbb{E}[|\psi_t\rangle\langle\psi_t|]$:
\begin{equation}
\label{eq:me}
\fl
\frac{\rmd}{\rmd t}\hat{\rho}_{t} =
-\frac{\rmi}{\hbar}[\hat{H},\hat{\rho}_{t}]  - \frac{\lambda}{2} \! \int\!\rmd^3 x\!\int\!
\rmd^3 {y} \, {\cal G}({\bf x}-{\bf y}) \left[
\hat{m}(\mathbf{x}), \left[ \hat{m}(\mathbf{y}), \hat{\rho}_{t} \right] \right], 
\end{equation}
which again is of the Lindblad type. We will refer to the second term---describing the collapse effect---as $\mathcal{L}[\hat{\rho}_{t}]$. For completeness, we write also the Fourier transformed equation: 
\begin{equation}
\label{eq:mek}
\fl
\frac{\rmd}{\rmd t}\hat{\rho}_{t} =
-\frac{\rmi}{\hbar}[\hat{H},\hat{\rho}_{t}]
+ \lambda \!\! \int\!\! \rmd^3 {k} \, \frac{{\cal G}({\bf k})}{(2\pi)^3} \!\!\left[ \hat{m}(\mathbf{k})\,\hat{\rho}_t\,\hat{m}^{\dagger}(\mathbf{k})
-\frac12\!\left\{\hat{m}^{\dagger}(\mathbf{k})\,\hat{m}(\mathbf{k}),\hat{\rho}_t\right\}
\right], 
\end{equation}
where $\hat{m}(\mathbf{k})$ and ${\cal G}({\bf k})$ are the Fourier transforms of $\hat{m}(\mathbf{x})$ and ${\cal G}({\bf x})$, respectively. 

The relevant open question for the present review is what causes the collapse, or equivalently, what is the origin of the noise $W_t({\bf x})$. A tempting answer is that it has a {\it gravitational} origin, and this idea is supported by two considerations. The first one  is that gravity naturally couples to the local mass density of each physical system, which is the natural choice for a collapse model, as discussed here above. The second one is that there is no proof so far that gravity is quantum, therefore the possibility is still open for it to provide the anti-hermitian coupling (see~\eref{eq:a3}), which is necessary to have a collapse equation. We will review four proposals, which connect  gravity to the collapse of the wave function: The Di{\'o}si--Penrose model~\cite{Diosi:1989,Diosi:2007a,Diosi:2013,Diosi:2014}, Adler's proposal~\cite{Adler:2016a}, K\'arolyha\'zy's model~\cite{Karolyhazy:1966,Karolyhazy:1986,Karolyhazy:1974,Karolyhazy:1990,Karolyhazy:1995,Karolyhazy:1982},
and the \sne~\cite{Diosi:1984,Penrose:1996,Penrose:1998}.

\subsection{The Di{\'o}si--Penrose model}
\label{sec:dp}

Di{\'o}si~\cite{Diosi:1989,Diosi:2007a,Diosi:2013,Diosi:2014} postulated~\eref{eq:sch} and assumed that the spatial correlator of the noise is proportional to the Newtonian gravitational potential~\cite{Diosi:1986,Diosi:1987a}:
\begin{equation}
{\cal G}({\bf x})= \frac{G}{\hbar}\,\frac{1}{|\mathbf{x}|};
\label{eq:ww2}
\end{equation}
this is the connection to gravity. Having defined ${\cal G}({\bf x})$ this way, $\lambda$ is a dimensionless constant, which Di{\'o}si set equal to 1. Equivalently, one can define ${\cal G}({\bf x}) = 1/|\mathbf{x}|$  and $\lambda = G/\hbar$. The structure of the Di{\'o}si--Penrose equation being the same as that of the CSL equation, the collapse mechanism works the same way, with a different strength due to the different choice of the correlation function and collapse rate.

One advantage of this equation,  which was one of Di{\'o}si's motivations for proposing it, is that it is defined without free parameters, contrary to the CSL model, which contains two phenomenological parameters, the collapse rate $\lambda$ and the spatial correlation length of the noise $r_C$. Here the collapse rate is governed by Newton's constant $G$, while the spatial correlation function is given by the ``width'' of the $1/|{\bf x}|$ function. However the model needs to be regularized,  because $1/|{\bf x}|$ does not vanish for $|{\bf x}| \rightarrow \infty$ fast enough to make integrals converge. This is best seen by considering the Lindblad term in~\eref{eq:mek}, which for Di{\'o}si's choice of the correlation function and for a fixed number $N$ of particles becomes (${\bf \mom} = \hbar {\bf k}$):
\begin{eqnarray}
\label{eq:den_manyN}
\fl
\mathcal{L}_{\text{\tiny DP}}^{\text{N}}[\hat{\rho}_t] =
\frac{G}{2\pi^2\hbar^2}\,\sum_{j,l=1}^Nm_j\,m_l\,
\int\,\frac{\rmd^3 Q}{\mom^2}\, f(\mom) \nonumber \\
\times \left[ \rme^{\frac{\rmi}{\hbar}\,\mathbf{\mom}\cdot\hat{\bf r}_j}\,\hat{\rho}(t)\,
e^{-\frac{\rmi}{\hbar}\,\mathbf{\mom}\cdot\hat{\bf r}_l} 
-\frac12\left\{e^{\frac{\rmi}{\hbar}\,\mathbf{\mom}\cdot\hat{\bf r}_j}\rme^{-\frac{\rmi}{\hbar}\,\mathbf{\mom}\cdot\hat{\bf r}_l} , \hat{\rho}(t)\right\}\right],
\end{eqnarray}
with $f(\mom) = 1$ at this stage; for $j=l$ integrals clearly diverge.

A cut-off can be introduced at the level of~\eref{eq:me}, by replacing the point-like density operator with a coarse-grained mass density operator, with a spatial resolution $R_0$. Di\'osi originally introduced the coarse-grained mass density as follows:
\begin{equation}\label{eq:dsp}
\hat{m}'({\bf x}) = \frac{3}{4 \pi R^3_0} \int\,\rmd^3 y\,
 \theta(R_0-|{\bf x}-{\bf y}|)
\,\hat{m}({\bf y}),
\end{equation}
where $\theta(x)$ is the Heaviside step function. Subsequently, 
Ghirardi \etal~\cite{Ghirardi:1990} introduced the coarse-graining as follows:
\begin{equation}\label{eq:co}
\hat{m}'({\bf x})=(2\pi R_0^2)^{-3/2}\int\,\rmd^3 y\,
\exp\left(-\frac{|{\bf x}-{\bf y}|^2}{2R_0^2}\right)\,\hat{m}({\bf y}).
\end{equation}
Note that $\hat{m}'({\bf x})$ is meant to replace $\hat{m}({\bf x})$ in~\eref{eq:me}. These two cut-offs are practically equivalent, and in the following we will consider only  the second one. \eref{eq:den_manyN}~remains the same, with now
\begin{equation}
\label{f2}
f({\mom}) =  \exp\left(-\frac{\mom^2R_0^2}{\hbar^2}\right).
\end{equation} 
Now integrals  converge. Di{\'o}si proposed to set $R_0 \simeq 10^{-15}~\meter$, which is the nucleon's radius. It is a reasonable assumption, given that this model is intrinsically nonrelativistic. 

The cut-off can be also introduced directly at the level of~\eref{eq:den_manyN}, by limiting the momentum integral to $|{\bf \mom}| \leq \mom_{\text{max}} = \hbar/R_0$.
It is easy to see that the new equation is mathematically equivalent to~\eref{eq:den_manyN} with 
$f(\mom)=\theta(\mom-\mom_{\text{max}})$, which eventually can be approximated by a damping Gaussian function, $f(\mom)\approx \rme^{-\mom^2/\mom^2_{\text{max}}}$, such as~\eref{f2}.
In this way one can assign an interpretation to the cut-off and even justify a specific value for it. In fact, $\mom_{\text{max}}$ can be read as an upper limit for the modes of the collapse field that are the dominant modes contributing to the collapse. These modes are also small enough to justify the nonrelativistic approach. This interpretation is very similar to Bethe's nonrelativistic computation of the Lamb shift (see~\cite{Mandel:1995}). Thus, if we argue that $\mom_{\text{max}}$ is the bound justifying the nonrelativistic approach, then one can replace $R_0$ by the Compton wavelength; that is to say: $R_0=\frac{2\pi\hbar}{mc}$, which is $R_0\approx10^{-15}~\meter$ for a nucleon. Accordingly, in this way, one can provide a justification for the cut-off and its chosen value.
This choice, however, is incompatible with experimental data. We will discuss this issue soon.

We now discuss how efficient the collapse is, one measure of it being how fast the off-diagonal elements of the density matrix decay in time. Let us consider the one-particle version of \eref{eq:den_manyN}:
\begin{equation}
\label{eq:den}
\mathcal{L}_{\text{\tiny DP}}^{\text{1}}[\hat{\rho}_t]=
\int\, \rmd^3 {\mom}\, \Gamma_{\text{\tiny DP}}({\bf \mom})
\left(
\rme^{\frac{\rmi}{\hbar}\,\mathbf{\mom}\cdot\hat{\mathbf{r}}}\,\hat{\rho}(t)\,
\rme^{-\frac{\rmi}{\hbar}\,\mathbf{\mom}\cdot\hat{\mathbf{r}}}
-\hat{\rho}(t)\right),
\end{equation}
with 
\begin{equation}
\label{rate}
\Gamma_{\text{\tiny DP}}({\bf \mom}) = \frac{G m^2}{2\pi^2\hbar^2}\,\frac{1}{\mom^2} \exp\left(-\frac{\mom^2R_0^2}{\hbar^2}\right).
\end{equation} 
A measure of the collapse rate is given by:
\begin{equation}
\label{eq:rate_0}
\Lambda_{\text{\tiny DP}}=\int\,\rmd^3 Q\,\Gamma_{\text{\tiny DP}}({\bf \mom})=\frac{Gm^2}{\sqrt{\pi}\hbar\,R_0},
\end{equation}
with $[\Lambda_{\text{\tiny DP}}]=\second^{-1}$. For a nucleon it gives: $\Lambda_{\text{\tiny DP}} \simeq 10^{-15}~\second^{-1}$, implying that coherences for microscopic particles are stable over extremely long times. On the other hand, as anticipated, for macroscopic objects the collapse is amplified. 
Assuming  a rigid many-body system and tracing out the relative coordinates, the dynamical equation for the center-of-mass density-matrix $\hat{\rho}^{\text{M}}_t$ takes
the same form as in~\eref{eq:den}, where $\Gamma_{\text{\tiny DP}}({\bf \mom})$ is replaced by~\cite{Diosi:1987a}:
\begin{equation}
\Gamma^{\text{M}}_{\text{\tiny DP}}({\bf \mom})=
\frac{G}{2\pi^2\hbar^2\mom^2}\,
|\tilde{\mu}_{\text{rel}}({\bf \mom})|^2\,\exp\left(-\frac{\mom^2R_0^2}{\hbar^2}\right) \,,
\end{equation}
with
$\tilde{\mu}_{\text{rel}}({\bf \mom})=\int\,\rmd^3 x\,\rme^{\rmi{\bf \mom}\cdot{\bf x}/\hbar}\,\mu_{\text{rel}}({\bf x})$, where $\mu_{\text{rel}}({\bf x})$ is the internal mass density. For example, for a homogeneous rigid sphere of mass $M$ and radius $R$, we get: 
$\tilde{\mu}_{\text{rel}}({\bf \mom})\approx M\,\exp(-\mom^2R^2/8\hbar^2)$. Accordingly, we find:
\begin{equation}
\Gamma^{\text{M}}_{\text{\tiny DP}}({\mom})\simeq \frac{G\,M^2}{2\pi^2\hbar^2}\,\frac{1}{\mom^2}
\exp\left(-\frac{\mom^2(R^2+4R^2_0)}{4\hbar^2}\right).
\end{equation}
Similar to~\eref{eq:rate_0}, here the total collapse rate becomes:
\begin{equation}
\label{eq:rate_M}
\Lambda^{\text{M}}_{\text{\tiny DP}} =
\int\,\rmd^3 Q\,\Gamma_{\text{M}}({\bf \mom})\simeq
\frac{2GM^2}{\,\hbar\,\sqrt{\pi(R^2+4R^2_0)}}.
\end{equation}
For example, $\Lambda^{\text{M}}_{\text{\tiny DP}}$ is of order $10^{-5}~\second^{-1}$ for a typical optomechanical nanosphere with $M\simeq 10^{9}~\atomicmass$ and $R \simeq 50~\nano\meter$~\cite{Chang:2010}. Evidently, the collapse rate of a nanosphere is ten orders of magnitude larger than that of a nucleon, and would be even larger for a truly macroscopic system.
Therefore, the model is capable of describing both the quantum properties of microscopic systems
and the classical properties of macroscopic objects.

It is interesting to write down the explicit time evolution of the off-diagonal elements; this will give the connection between Di{\'o}si's model and Penrose's idea, which eventually justifies why the model is called the Di{\'o}si--Penrose model. Given the master equation~\eref{eq:me}, with Di{\'o}si's choice for the collapse rate and spatial correlator of the noise, and
using the characteristic function~\cite{Savage:1985,Smirne:2010}, the one-particle state at time $t$ in the position representation, $\langle{\bf x}|\hat{\rho}^{\text{1}}(t)|\mathbf{x}'\rangle = \rho(\mathbf{x}, \mathbf{x}',t)$, is found to be
\begin{eqnarray}
\label{eq:sol}
\fl
\rho(\mathbf{x}, \mathbf{x}', t)  =  \int\rmd^3 y\,\int\,
\frac{\rmd^3 p}{(2 \pi \hbar)^3} 
\,\rho_0 (\mathbf{x}+\mathbf{y}, \mathbf{x}'+\mathbf{y}, t)\nonumber \\
\times\exp\left(-\frac{\rmi}{\hbar}{\bf y} \cdot \mathbf{p}
- \frac{1}{\hbar}\int^t_0\,\rmd\tau\left(
U \left(\frac{{\bf p}\,\tau}{ m} +\mathbf{x}-\mathbf{x}'\right)
-U(\vec 0)\right)\right),
\end{eqnarray}
where $ \rho_0(\mathbf{x}, \mathbf{x}',t)$ is the solution of the free \schr\ dynamics and
\begin{equation}\label{eqn:grav-interaction-energy}
U(\mathbf{x})= - G\,\int\rmd^3 r\,\int\rmd^3 r'\, \frac{m'(\mathbf{r}) \,m'(\mathbf{r}')}
{|\mathbf{x}+\mathbf{r}-\mathbf{r}'|}
=
-G m^2\,\frac{\erf(|{\bf x}|/R_0)}
{|{\bf x}|},
\end{equation}
is the Newtonian self-interaction where $\erf(x)$ is the Gauss error function. 
If one neglects the pure \schr\ contribution in~\eref{eq:me}, which is justified on the short time scale, then~\eref{eq:sol} reduces to an exponential
decay of the form: 
\begin{equation}
\label{eq:damp}
\rho(\mathbf{x}, \mathbf{x}', t) = \exp\left(- \frac{t}{\tau(\mathbf{x}, \mathbf{x}')}\right)\rho(\mathbf{x}, \mathbf{x}', 0),
\end{equation}
where the characteristic damping time $\tau$ is:
\begin{equation}
\label{eq:td}
\tau(\mathbf{x}, \mathbf{x}') = \frac{\hbar}
{U \left(\mathbf{x}-\mathbf{x}'\right)-U(\vec 0)}.
\end{equation}
Again, \eref{eq:damp} implies that spatial superpositions of positions ${\bf x}$ and ${\bf x'}$ decay with the rate $\tau({\bf x},{\bf x}')$.
This is precisely the endpoint of Penrose's idea~\cite{Penrose:1994,Penrose:1994a,Penrose:1998a,Penrose:1996,Penrose:1998,Penrose:2014}. 

According to Penrose,  given two macroscopic lumps of matter  at different locations, each being in a stationary state, i.\,e.~an energy eigenstate, with the same eigenvalue, then also their superposition is a stationary state and persists in time, according to quantum theory. This is another way to express the \schr's cat problem. However,  if one takes also gravity into account, Penrose argues, things change. Each state will give rise to a stationary spacetime geometry, and the two differ macroscopically from each other. When a superposition of the two lumps is considered, then also the two spacetime geometries will be superimposed, and this will not be stationary any longer. Penrose writes: \textit{``We have to consider carefully what a `stationary state' means in a context such as this. In a stationary spacetime, we have a well-defined concept of `stationary' for a quantum state in that background, because there is a Killing vector $T$ in the spacetime that generates the time-translations. Regarding $T$ as a differential operator (the `$\partial/\partial t$' for the spacetime), we simply ask for the quantum states that are eigenstates of $T$, and these will be the stationary states, i.\,e.~states with well-defined energy values. [\dots] However, for the superposed state we are considering here we have a serious problem. For we do not now have a specific spacetime, but a superposition of two slightly differing spacetimes. How are we to regard such a `superposition of spacetimes'? Is there an operator that we can use to describe `time-translation' in such a superposed spacetime? Such an operator would be needed so that we can identify the `stationary states' as its eigenvectors, these being the states with definite energy. It will be shown that there is a fundamental difficulty with these concepts, and that the notion of time-translation operator is essentially ill defined''}~\cite{Penrose:1996}.

This ill-definedness leads to an uncertainty in the energy, which makes the superposition unstable. Penrose provides a quantitative  measure of the energy uncertainty: it is the gravitational self-energy of the \emph{difference} between the two mass distributions $\rho$ and $\rho'$ of the two superposed states,
\begin{equation}
\Delta E = -4\pi\,G\,\int\,\rmd^3 r \, \int\,\rmd^3 r'\,
\frac{\left(\rho(\vec r)-\rho'(\vec r)\right) \left(\rho(\vec r')-\rho'(\vec r')\right)}{\abs{\vec r - \vec r'}} \,.
\end{equation}
As Penrose points out, this coincides with the use of the Newtonian gravitational interaction energy of the corresponding mass distribution at two positions according to \eref{eqn:grav-interaction-energy},
as long as both superposed states have the same gravitational self-energy (i.\,e.~for a displacement, but not for a shape-changing superposition). Using the energy-time uncertainty relation, this can imply that the superposition decays to one of the localized states with the lifetime $\tau=\hbar/\Delta E$, which is equivalent to~\eref{eq:td}. Therefore, Di\'osi's dynamical equation implements Penrose's idea. This is why one speaks of the Di\'osi--Penrose model.

We have seen that collapse models couple quantum systems to an external classical noise. As such, the system undergoes a Brownian motion and energy is exchanged. In particular, the energy of the quantum system is not conserved anymore. The CSL model, and the Di{\'o}si--Penrose model as well, do not include dissipative effects, therefore the energy of the system steadily increases and eventually diverges, as if the noise is at infinite temperature. The rate of energy increase can be easily calculated
and turns out to be:
\begin{equation}\label{eq:hh}
\frac{\rmd E_{\text{\tiny DP}}(t)}{\rmd t}
= \frac{2 \pi}{m} \int^{\infty}_0 \rmd^3 Q \, \Gamma_{\text{\tiny DP}}(\mom)\, \mom^4 = \frac{m \,G \,\hbar}{4\sqrt{\pi}\,R_0^3},
\end{equation}
with $E_{\text{\tiny DP}}(t)=\Tr[\hat{\rho}(t) \hat{H}]$ where $\hat{\rho}(t)$ satisfies the Di{\'o}si--Penrose dynamics. From this relation, one can easily  evaluate the different implications of the cut-off proposed, respectively,
by Di\'osi~\cite{Diosi:1986} and Ghirardi \etal~\cite{Ghirardi:1990}. In the former case, $R_0 = 10^{-15}~\meter$, one gets a rate for the energy increase of order $10^{-4}~\kelvin\per\second$ for a proton, which means a thermal catastrophe!
Precisely this fact induced Ghirardi \etal to introduce a much larger value, $R_0 = 10^{-7}~\meter$, in which case the rate is of order $10^{-28}~\kelvin\per\second$ for a proton, which is indeed a much more reasonable value.  Although in this way the problem of overheating has been partially resolved, it is clear that the introduction of a cut-off $R_0 = 10^{-7}~\meter$ is much less justified than the one originally proposed by Di{\'o}si. One of the main motivations of the Di{\'o}si--Penrose model, namely, to provide a phenomenological model without free parameters, is in this way lost.

In~\cite{Bahrami:2014a} the possibility has been explored, whether the original value $R_0 = 10^{-15}~\meter$ can be retained, while including dissipative effects, which mitigate the fast energy increase. This would be a rather natural resolution of the problem: physical noises cannot pump too much energy in the system, which should eventually thermalize to the (finite) temperature of the noise. This procedure successfully works for the CSL model~\cite{Smirne:2015} (see also~\cite{Bassi:2005,Smirne:2014}), however fails for the Di{\'o}si--Penrose model: in this case, thermalization can be formally achieved, but at the price of introducing very strong dissipation, causing with high probability sudden flips in momentum of the system. The momentum flip induced by the universal noise field causing the collapse of the wave function 
would mean that the latter can transfer an energy of the order
of tens of MeV to a nucleon in a nucleus
(corresponding to the average kinetic energy of a nucleon in a Fermi-gas model~\cite{Bertulani:2004}), which would induce instantaneous matter dissociation.
Accordingly, with $R_0=10^{-15}~\meter$, the Di{\'o}si--Penrose model is only an effective model for the mass ranges larger or comparable with $m_r\sim10^{11}~\atomicmass$.
In other words, the Di{\'o}si--Penrose model should be applied only to mesoscopic and macroscopic systems. It is worth mentioning that the value $m_r$ is also very different from the Planck mass $\mplanck\sim10^{19}~\atomicmass$, which is sometimes considered as a borderline between quantum and classical masses~\cite{Lamine:2006}.

\subsection{Adler's proposal}
\label{sec:a}

Di{\'o}si assumes~\eref{eq:sch} and postulates~\eref{eq:ww2} for the correlation function of the noise. Adler's point of view is different~\cite{Adler:2016a}. He starts with~\eref{eq:a3}, which can be taken as the starting point to justify the collapse~\eref{eq:fghr} as we showed in the introduction to this section, and thus of~\eref{eq:sch} and~\eref{eq:me} with the appropriate choice of the collapse operators. Next he notices that one can naturally introduce a {\it mass-proportional anti-hermitian} coupling to the wave function by assuming that 
gravity is fundamentally {\it classical} and its metric contains a rapidly fluctuating {\it complex} component. In other words:
\begin{equation}
g_{\mu\nu} = \overline{g}_{\mu\nu} + h_{\mu\nu},
\end{equation}  
with $\overline{g}_{\mu\nu}$ the conventional real spacetime metric, and $h_{\mu\nu}$ an irreducibly complex fluctuation. For simplicity, we will set $\overline{g}_{\mu\nu} = \eta_{\mu\nu}$, i.\,e.~we will consider only perturbation around the flat Minkowski metric. It is easy to show how, under these premises, one obtains the desired coupling term.

The standard action of a matter field coupled to gravity is:
\begin{equation}
S=  \int \rmd^{4}x\sqrt{-g}\,\mathcal{L}_\text{matter},
\label{eq:ma}
\end{equation}
where $\mathcal{L}_\text{matter}$ is the matter Lagrangian, $g_{\mu\nu}$ is the metric tensor and $\sqrt{-g}=\sqrt{-\det[g_{\mu\nu}]}$.  The Taylor expansion around $ \eta_{\mu\nu}$ gives:
\begin{equation}
S=\int \rmd^{4}x\left[\mathcal{L}_\text{matter}^{(0)}-\frac{1}{2}h^{\mu\nu}T_{\mu\nu}^{(0)}\right] +\text{higher order terms};
\label{eq:interact_action}
\end{equation}
the apex $(0)$ denotes quantities with respect to the flat spacetime $\eta_{\mu\nu}$, and $T_{\mu\nu}$ is the stress energy tensor associated to the Lagrangian $\mathcal{L}_\text{matter}$.  In the weak field limit, gravity couples to matter linearly through the stress energy tensor. 

In the nonrelativistic limit, considering for example the Lagrangian of a Klein-Gordon field $\phi$, one can show that $h^{\mu\nu}T_{\mu\nu}^{(0)} \simeq mc^2 h^{00}\phi^{*}\phi$. Second quantization of the field then immediately gives the one-particle Hamiltonian $\vec{p}^2/2m + mc^2 h^{00}/2$, therefore the $N$-particle Hamiltonian can be written as:
\begin{equation}
\label{eq:puc}
H = H_0 + c^2\xi \int \rmd^{3}x\, \hat{m}({\bf x})w({\bf x},t),
\end{equation}
where $H_0$ is the kinetic term, $\hat{m}({\bf x})$ is the mass density operator introduced in the introduction to this section, and we have renamed $h^{00}({\bf x},t) /2= \xi w({\bf x},t)$, where $w({\bf x},t)$ is an a-dimensional complex random field, whose fluctuations are of order 1, while the constant $\xi$ measures the size of the fluctuations. The real part of $w({\bf x},t)$ contributes to the Hamiltonian as a standard random potential, giving rise to decoherence effects as discussed in \sref{sec:gd-class}. Its imaginary part provides the anti-hermitian random coupling, which is the starting point for deriving the collapse dynamics.

For the purposes of this review, it is not particularly relevant to write down the collapse equation arising from~\eref{eq:puc}. It is the generalization of~\eref{eq:sch} to {\it complex} and {\it non-white} noises, and in general it can be written down explicitly only as a perturbative expansion. More details can be found in~\cite{Gasbarri:2017}. One can show that the usual properties hold: wave functions are localized in space and the amplification mechanism assures that the collapse rate grows with the mass/size of the system. 

It is more instructive to consider the corresponding master equation for the density matrix $\rho_t$. The Hamiltonian evolution is the usual one, while the Lindblad term reads:
\begin{eqnarray}
\fl
{\cal L}_{\text{A}}[\hat{\rho}_t]
 = -\frac{\xi^{2}c^4}{\hbar^{2}}\int\!\rmd^3 {x}\,\int\!  \rmd^3 {y}\!\int_{0}^{t}\!\rmd\tau\,{\cal G}_{t-\tau}^{\text{R}}(\mathbf{x}-\mathbf{y})\com{\hat{m}(\mathbf{x})}{\com{\hat{m}_{\tau-t}(\mathbf{y})}{\rho_{t}}}+ \nonumber \\
 -\frac{\rmi\xi^{2}c^4}{\hbar^{2}}\int\!\rmd^3 {x}\,\int\!  \rmd^3 {y}\!\int_{0}^{t}\!\rmd\tau\,{\cal G}_{t-\tau}^{\text{I}}(\mathbf{x}-\mathbf{y})\com{\hat{m}(\mathbf{x})}{\acom{\hat{m}_{\tau-t}(\mathbf{y})}{\rho_{t}}}, 
\label{megrav-1}
\end{eqnarray}
where ${\cal G}^{\text{R}}$ and ${\cal G}^{\text{I}}$ are the real and the imaginary parts
of the correlation function of the noise field: ${\cal G}_{t,\tau}(\vec{x},\vec{y})=\mathbb{E}[w^{*}(\vec{x},t)w(\vec{y},\tau)]$.
In writing the above equation, we assumed that the noise 
is statistically homogeneous over space and time: ${\cal G}_{t,\tau}(\vec{x},\vec{y}) = {\cal G}_{t-\tau}(\vec{x}-\vec{y})$. Also, $\hat{m}_{\tau-t}(\mathbf{y})$ is the time evolved mass density operator, the time evolution generated by the Hamiltonian $H_0$. 
The equation, which is valid up to second perturbative order in the parameter $\xi$, generalizes~\eref{eq:me} to the case of complex and non-white noises.  Here we can see a rather important difference between Adler's model and the Di{\'o}si--Penrose model. While both rely on the same structure for the collapse equation, the latter gives a specific form for the correlation function of the noise---the Newtonian classical gravitational potential---while Adler's model does not specify the form of the correlation function, only requiring the gravitational fluctuations to have a complex fluctuating component. 

At this stage, one is not able to characterize the noise correlator in Adler's model. One might guess that, if the noise has a cosmological origin, its correlation function is highly non trivial, but apart from this nothing more can be said unless one establishes the origin of the noise, which is unknown. What one can do is to analyze bounds on the strength of the fluctuations, by comparing the predictions of the model with experimental data. To this purpose, let us consider the center-of-mass equation of a composite system. Under suitable approximations, which are typically valid for crystalline structures forming macroscopic objects, the center of mass and relative motions decouple and the Lindblad term for the center-of-mass  density-matrix $\hat{\rho}^{\text{M}}_t$ takes the form~\cite{Gasbarri:2017}:
\begin{eqnarray}
\fl
{\cal L}_{\text{A}}[\hat{\rho}^{\text{M}}_t]
  = -\frac{\xi^{2}c^4}{\hbar^{2}}\frac{1}{(2\pi\hbar)^{3}} \int_{0}^{t}\! \rmd\tau\! \int\! \rmd^3 {Q}\,\Gamma_{t-\tau}^{\text{M,R}}(\mathbf{Q})\left[\rme^{-\frac{\rmi}{\hbar}\mathbf{Q}\cdot\hat{\mathbf{r}}},\left[\rme^{\frac{\rmi}{\hbar}\mathbf{Q}\cdot\hat{\mathbf{r}}_{\tau-t}},\hat{\rho}^{\text{M}}_{t}\right]\right]+ \nonumber \\
 -\frac{\rmi\xi^{2}c^4}{\hbar^{2}}\frac{1}{(2\pi\hbar)^{3}}\int_{0}^{t}\! \rmd\tau\! \int\! \rmd^3 {Q}\,\Gamma_{t-\tau}^{\text{M,I}}(\mathbf{Q}) \left[\rme^{-\frac{\rmi}{\hbar}\mathbf{Q}\cdot\hat{\mathbf{r}}},\left\{ \rme^{\frac{\rmi}{\hbar}\mathbf{Q}\cdot\hat{\mathbf{r}}_{\tau-t}},\hat{\rho}_{t}\right\} \right],
\label{megrav-1-1-2-1}
\end{eqnarray}
where $\Gamma_{t-\tau}^{\text{M,R/I}}(\mathbf{Q}) = \Gamma_{t-\tau}^{\text{R/I}}(\mathbf{Q}) |\tilde{\mu}_{\text{rel}}({\bf \mom})|^2$, with $\Gamma_{t-\tau}^{\text{R/I}}(\mathbf{Q})$ the Fourier transform of ${\cal G}^{\text{R/I}}_{t-\tau}(\vec{x})$ and $\tilde{\mu}_{\text{rel}}({\bf \mom})$ the Fourier transform of the internal mass density $\mu_{\text{rel}}({\bf x})$. The above expression should be compared with the analogous expression \eref{eq:den} of the Di{\'o}si--Penrose model, and again it represents its generalization to complex and non-white noises.

As discussed in \sref{sec:dp}, one prediction of~\eref{megrav-1-1-2-1} is that the collapse noise induces a Brownian diffusion, which  implies of a variety of non-standard effects, like an excess diffusion of center-of-mass motion~\cite{Bahrami:2014b,Nimmrichter:2014,Diosi:2015a,Li:2016,Vinante:2016}, spontaneous heating~\cite{Laloe:2014,Bilardello:2016}, and spontaneous photon emission~\cite{Adler:2013,Donadi:2014, Curceanu:2016a} from charged particles. Available experimental data allow to sets bounds on the noise correlator, based on these effects.

\begin{figure}[t]
\centering
\includegraphics[width=0.6\linewidth]{{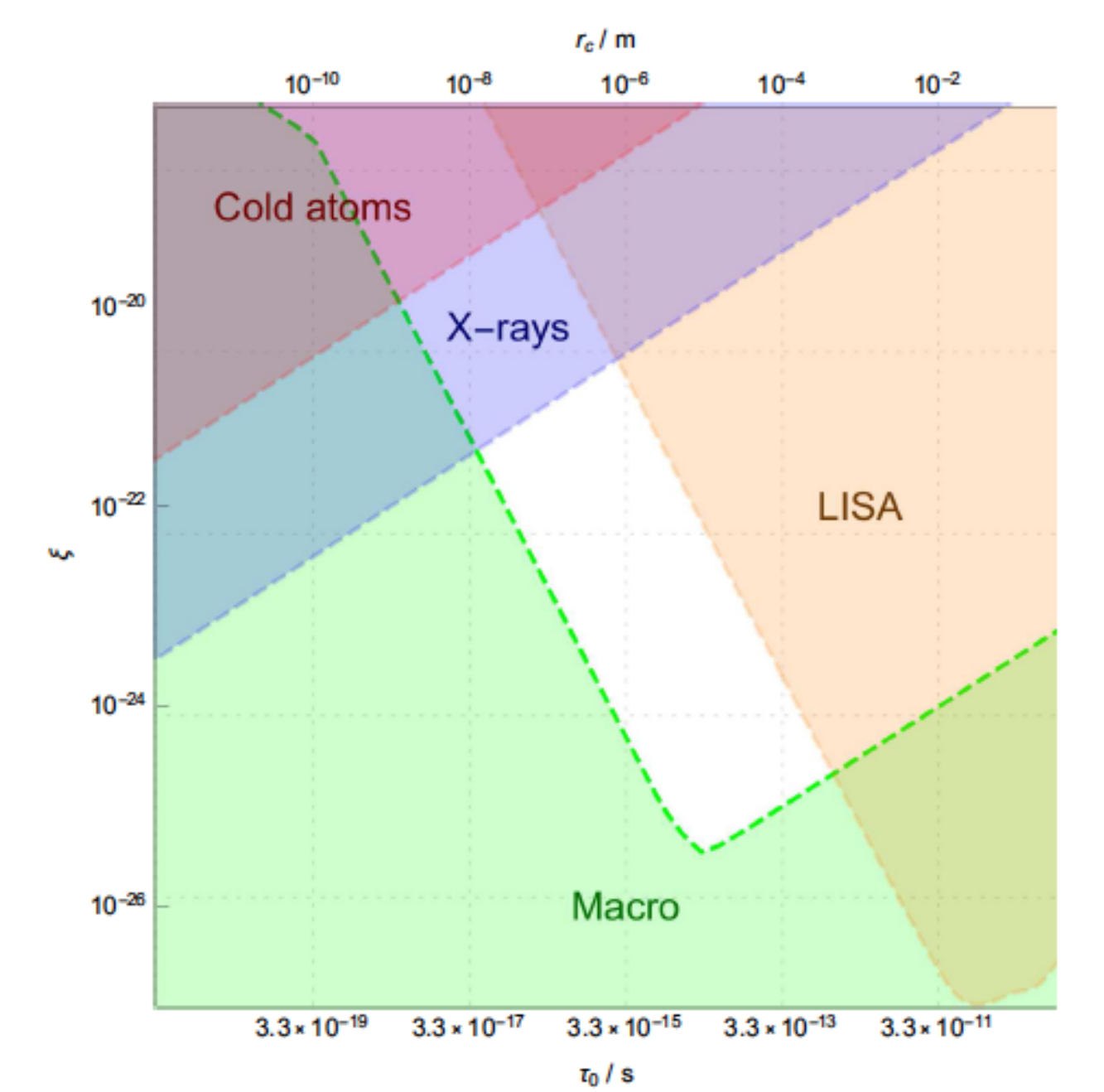}}
\caption{$(\xi,r_C)$ or equivalently $(\xi,\tau_0)$ parameter diagram of the gravity-induced collapse model. The white area is the allowed region. The blue shaded region ({\it X-rays}) is excluded by data analysis of X-rays measurements~\cite{Curceanu:2016}. The orange shaded region ({\it LISA}) is excluded from data analysis of LISA Pathfinder~\cite{Carlesso:2016}. The green shaded region (\textit{Macro}) is an estimate of the region excluded by the requirement that the collapse is strong enough to localize macroscopic objects~\cite{Toros:2016,Toros:2016a}. Note that X-ray measurements  sample the high frequency region of the spectrum ($\sim  10^{18}$ Hz) and would disappear if the noise correlator has  a cut-off below such frequencies, which is plausible. In such a case, the stronger upper bound on the left part of the plane is given by data analysis with cold atom experiments ({\it Cold atoms})~\cite{Bilardello:2016}.}
\centering
\label{Fig_parameters}
\end{figure}

Discussing the experimental constraints on the noise correlator in its full generality is too difficult. We will limit the discussion to a restricted class of Gaussian correlations functions, in such a way that the collapse dynamics is controlled by only two parameters. Specifically, we consider the Markovian limit by imposing:
\begin{equation}
\Gamma_s^{\text{R/I}}(\mathbf{Q})	\simeq \Gamma^{\text{R/I}}(\mathbf{Q})\,\tau_0\delta(s)
\label{Markovian_limit}
\end{equation}
with $[\tau_0]=[T]$. One can show that in this limit $\Gamma^{\text{I}}(\mathbf{Q}) = 0$. In addition,  we assume that $\Gamma^{\text{R}}(\mathbf{Q})$ has the a Gaussian shape:
\begin{equation}
\Gamma^{\text{R}}(\mathbf{Q}) = r_C^3 \exp(-r_C^2\mathbf{Q}^{2}/\hbar^{2}),
\end{equation}
as in the CSL model, where $[r_C]=[L]$. With these assumptions, after some algebra, the Lindblad operator in~\eref{megrav-1-1-2-1} reduces to
\begin{equation}
\label{almostcslme}
\fl
{\cal L}_{\text{A}}[\hat{\rho}^{\text{M}}_t] = -\frac{\xi^{2}c^{4}r_{C}^{3}\tau_{0}}{(2\pi\hbar)^{3}2\hbar^{2}}\int\! \rmd^3 {Q}\,|\tilde{\mu}_{\text{rel}}({\bf \mom})|^2 \rme^{-r_{C}^{2}Q^{2}/\hbar^{2}}\left[\rme^{-\frac{\rmi}{\hbar}\mathbf{Q}\cdot\hat{\mathbf{r}}},\left[\rme^{\frac{\rmi}{\hbar}\mathbf{Q}\cdot\hat{\mathbf{r}}},\hat{\rho}_{t}\right]\right].
\end{equation}
This equation should be compared with the CSL master equation~\cite{Ghirardi:1990a}:
\begin{equation}\label{cslme}
\fl
{\cal L}_{\text{\tiny CSL}}[\hat{\rho}^{\text{M}}_t] = -\frac{\lambda(4\pi r_C^2)^{3/2}}{(2\pi\hbar)^{3}m_0^2}\!\int\! \rmd^3 {Q}\,|\tilde{\mu}_{\text{rel}}({\bf \mom})|^2 \rme^{-r_C^2\mathbf{Q}^{2}/\hbar^{2}}\!\left[\rme^{-\frac{\rmi}{\hbar}\mathbf{Q}\cdot\hat{\mathbf{r}}},\left[\rme^{\frac{\rmi}{\hbar}\mathbf{Q}\cdot\hat{\mathbf{r}}},\hat{\rho}_{t}\right]\right]\!.
\end{equation}   
In particular, \eref{almostcslme} reduces to the CSL master equation given in~\eref{cslme} by setting:
\begin{equation}
\xi=\frac{4\hbar\pi^{3/4}}{m_{0}c^{2}}\sqrt{\frac{\lambda}{\tau_{0}}}.
\end{equation}
With this approximations the model is fully characterized by: the magnitude $\xi$ of the metric fluctuations, the time cut-off $\tau_0$ and the space cut-off $r_C$. To further simplify the discussion, as first proposed in~\cite{Bassi:2002} we assume the time cut-off to be related to the space cut-off via $\tau_0=r_C/c$. One can then give bounds on $\xi$ by using the bounds already set for the CSL parameters $\lambda$ and $r_C$. The most recent bounds are summarized in \fref{Fig_parameters}.

It is interesting to
compare these results with the recent discovery of gravitational waves~\cite{Abbott:2016}, observed in frequency range from 35 to 250 Hz and with a peak strain of $1.0 \times 10^{-21}$. Clearly, gravitational waves are real, while here the claim is that the collapse is caused by complex fluctuations of the metric. Also, significant gravitational waves typically have long wave lengths, while here the relevant part of the  spectrum  is at high frequencies. However it is interesting to see that the order of magnitude of real waves and complex fluctuations---which allow for an efficient collapse and are compatible with experimental data---are not so far away from each other.

\subsection{K{\'a}rolyh{\'a}zy model}

The key idea of K{\'a}rolyh{\'a}zy~\cite{Karolyhazy:1966,Karolyhazy:1986,Karolyhazy:1974,Karolyhazy:1990,Karolyhazy:1995,Karolyhazy:1982}, later developed by Frenkel~\cite{Frenkel:2002,Frenkel:1990,Frenkel:1995,Frenkel:1997}, is that lengths and times cannot be measured with arbitrary precision, given that measuring devices are made of atoms and therefore are ultimately quantum and must obey the Heisenberg uncertainty principle. Thus there is a fundamental uncertainty in the structure of spacetime. By using the uncertainty principle and some basic arguments, he arrives at  the following relation~\cite{Karolyhazy:1966}:
\begin{equation}
\label{eq:xfgjgj}
\Delta s^2 = \lplanck^{4/3} s^{3/2},
\end{equation}
where $\lplanck = \sqrt{\hbar G/c^3}$ is the Planck length. The above relation expresses the best precision with which a length $s$ can be measured. 

So spacetime is undetermined beyond some scale, according to K{\'a}rolyh{\'a}zy. He expresses this uncertainty by promoting the metric $g_{\mu\nu}(x)$ to a stochastic metric $g_{\mu\nu}(\omega,x)$, and averages are taken in the end. In the weak field limit we have:
\begin{equation}\label{eqn:lingrav}
 g_{\mu \nu}(\omega,x) = \eta_{\mu \nu} + h_{\mu \nu}(\omega, x) \,,
\end{equation}
where $h_{\mu \nu}(\omega, x)$, as usual, represent random perturbations around the flat spacetime metric (from now on we will omit the random variable $\omega$), which are assumed to obey the wave equation, in the appropriate gauge:
\begin{equation}
\label{eq:dfghdfg}
 \Box h_{\mu \nu} =  0 \,,
 \end{equation}
where $\Box$ denotes the d'Alembert operator.
This is true if we neglect the effect of matter on spacetime fluctuations. \Eref{eq:dfghdfg}~allows for a plane-wave expansion:
\begin{equation}
h_{\mu \nu}(x) = \frac{1}{\sqrt{L^3}} \sum_{\bf k} \left[ c_{\mu\nu}({\bf k}) \rme^{\rmi {\bf k} \cdot {\bf x} - \omega_{\bf k} t} + \text{c.c.} \right],
\end{equation}
with $x = {\bf x}, t$ and $\omega_{\bf k} = c |{\bf k}|$. Now the Fourier coefficients $c_{\mu\nu}({\bf k})$ are random, and they are assumed to fluctuate around zero: ${\mathbb E}[c_{\mu\nu}({\bf k})] = 0$,
with ${\mathbb E}[\cdot]$ denoting the stochastic average.
Box normalization (of side  $L$) has been assumed.

Given the standard definition of length,
\begin{equation}
s^2 = \int \left( -g_{\mu\nu} \frac{d x^{\mu}}{\rmd t} \frac{d x^{\nu}}{\rmd t} \right) \rmd t,
\end{equation}
which again is random since $g_{\mu\nu}$ is, the fundamental relation~\eref{eq:xfgjgj}, where we identify  $\Delta s^2 = \mathbb{E}[(s - {\mathbb E}[s])^2]$, this yields the following spectrum for the fluctuations $c({\bf k}) \equiv c_{00}({\bf k})$, the only ones which will be relevant for the following discussion:
\begin{equation}
{\mathbb E}[c({\bf k}) c^*({\bf k}')]  = \delta_{{\bf k}, {\bf k}'}\, \lplanck^{4/3} k^{-5/3}.
\end{equation}
So~\eref{eq:xfgjgj} selects a specific stochastic behavior of the metric fluctuations.  

Now calculations proceed in the standard way. In the nonrelativistic and Newtonian limit, the metric perturbation enters the \schr\ equation as a random potential $V({\bf x}, t) = mc^2 h_{00}({\bf x}, t)/2$, 
the generalization to many particles being straightforward.  The resulting  dynamics is stochastic and leads to decoherence, as different terms of a {\it spatial} superposition will acquire different phases, which wash away the coherent behavior.

The decoherence time $\tau$ has been computed and it can be expressed as: $\tau = m a_c^2 / \hbar$, where $a_c$ is the shortest coherence distance, to which the stochastic dynamics assigns a phase mismatch $\simeq \pi$, which gives full decoherence. Calculations give:
\begin{equation}
a_c \simeq \frac{\hbar}{G} \frac{1}{m^3},
\end{equation}
for an elementary particle of mass $m$. For example, for one electron one gets: $a_c \simeq 10^{35}~\centi\meter$ and correspondingly $\tau \simeq 10^{70}~\second$. This gravitational decoherence effect is fully negligible for microscopic systems. On the other hand, for a macroscopic object of mass $M$ and linear dimensions $R$ one has:
\begin{equation}
a_c \simeq \left( \frac{\hbar^2}{G} \right)^{\!\!1/2} \frac{R^{2/3}}{M};
\end{equation}
for an object of size $\simeq 1~\centi\meter$ and normal density, one gets $a_c \simeq 10^{-16}~\centi\meter$ and $\tau \simeq 10^{-4}~\second$. Macroscopic superpositions rapidly decay. 

K{\'a}rolyh{\'a}zy's model assumes the validity of the \schr\ equation, supplemented by a random gravitational potential with a specific correlation function. As such, although often quoted as a collapse model as those described so far in this section, it is not: here the superposition principle remains valid and macroscopic superpositions are still solutions of the equations of motions. As such, the model does not represent a resolution of the quantum measurement problem. It is, however, interesting because it shows that when quantum systems couple to gravity, even if gravity is treated classically, \emph{intrinsic} fluctuations can exist, which cause decoherence preventing to fully control the quantum state of the system.

\subsection{The Schr\"odinger--Newton equation}\label{SNtheo}

The \sne\ was first proposed by Di{\'o}si~\cite{Diosi:1984} and Penrose~\cite{Penrose:1996,Penrose:1998}, who were interested in its stationary solutions~\cite{Moroz:1998,Tod:1999,Harrison:2003} as the final states of the wave function collapse.
Its dynamical behavior received renewed attention, mainly due to its connection with the question whether gravity should really be quantized~\cite{Carlip:2008}, and due to its falsifiability in envisaged
experiments~\cite{Giulini:2011,Giulini:2013,Yang:2013,Grossardt:2016} (cf. \sref{sec:experiments}).
The  one particle \sne\ reads:
\begin{equation}
\label{eqn:sn}
\rmi \hbar \frac{\partial}{\partial t} \psi(t, {\bf r}) = \left( -\frac{\hbar^2}{2 m} \nabla^2
- G m^2 \int \rmd^3 r' \, \frac{\abs{\psi(t,{\bf r}')}^2}{\abs{{\bf r} - {\bf r}'}}
\right) \psi(t, {\bf r}) \,.
\end{equation}
It is a nonlinear equation, the nonlinearity arising from the self-gravitational interaction among different parts of the wave function, as if the (absolute value squared of the) wave function  represents the mass distribution of the system. This is an attractive effect, which opposes the quantum diffusion originating from the Laplacian. Therefore one expects a slower spreading of the wave function over time, with respect to the usual quantum dynamics, and possibly a collapse for large masses. Asymptotically, a stable solitonic state should remain, where attraction and diffusion exactly compensate. 

In particular, given a superposition of two states displaced in space, the \schr--Newton term gives a mutual attraction, which turns the superposition into a localized state. This is why this equation is discussed in connection with the quantum measurement problem as a possible way to explain why macroscopic superpositions are not observed: for massive systems, the \schr--Newton attractive term becomes dominant and kills any possible superposition in space. We will come back on this issue.

One relevant question is the justification of the \sne, which has given rise to a rather lively debate~\cite{Anastopoulos:2014,Anastopoulos:2014a,Adler:2007,Christian:1997}. The claim of the supporters of this equation is that  it
follows from semi-classical gravity. Here one has to be clear about the meaning of {\it semi-classical gravity}.
The majority of the physics community assumes that gravity must be
quantized in some way, and that semi-classical gravity is only an \emph{effective}
theory, which holds in situations where matter is treated quantum mechanically and gravity
classically (although it is fundamentally quantum). This effective description comprises so far all observable physical phenomena.
From this perspective, from which the fundamental theory is fully quantum, the evolution of the wave function is linear,  
and one cannot expect any nonlinear term in the dynamics. The self-interactions present in the classical theory would then be treated in the same
way as in Quantum Electrodynamics, namely, through the normal ordering and renormalization prescription.
They would lead to a mass-renormalization of the theory rather than a potential term in the \schr\ 
equation~\cite{Anastopoulos:2014,Anastopoulos:2014a}. Within this framework, the \sne\ arises as a {\it mean-field} equation, like the Gross--Pitaevskii equation~\cite{Gross:1961,Pitaevskii:1961} commonly used in condensed matter theory, whose validity is restricted
to the case of large numbers of particles.

However, the key assumption of the proponents of the \sne~is different.
The claim is that it does follow from a theory in which only
matter fields are quantized, while the gravitational field remains {\it fundamentally} classical. If one refers to this theory as semi-classical gravity~\cite{Kibble:1981,Mattingly:2005,Kiefer:2007}, 
then the \sne~does follow from it, as we now review~\cite{Bahrami:2014}.

The most natural dynamical equation of such an hybrid theory is provided by the semi-classical Einstein equations:
\begin{equation}
\label{eqn:sce}
R_{\mu \nu} - \frac{1}{2} g_{\mu \nu} R = \frac{8 \pi G}{c^4} \,
\bra{\Psi} \hat{T}_{\mu \nu} \ket{\Psi} \,,
\end{equation}
where the classical energy-momentum tensor in Einstein's equations is replaced by the expectation value
of the corresponding quantum operator in a given quantum state $\Psi$. It is clear that this equation, taken seriously, gives a nonlinear evolution for the wave function. \eref{eqn:sce}~has a long history, dating
back to the works of M\o{}ller~\cite{Moller:1962} and Rosenfeld~\cite{Rosenfeld:1963}.
It has been commented repeatedly that such a theory would be incompatible with established principles of
physics~\cite{Eppley:1977,Page:1981} but these arguments can be refuted~\cite{Mattingly:2005,Kiefer:2007,Mattingly:2006,Albers:2008}.

It is important to note that~\eref{eqn:sce} implies the conservation of energy-momentum, $\partial^\mu \bra{\Psi} \hat{T}_{\mu \nu} \ket{\Psi} = 0$, even if nonlinear terms are present, which ``collapse'' the wave function. 
This is not the case for the standard instantaneous collapse in quantum mechanics, as well as of spontaneous wave function collapse models, as discussed in \sref{sec:dp} and~\ref{sec:a}.

Let us expand the metric around the flat metric:
\begin{equation}
 g_{\mu \nu} = \eta_{\mu \nu} + h_{\mu \nu} \,,
\end{equation}
the expansion in $h_{\mu \nu}$ is well-known to yield the gravitational wave equations at leading order.
\eref{eqn:sce}~approximates to~\cite{Misner:1973p435}
\begin{equation}
\label{eqn:grav-wave}
 \Box h_{\mu \nu} = - \frac{16 \pi G}{c^4}  \left(\bra{\Psi} \hat{T}_{\mu \nu}^{(0)} \ket{\Psi}
- \frac{1}{2} \eta_{\mu \nu} \bra{\Psi} \eta^{\rho \sigma} \hat{T}_{\rho \sigma}^{(0)} \ket{\Psi} \right) \,,
\end{equation}
where the de Donder gauge-condition $\partial^\mu(h_{\mu\nu}-\frac{1}{2}\eta_{\mu\nu} \eta^{\rho \sigma} h_{\rho \sigma}) = 0$ has been imposed.
Note that the energy-momentum tensor at this order of the linear approximation is that for flat spacetime, while in~\eref{eqn:sce}
it was still in curved spacetime.
In the Newtonian limit, where $\bra{\Psi} \hat{T}_{00}^{(0)} \ket{\Psi}$ is large compared to the other nine
components of the energy-momentum tensor, \eref{eqn:grav-wave}~becomes the Poisson equation
\begin{equation}
\label{eqn:poisson}
 \nabla^2 V = \frac{4 \pi G}{c^2} \, \bra{\Psi} \hat{T}_{00}^{(0)} \ket{\Psi} 
\end{equation}
for the potential $V = -\frac{c^2}{2} h_{00}$.
This is of course a well known result: in the Newtonian limit, Einstein's equations tell that the matter distribution in space generates a Newtonian gravitational potential. In this case, however, the matter distribution is the expectation value of the quantized form of the stress-energy tensor.

The gravitational potential enters the matter Hamiltonian, which eventually drives the quantum dynamics. Within the linearized theory, the interaction Hamiltonian is (see \sref{sec:a}):
\begin{equation}
\label{eq:H_int}
 \hat{H}_\text{int} = -\frac{1}{2} \int \rmd^3 r \, h_{\mu \nu} \,\hat{T}^{\mu \nu} \,.
\end{equation}
It is important to point out the difference with respect to a standard quantized theory of gravity. In the latter,
$h_{\mu\nu}$ becomes an operator as well, simply by applying the correspondence principle to the
perturbation $h_{\mu\nu}$ of the metric---and thereby treating the classical $h_{\mu\nu}$ like a field living
\emph{on} flat spacetime rather than a property \emph{of} spacetime.
In contrast to this, $h_{\mu\nu}$ here remains fundamentally classical.
It is determined by the wave equation~\eref{eqn:grav-wave},
which is to be understood as classical equation of motion.

In the Newtonian limit, where $\hat{T}_{00}^{(0)}$ is the dominant term of the energy-momentum tensor, the
interaction Hamiltonian then becomes
\begin{equation}
\label{eqn:hint}
 \hat{H}_\text{int} = \int \rmd^3 r \, V \,\hat{T}^{00} = -G \int \rmd^3 r \, \int \rmd^3 r'\, \frac{\bra{\Psi}
\hat{\varrho}({\bf r}')\ket{\Psi}}{\abs{{\bf r} - {\bf r}'}} \, \hat{\varrho}({\bf r}) \,,
\end{equation}
where~\eref{eqn:poisson} has been integrated, and  $\hat{T}_{00}^{(0)} = c^2 \hat{\varrho}$
in the nonrelativistic limit.
The second-quantized mass density operator $\hat{\varrho}$ is simply $m \hat{\psi}^\dagger \hat{\psi}$ when only one kind
of particle is present.
Therefore, following the standard procedure we end up with the \sne\ in Fock space:
\begin{eqnarray} 
\label{eqn:fock}
\fl
\rmi \hbar \frac{\partial}{\partial t} \ket{\Psi}
=  \Bigg[ \int \rmd^3 r \, \hat{\psi}^\dagger({\bf r}) \left(-\frac{\hbar^2}{2m} \nabla^2 \right) \hat{\psi}({\bf r}) \nonumber \\
 - G\, m^2 \int \rmd^3 r \, \int\rmd^3 r' \, \frac{\bra{\Psi}
\hat{\psi}^\dagger({\bf r}') \hat{\psi}({\bf r}')\ket{\Psi}}{\abs{{\bf r} - {\bf r}'}} \, \hat{\psi}^\dagger({\bf r}) \hat{\psi}({\bf r}) \Bigg] \ket{\Psi}\,.
\end{eqnarray}

In the nonrelativistic limit the number of particles is conserved and we can immediately write the $N$-particle first-quantized \sne~\cite{Diosi:1984}:
\begin{eqnarray}
\label{eq:n-particleSN}
\fl
\rmi\hbar \frac{\partial}{\partial t} \psi_N(t, \{ {\bf r}\})
= \left[ -\sum_{i=1}^N\frac{\hbar^2}{2m}\nabla^2_i \right. \nonumber \\
\left. -G m^2 \sum_{i,j=1}^N \int\prod_{k=1}^N  \rmd^3 r'_k 
 \frac{\abs{\psi_N(t, \{ {\bf r}' \})}^2}{\abs{{\bf r}_i - {\bf r}'_j}}
\right]\psi_N(t, \{ {\bf r} \})\,,
\end{eqnarray}
where $\{ {\bf r} \} = {\bf r}_1, \ldots , {\bf r}_N$. The one-particle \sne~\eref{eqn:sn} follows directly by taking $N = 1$. One should stress once again that this nonlinear equation is the weak-field, Newtonian and nonrelativistic limit of~\eref{eqn:sce}, which itself is nonlinear with respect to the wave function and is assumed to be a fundamental equation.

Coming back to the attractive, ``collapse-like'' effect of the \sne, one should point out an important difference with respect to the von Neumann collapse postulate, and with respect to collapse models. 
Let us consider an experiment where a particle's position is measured. Take an initial superposition state for
the particle
\begin{equation}
 \psi({\bf r})=\frac{1}{\sqrt{2}}(\psi_1({\bf r})+\psi_2({\bf r})) \,,
\end{equation}
where $\psi_1({\bf r})$ and $\psi_2({\bf r})$ are wave packets well localized around ${\bf r}_1$ and ${\bf r}_2$,
respectively. 
During the measurement, this state couples with the massive measuring instrument (say, a pointer) as follows:
\begin{equation}\label{eqn:spatial-superpos-pointer}
 \Psi({\bf r},{\bf R})=\frac{1}{\sqrt{2}}(\psi_1({\bf r})\Phi_1({\bf R})+\psi_2({\bf r})\Phi_2({\bf R})) \,,
\end{equation}
where $\Phi_{1}({\bf R})$ and $\Phi_{2}({\bf R})$ are two localized wave functions of the pointer, centered
around ${\bf R}_1$ and ${\bf R}_2$, respectively.
The positions ${\bf R} = {\bf R}_{1,2}$
correspond to the particle being around positions ${\bf r}_{1,2}$.
Since the pointer is a classical system, according to the orthodox interpretation, the wave function collapses
at $ {\bf R} = {\bf R}_{1}$ or ${\bf R} = {\bf R}_2$, revealing in this way the outcome of the measurement. This
means that the particle is found half of the times around the position ${\bf r}_1$ and half of the times around
the position ${\bf r}_2$. Collapse models provide a similar description.

According to the \sne\ without the standard collapse postulate, on the other hand,
a superposition state as in~\eref{eqn:spatial-superpos-pointer} implies a gravitational attraction 
between the spatial wave packets $\Phi_{1}$ and $\Phi_2$ representing the massive pointer, on top of the standard quantum dynamics. The wave function of
the pointer  ``collapses'' towards the average position $({\bf R}_1 + {\bf R}_2)/{2}$, simply due to the
symmetry of the deterministic dynamics and the initial state.
Numerical simulations confirm this behavior of spatial superpositions collapsing to an average 
position~\cite{Harrison:2003}. Such a behavior is however in obvious
contradiction with the standard collapse postulate, as well as with our everyday experience, where the pointer
is found with equal probability either at
${\bf R} = {\bf R}_1$ or at ${\bf R} = {\bf R}_2$, and never in the middle.
Moreover the \sne\ is deterministic and as such it cannot explain why quantum measurements occur randomly,
distributed according to the Born rule. 

There is another undesired feature of the \sne, connected to the previous result. Being a nonlinear and deterministic equation, it allows for superluminal signaling~\cite{Gisin:1989,Polchinski:1991}. This can be seen explicitly by considering Alice and Bob, apart from each other, each having a particle, the two being in  a singlet spin state~\cite{Bahrami:2014}. If Alice measures the spin along a chosen $z$ direction, then in a way or another Bob's particle will have spin either up or down along the $z$ direction (if this does not happen, then the situation violates experimental evidence). If he uses a Stern-Gerlach setup to decide in which spin state his particle is, the particle will be deflected either upwards {\it or} downwards. A spot will appear either in the upper or lower part of a screen positioned right after the setup. 

If on the other hand Alice measures the spin along the $x$ direction, then Bob's particle will have definite spin along the same direction and if he uses the previous Stern--Gerlach setup to decide the spin state of the particle (Alice is far away, he does not know which type of measurement she performed), then the particle's state will become a superposition of being  deflected either upwards {\it and} downwards. Then the \schr--Newton term will (slightly) attract the two pieces of the superposition towards each other, and in the end spots appear on the screen, but not in the same position as in the previous situation. In this way Bob can understand which type of measurement Alice performed, and the two can establish a protocol for superluminal signaling. 

An important remark is at order. The kind of faster-than-light signaling here discussed  is an effect
of the (more or less) instantaneous collapse of the wave function (as a result of Alice's measurement), together with the nonlinear
character of the dynamics described by the \sne. Therefore, even if one describes the whole situation in a fully
relativistic way (i.\,e.~by some sort of ``Dirac--Newton equation'', which one could eventually obtain by
applying~\eref{eqn:sce} to a Dirac field), one would not get rid of the instantaneous collapse of the wave function
upon measurement, nor of the nonlinear character of the dynamics. What would change is the way the two parts of
the superposition attract each other: in the \sne\ this attraction is instantaneous, while in the relativistic
framework it would likely have a finite speed. This amounts in slight differences in the self-gravitation
effects, which do not play any important role for the argument proposed here. As long as there is some measurable
effect of self-gravitation, Bob can always exploit it to figure out Alice's measurement setting, and thereby
receive a signal with the ``speed of collapse'' (which is infinite in the standard collapse prescription and
has been shown to exceed the speed of light by orders of magnitude in a multitude of 
experiments~\cite{Stefanov:2002,Scarani:2014,Bancal:2012}). Note that collapse models do not have this problem, as discussed before in this section.

Given the above arguments, the \sne~poses serious problems if  regarded as a resolution of the measurement problem in quantum mechanics, equivalently as a way to describe the quantum-to-classical transition. It must be heavily modified in order to achieve this goal, perhaps to the point of betraying the original arguments that led to proposing it in the first place. On the other hand, it is an alternative to {\it quantum gravity}, meaning with it the program of quantizing gravity, and differs from it in a quantitative and computable way. It can serve as a model to decide whether gravity is quantum or not. For example, consider the question: what is the gravitational field generated by a quantum superposition in space? Is it the superposition of the two gravitational fields, as predicted by quantum gravity, or something else? The \sne~predicts it is something else---the classical sum of the two fields generated by the two terms of the superposition. Then matter in such a state should behave differently than what predicted by quantum gravity, even in the nonrelativistic and weak field limit. These differences can be quantified by solving~\eref{eqn:sn} and detected, as discussed in \sref{SNtest}. The \sne\ has the merit of providing a guideline for challenging quantum gravity at low energies, where experiments are cheaper and more flexible.

This is the subject of the next section, where we will review experiments probing this interplay between quantum mechanics and gravity. 

Before concluding this section, it is worth remarking that spontaneous wave function collapse models, as extensions of the standard unitary quantum evolution, open a new possibilities to tackle some of the open problems in cosmology. For example, it has been suggested that their non-unitary dynamics offers a natural resolution to the black hole information paradox~\cite{Okon:2015,Modak:2014}, to the dark energy problem~\cite{Josset:2017} (as they violate energy-momentum conservation) and to the measurement problem in the cosmological context~\cite{Martin:2012}. A cosmological constant term can be obtained also by classical models of gravity~\cite{Altamirano:2017}. 

\section{Table-top experiments to probe gravity}\label{sec:experiments}
\begin{figure}[htbp]
\begin{center}
\includegraphics[width=1\linewidth]{{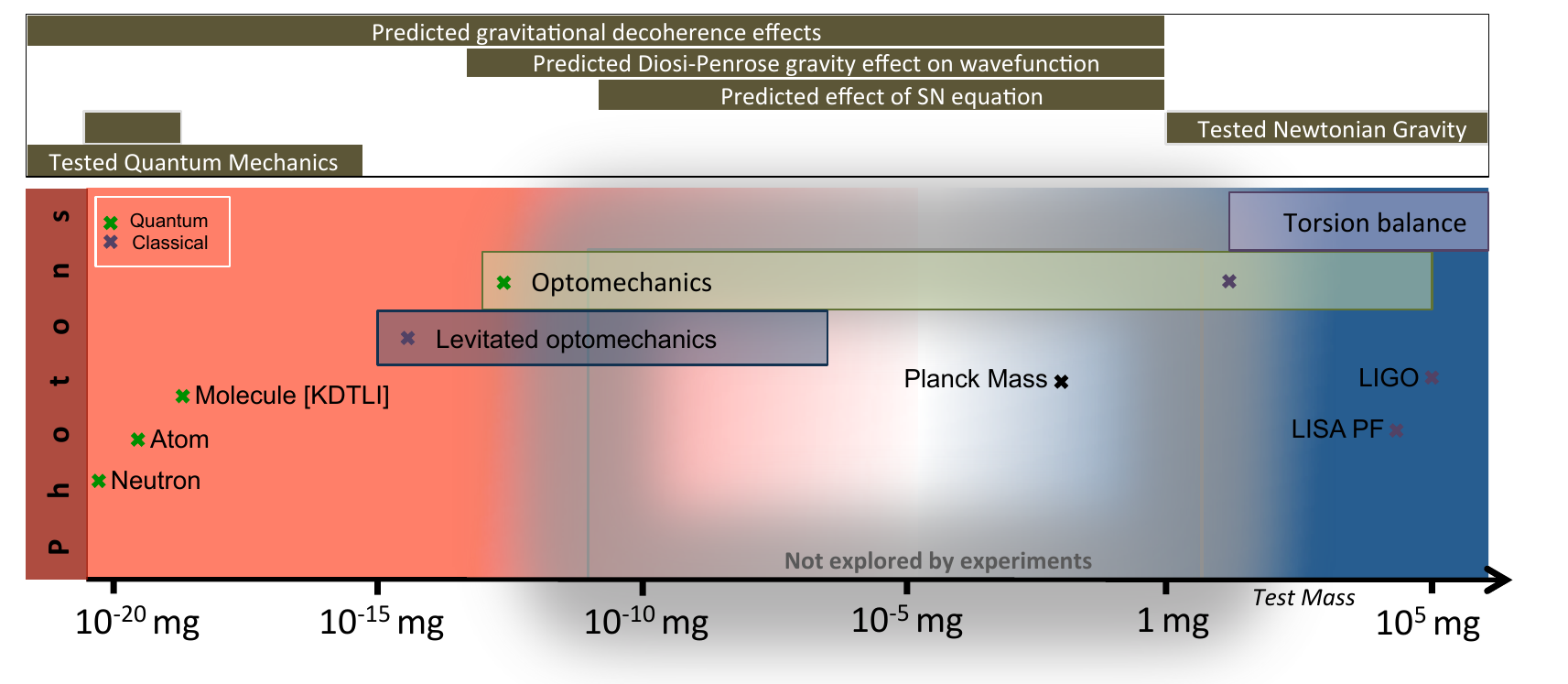}}
\caption{{\bf Exploration map of mass:} Mass range of the test mass as explored by experiments. Experiments to detect gravity have been done in the classical domain, {\it right hand side of picture}, with comparable large masses. Quantum experiments are routinely performed by using objects of much smaller masses so that gravity effects do not become visible or relevant. Neutron and atom matter-wave interferometers are different as the test mass there is very small (the mass of a single neutron or atom), but in a spatial superposition state. The desired mass range for---at least some of---the experiments summarized in this review article is at the overlap between sufficiently large mass to see significant effects of gravity of the particle itself, while the particle can be maintained in a non-classical state. The domain where massive particles can be prepared in such non-classical states is {\it on the left hand side} of the picture.}
\label{massrange}
\end{center}
\end{figure}

Here we summarize the state-of-the art of precise Newtonian and non-Newtonian gravity measurement and the feasibility of experimental tests of gravity effects in the quantum domain. We will mainly be concerned with table-top experiments in the nonrelativistic regime as such experiments may provide a new access to shine light on the quantum and gravity interplay. Therefore the main emphasis is to explore possible routes to enter the new parameter regime, where both quantum mechanics and gravity are significant, see \fref{massrange}. This means the mass of the object has to be large enough to show gravity effects while also not being too large to still allow for the preparation of a non-classical behavior of that massive object. That regime where both physical effects, the quantum and the gravity, could be expected to be relevant is at around the Planck mass~\cite{planck1906vorlesungen}, which is derived from the right mixture of fundamental constants ($\hbar$ Planck's constant, $c$ speed of light, $G$ gravitational constant) $\mplanck= \sqrt{\hbar c/G} =2.176470(51) \times 10^{-8}~\kilogram$ (based on the official CODATA value for all fundamental constants as announced by NIST~\cite{mohr2016codata}) or below. No quantum experiment has been performed in that mass range, but neither has an experiment to probe gravity. The smallest source masses which have been used to verify gravity are of the order of $10^{-2}~\kilo\gram$ which was at a distance of $10^{-2}~\meter$ from the test mass~\cite{ritter1990experimental}.

When we refer to quantum mechanical behavior of massive systems, we mean the center-of-mass motion of such a system, which may consist of many atoms. Surely, there are many other (we call those internal) degrees of freedom of the same system such as electronic states or vibrations and rotations which are described as relative motions of the atoms forming the large object, but here we are not concerned with those. When we talk about superpositions, we mean spatial superpositions, in the sense of the center of mass of a single particle, which can be elementary or composite, being {\it here} and {\it there} at a given time (the \schr\ cat state). The most massive complex quantum systems, which have been experimentally put in such a superposition state, are complex organic molecules of a mass on the order of $m_\text{max}=10^{-22}~\kilo\gram$~\cite{Gerlich:2007, gerlich2011quantum, Eibenberger:2013}.

While the emphasis of this review is not so much to explore the regime where a quantum system is coupled to a large external mass as in the seminal neutron interferometer experiment by Colella, Overhauser and Werner (COW)~\cite{Colella:1975} or atom interferometry~\cite{peters1999measurement} experiments, we will also mention the state of the art of those. Strikingly, those matter-wave interferometry experiments show that gravity effects of comparably large external masses can be detected by quantum states of much lower mass objects, which may be used as motivation for perspectives of table-top experiments to investigate the interface of quantum and gravity, while at the same time it is clear that the type of gravity investigated in such cases is of Newtonian type in the first place. In any case we will try to work out the difference between such regimes---the low mass and the high mass regimes---to also emphasize the possibility to use photons, objects of zero rest mass, to watch out for gravity induced effects on quantum systems such as decoherence~\cite{zych2012general}.

One of the proposed ideas~\cite{Ralph:2009} takes a quantum information point of view on the propagation of an entangled photon pair in curved spacetime and predicts a detectable gravity induced decoherence effect, which was coined with the name gravity induced entanglement decorrelation. A space based satellite experiment involving the preparation and detection of a pair of polarization entangled photons is summarized in a recent review article on fundamental quantum optics experiments with satellites~\cite{Rideout:2012}.
\begin{figure}[t!]
\begin{center}
\includegraphics[width=0.45\linewidth]{{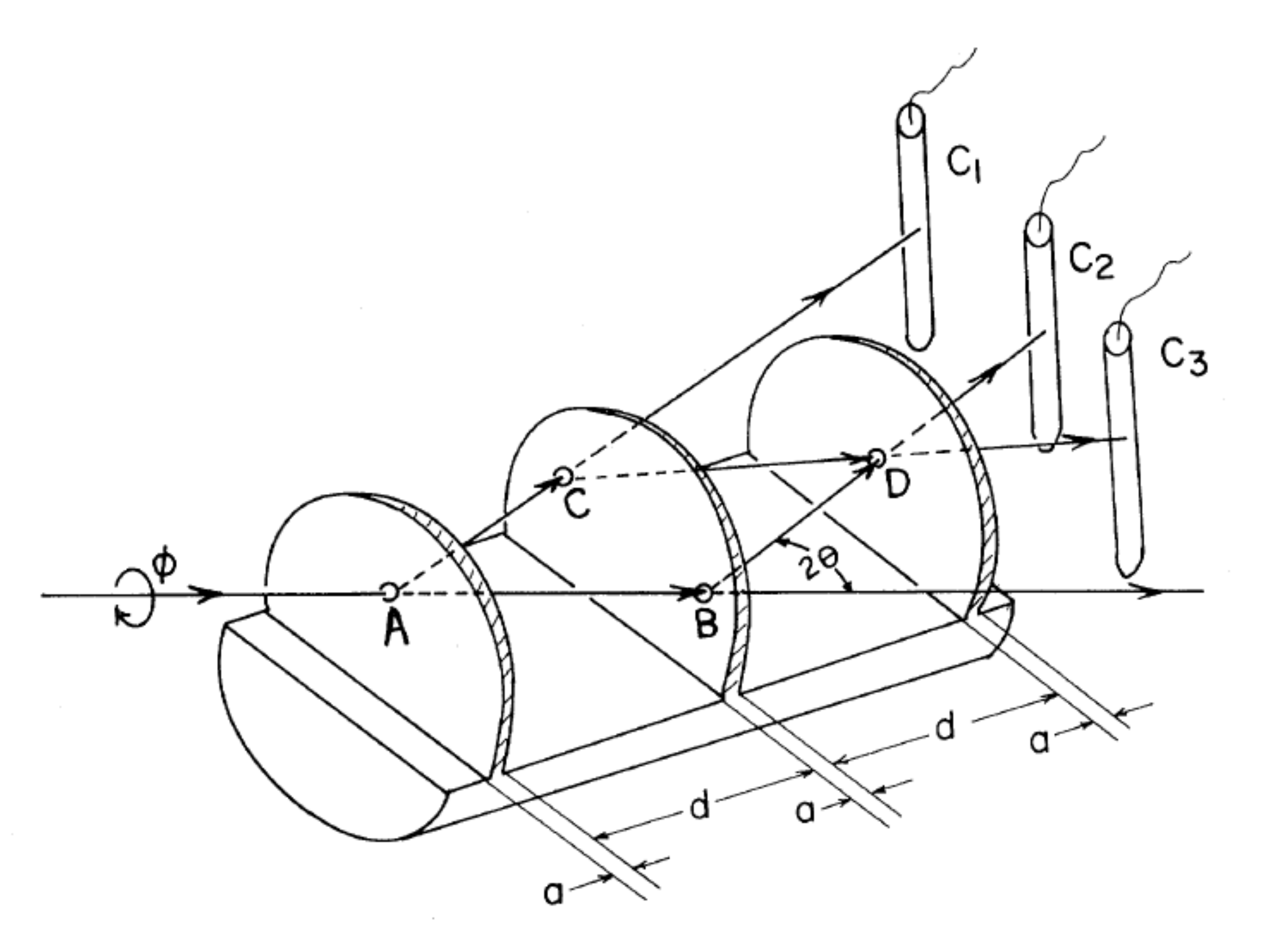}}
\hspace{0.05\linewidth}
\includegraphics[width=0.3\linewidth]{{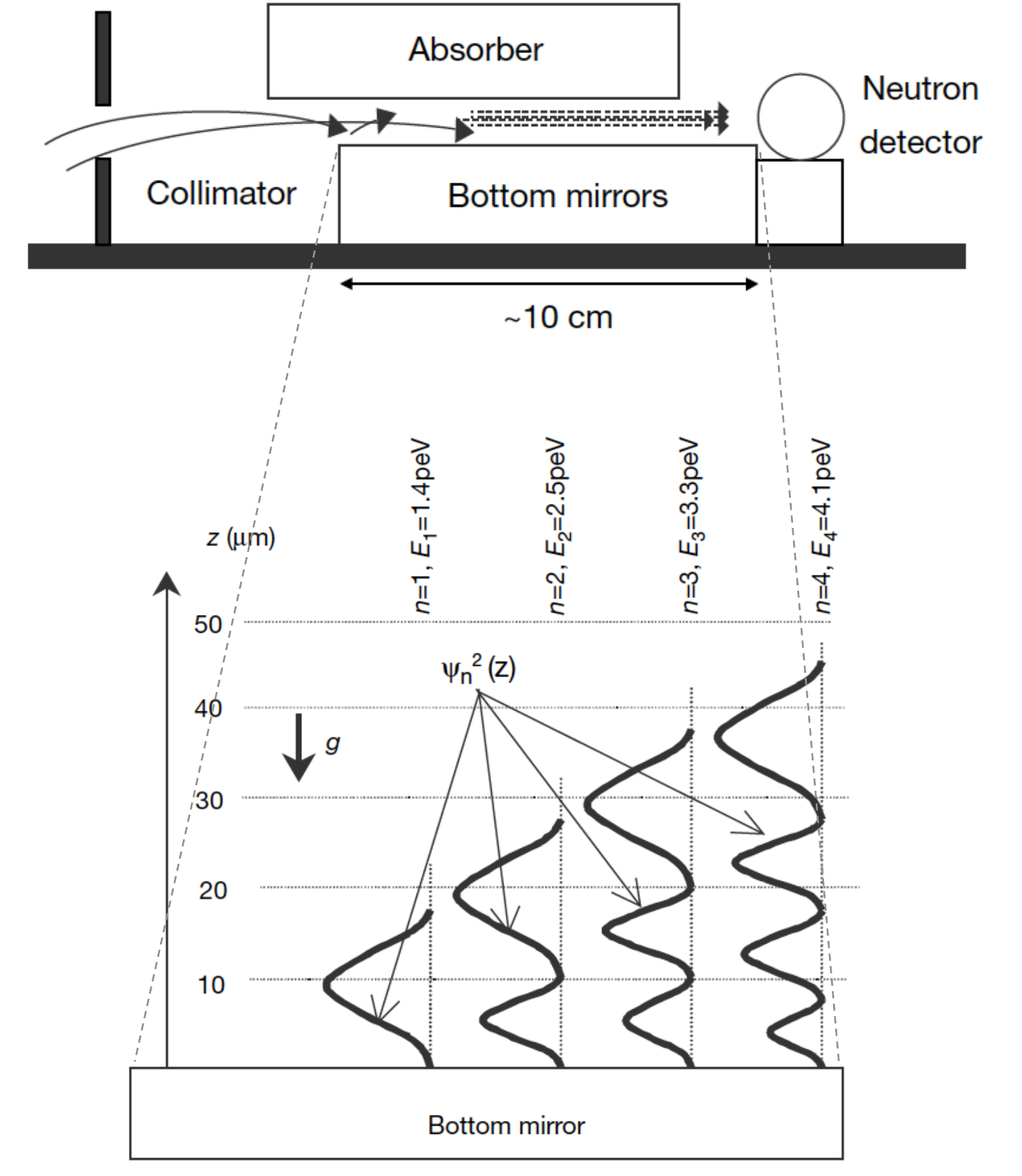}}
\caption{{\bf Neutrons to measure gravity:} {\it Left Panel}: The schematics of the seminal COW experiment. Neutrons matter-wave functions are split and recombined by diffraction at solid crystals to from a closed Mach-Zehnder interferometer. The dimensions of the experimental setup are on the order of cm. The whole setup is then rotated with respect to the $g$-field. The respective gravitational phase shift is detected as a lost of fringe visibility of the interference pattern. Reprinted figure with permission from~\cite{Colella:1975}. Copyright 1975 by the American Physical Society. {\it Right Panel}: A beam of neutrons is passes horizontally above a mirror surface. The neutron matter-waves are shown to be able to only occupy a discrete set of energy states which can be explained including the effect of Newtonian gravity. Reprinted by permission from Macmillan Publishers Ltd: Nature~\cite{nesvizhevsky2002quantum}, copyright 2002.}
\label{Neutrons}
\end{center}
\end{figure}

There are also other proposals for cosmological effects related to gravity in the context of dark matter~\cite{riedel2013direct, bateman2015existence} and dark energy~\cite{burrage2015probing} or for the detection of gravitational waves~\cite{dimopoulos2009gravitational, arvanitaki2013detecting}, which have been proposed and some experiments have been done already with atom interferometers~\cite{hamilton2015atom} and torsion balances~\cite{Kapner:2007}, which we will not explore in too much detail here. See \sref{sec:gd-class} for an account on the decoherence of matter-waves by gravitational waves.

Typically for gravity experiments there are two masses involved, the (usually large) source mass which generates a significant curvature of spacetime (i.\,e.~gravitational field or potential), and the test mass which is probing the gravity effect generated by the source mass. Torsion balances are the classic device for gravity experiments~\cite{cavendish1798experiments}. One can single out two regimes interesting for experimental investigation: i) The regime where a quantum system is the test mass and interacts with a large external source mass. This is the regime where neutron and atom interferometry are already very successful and provide tools for precise measurements of gravity effects. ii) The regime where the quantum system itself carries sufficient mass to be the source mass and to allow for related quantum gravity effects to become experimentally accessible. So far there has been no convincing experiment in the second regime. Any experiment performed in that second regime will ultimately give insight into the interplay between gravity and quantum mechanics. Test of the \sne\ and of quantum effects in gravity fall in the latter regime. It may very well be that there are surprises waiting for us if we become able to probe that regime by experiments.  

The outline of the experimental section is as follows. We first summarize experimental tests of classical gravity by neutron and atom interferometers as well as by torsion balances and optomechanics experiments in \sref{ExpCG}. Then we discuss experimental prospects for testing the \sne\ in \sref{SNtest}, then proposed tests of quantum gravity in \sref{QGtest}. We then discuss proposals for tests of gravitational decoherence in \sref{Gdeco} as well as the idea to directly measure the gravity generated by a quantum superposition state in \sref{Gsup}.


\begin{table}
\caption{\label{arttype}The table is adapted and reprinted from~\cite{mohr2016codata}, with the permission of AIP Publishing. It shows all $G$ measurements included for the CODATA value as well as their uncertainties ($\Delta G/G$). Most of the measurements are taken by torsion balances, while only one, LENS-14, is an atom interferometer. 
NIST-82: National Institute of Standards and Technology, Gaithersburg, Maryland, and Boulder, Colorado, USA; 
TR\&D-96: Tribotech Research and Development Company, Moscow, Russian Federation; 
LANL-97: Los Alamos National Laboratory, Los Alamos, New Mexico, USA; 
UWash-00: University of Washington, Seattle, Washington, USA; 
BIPM-01: International Bureau of Weights and Measures, Sèvres, France; 
UWup-02: University of Wuppertal, Wuppertal, Germany; 
MSL-03: Measurement Standards Laboratory, Lower Hutt, New Zealand; 
HUST-05 \& HUST-09: Huazhong University of Science and Technology, Wuhan, PRC; 
UZur-06: University of Zurich, Zurich, Switzerland; 
JILA-10: JILA, University of Colorado and National Institute of Standards and Technology, Boulder, Colorado, USA; 
LENS-14: European Laboratory for Non-Linear Spectroscopy, University of Florence, Florence, Italy; 
UCI-14: University of California, Irvine, Irvine, California, USA.
}
\footnotesize\rm
\begin{tabular*}{\textwidth}{@{}l*{15}{@{\extracolsep{0pt plus12pt}}l}}
\br
Identification&10$^{11} G$ (m$^3$kg$^{-1}$s$^{-2}$)& $\Delta G/G$ \\
\mr
NIST-82, Fiber torsion balance&6.672 48(43)&6.4$\times 10^{-5}$\\
TR\&D-96, Fiber torsion balance&6.672 9(5)&7.5$\times 10^{-5}$\\
LANL-97, Fiber torsion balance&6.673 98(70)&1.0$\times 10^{-4}$\\
UWash-00, Fiber torsion balance&6.674 255(92)&1.4$\times 10^{-5}$\\
BIPM-01, Strip torsion balance&6.675 59(27)&4.0$\times 10^{-5}$\\
UWup-02, Suspended body&6.674 22(98)&1.5$\times 10^{-4}$\\
MSL-03, Strip torsion balance&6.673 87(27)&4.0$\times 10^{-5}$\\
HUST-05, Fiber torsion balance&6.672 22(87)&1.3$\times 10^{-5}$\\
UZur-06, Stationary body&6.674 25(12)&1.9$\times 10^{-5}$\\
HUST-09, Fiber torsion balance&6.673 49(18)&2.7$\times 10^{-5}$\\
JILA-10, Suspended body&6.672 34(14)&2.1$\times 10^{-5}$\\
BIPM-14, Strip torsion balance&6.675 54(16)&2.4$\times 10^{-5}$\\
LENS-14, Double atom interferometer&6.671 91(99)&1.5$\times 10^{-4}$\\
UCI-14, Cryogenic torsion balance&6.674 35(13)&1.9$\times 10^{-5}$\\
\br
\end{tabular*}
\end{table}

\subsection{Table-top experimental tests of Newtonian gravity and general relativity} \label{ExpCG}

In order to reach the regimes where both gravity and quantum effects must be taken into account,
one can either start from microscopic quantum systems and attempt to increase their masses such that gravitational effects become relevant, or one approaches from the opposite direction of gravitating systems, reaching for ever smaller system sizes, in order to reach the quantum regime.

For the latter approach, we will summarize the state of the art to detect {\it classical} Newtonian gravity, as well as some general relativity effects, which amount essentially to observing the drop of different sorts of ``apples'' and measure the duration and distance of the drop with exceedingly precise methods. Effectively---while testing the universality of free fall, and therefore one component of Einstein's equivalence principle, which is the basis of the theory of general relativity---such experiments perform a measurement of the gravitational constants $G$ and $g$; the latter in the case of a drop in the gravitational field of Earth and the former in relation to a further large source mass close to the test mass. 

We should not forget that amongst the group of broadly accepted fundamental constants of Nature, the gravitational constant $G=6.67408(31)\times 10^{-11}$m$^3$kg$^{-1}$s$^{-2}$~\cite{mohr2016codata}, is the one which is known with lesser precision by many orders of magnitude, namely $\Delta G/G =4.7\times 10^{-5}$. This fact is always assumed to be reasoned by the comparably much weaker strength of the gravitational interaction relative to the other three known fundamental forces, which makes the determination of $G$ by experiment notoriously harder. While the interaction itself can be measured with high precision (as outlined in the following), to know the exact distribution of mass in the bodies involved in the gravity test turns out to be the limiting factor~\cite{rosi2014precision}.
See \tref{arttype} for a summary of all experiments to measure $G$ which are included in the evaluation of the official $G$-value according to CODATA~\cite{mohr2016codata}. The data show the surprising wide spread of the experiment values.
We briefly review how the different experimental schemes work, both for measurements of $g$ and $G$, in classical and quantum set-ups.

\begin{figure}[t!]
\begin{center}
\includegraphics[width=0.3\linewidth]{{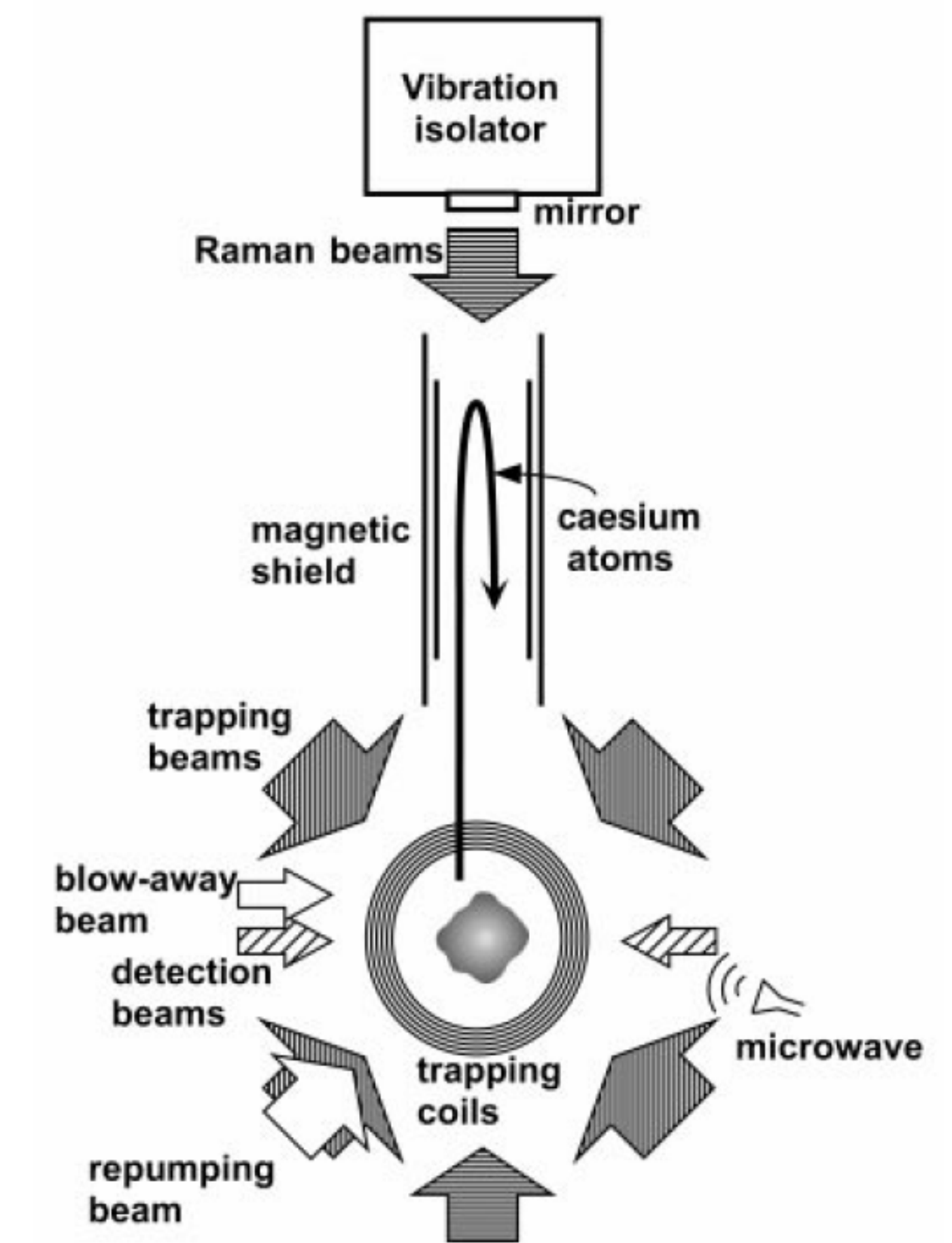}}%
\includegraphics[width=0.3\linewidth]{{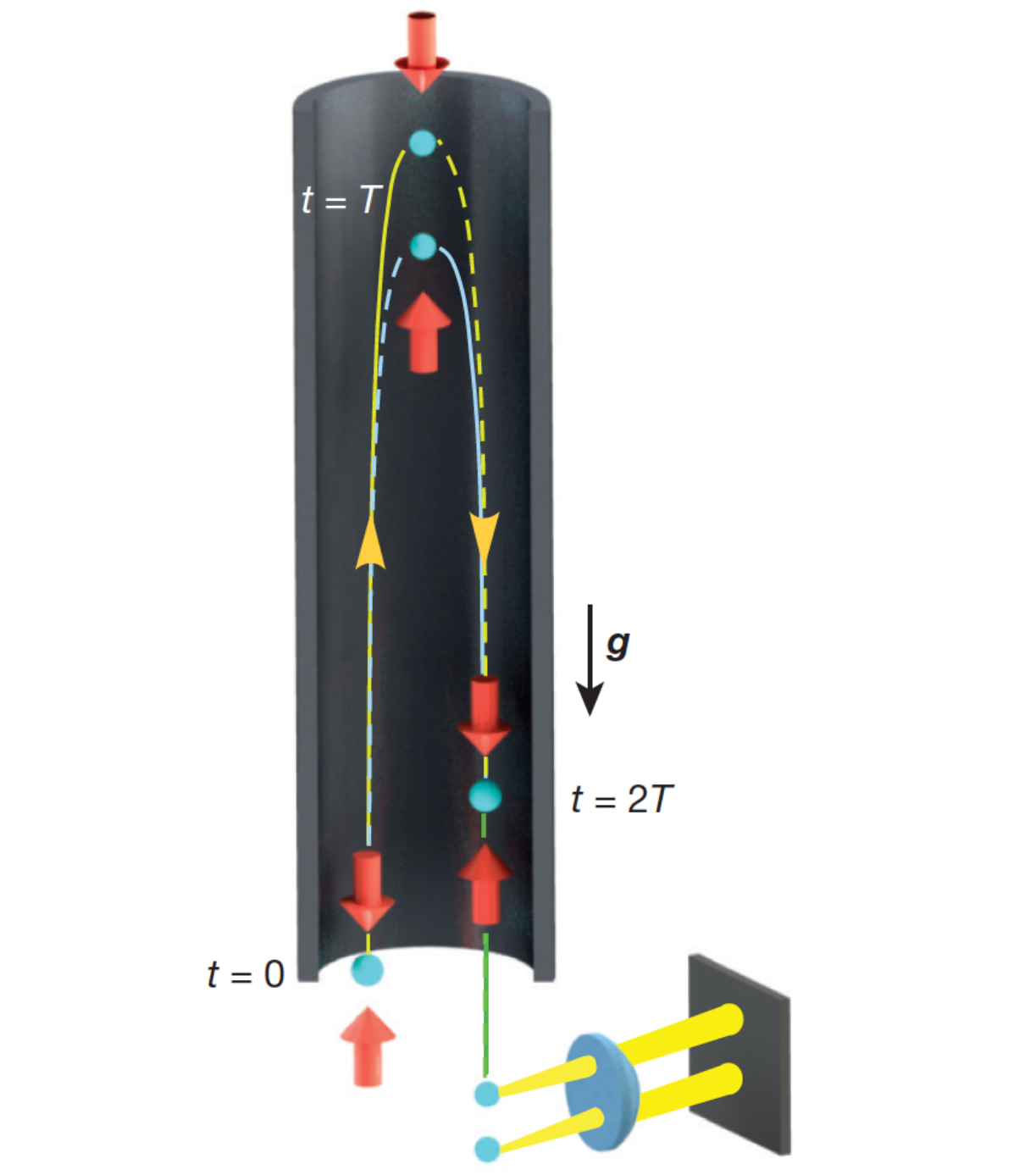}}%
\includegraphics[width=0.3\linewidth]{{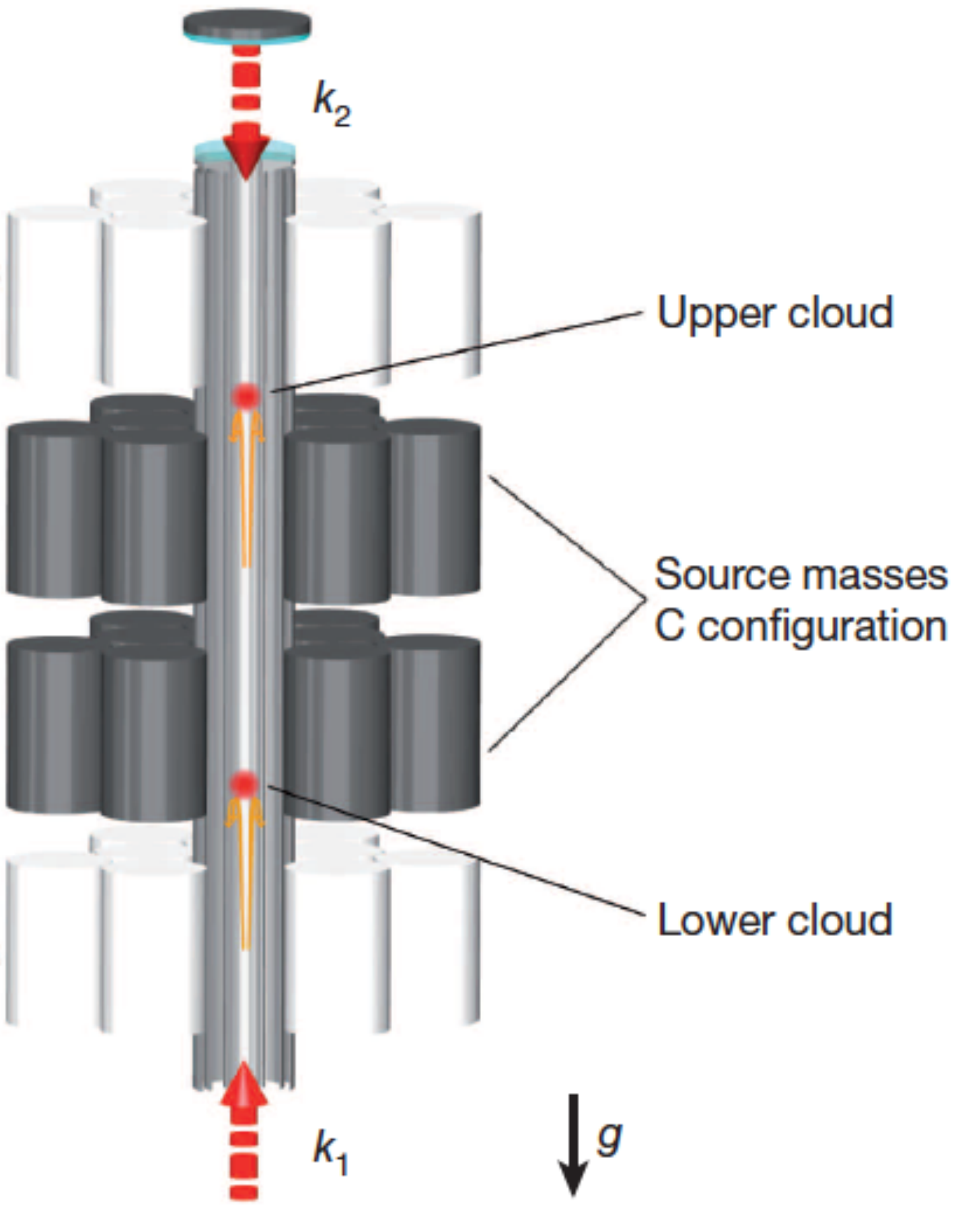}}
\caption{{\bf  Cold atoms to measure classical gravity:} {\it Left Panel}: First experiment to detect gravitational acceleration by Earth ($g$) in atom interferometry.
Reprinted by permission from Macmillan Publishers Ltd: Nature~\cite{peters1999measurement}, copyright 1999.
{\it Middle Panel}: The 10~\meter\ cold atom BEC fountain to generate single atom superposition states of 0.5 m size.
Reprinted by permission from Macmillan Publishers Ltd: Nature~\cite{kovachy2015quantum}, copyright 2015.
{\it Right Panel:} The assembly of big source masses is around the tube containing two clouds of cold atom test masses. The systematic comparison of the result of the atom interferometry is performed on both clouds while moving the source masses. The experiment has been used for a measurement of $G$.
Reprinted by permission from Macmillan Publishers Ltd: Nature~\cite{rosi2014precision}, copyright 2014.}
\label{Atoms}
\end{center}
\end{figure}

{\it A. Optical Interferometry of dropping corner cubes.---}The drop of reflective corner cubes is used to measure free fall time in Earth's gravity. Classical optical interferometry is used for very precise measurement of the drop/fall time. This technology is portable~\cite{zumberge1982portable} and a version has been used since the Apollo missions for lunar optical experiments~\cite{dickey1994lunar}.  The most precise version of a similar idea has been used in LISA pathfinder to realize a drag-free navigation of the satellite~\cite{armano2016sub} by mutually stabilizing two masses by means of optical interferometry, while the gravitational wave detectors such as LIGO~\cite{Abbott:2016} can be regarded as a static version of this experimental scheme.

{\it B. Neutron experiments.---}The experiment by COW~\cite{Colella:1975} is a recoverable matter-wave dephasing effect rather than the destructive effect of gravitational decoherence. Depending on the orientation of the two arms of the Mach Zehnder type neutron interferometer with respect to the Earth's gravitational potential, the matter-waves acquire different gravitational phases, which upon recombination result in an overall phase shift, see \fref{Neutrons} (left) for an illustration of the original experimental setup.

A further series of stunning experiment has been performed with neutrons traveling in the gravitational potential at the surface of the Earth~\cite{nesvizhevsky2002quantum}.
Such experiments reveal the existence of a discrete set of states to be occupied by the neutron wave function in the classical gravitational potential of the Earth, see \fref{Neutrons} (right). 

{\it C. Atom interferometers.---}The first atom interferometer experiment of the light-matter beam splitter type to measure a gravitational effect has been performed by Peters~\cite{peters1999measurement} by measuring the effect of acceleration $g$ on a cloud of cold atoms. The basic idea is similar to the COW neutron interferometer; the matter-wave couples to the external gravitational potential and acquires a related phase shift, which can in turn be measured very precisely by the interferometer, see \fref{Atoms} (left). That same experiment has been used for a high-precision gravity measurement~\cite{peters2001high}. Many revolutionary contributions to the field of detecting gravity with atom interferometers have been made by Kasevich~\cite{kasevich1991atomic, snadden1998measurement, chiow2011102, kovachy2015quantum}, see \fref{Atoms} (middle) for an illustration of the 10 m atomic fountain at Stanford. A very precise measurement of $g$ has also been performed by an atom interferometer experiment with Bose--Einstein condensed (BEC) atoms~\cite{poli2011precision}, while $G$ has also been measured by using a differential technique with two atomic clouds in different positions relative to an assembly of bigger source masses~\cite{rosi2014precision}, see \fref{Atoms} (right). The technology of cold atom and BEC interferometers for gravity sensing is very mature and is on the verge to commercialization~\cite{geiger2011detecting, bodart2010cold}. It has been demonstrated to work in extreme environments such as drop towers based on BEC interferometry on atom chip technology~\cite{van2010bose, Muentinga:2013} and has been proposed to be used in space experiments~\cite{tino2013precision} with first demonstrations of sounding rocket flights into space by the MAIUS collaboration~\cite{lachmann2017creating}.

\begin{figure}[t!]
\begin{center}
\includegraphics[width=0.6\linewidth]{{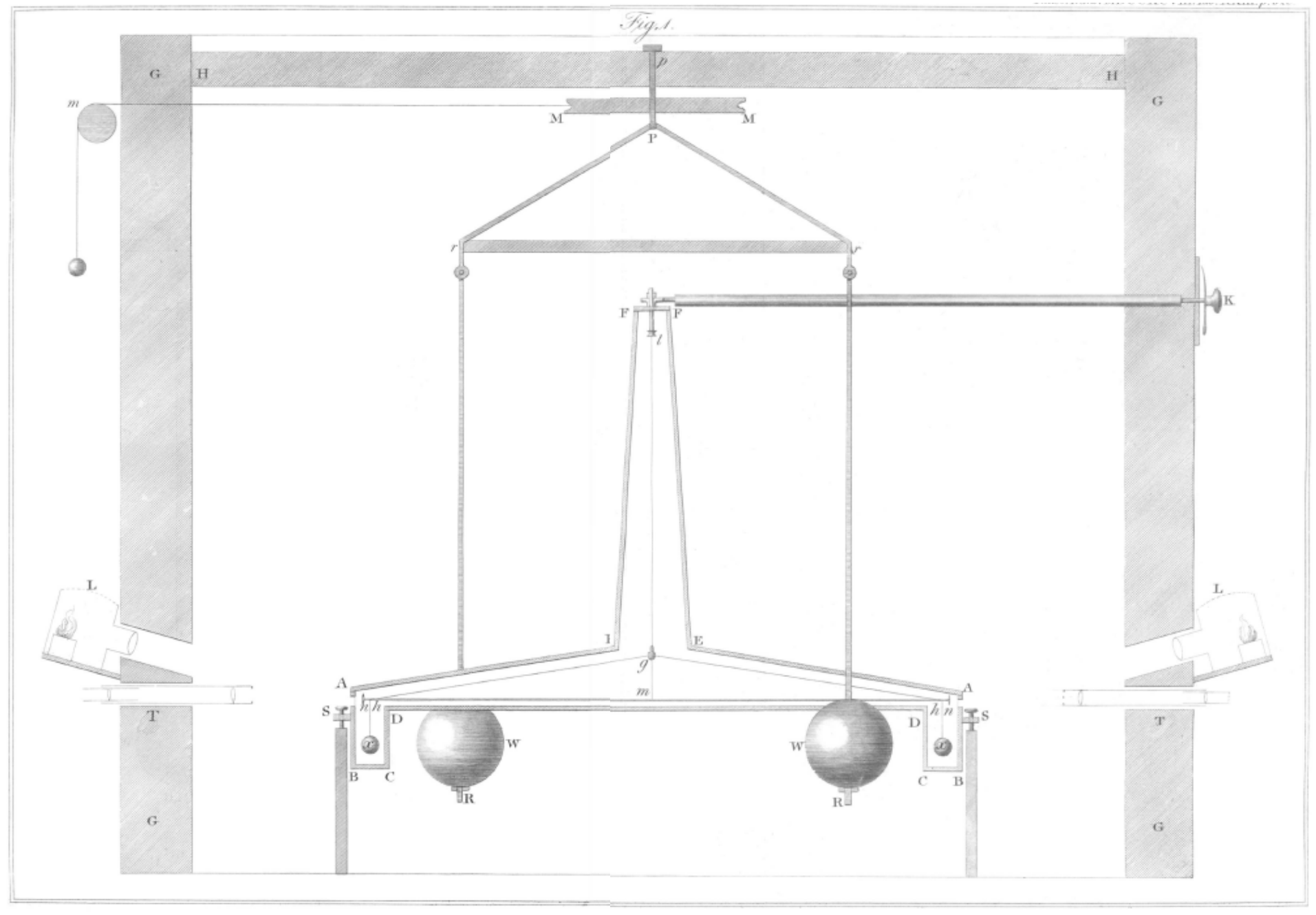}}%
\includegraphics[width=0.35\linewidth]{{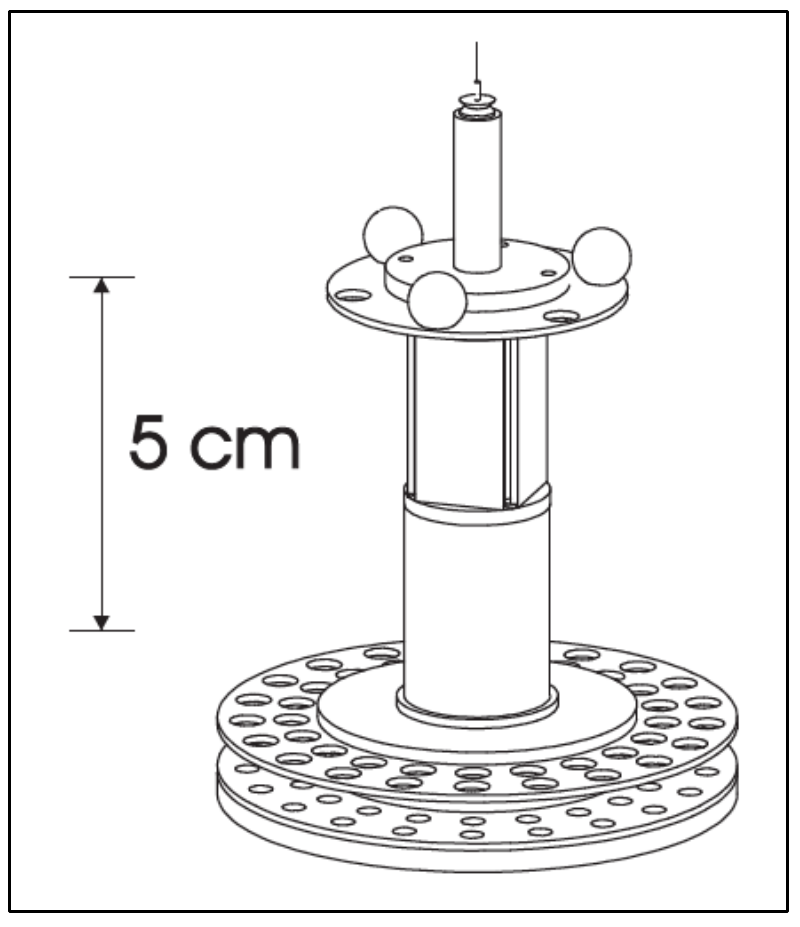}}
\caption{{\bf  Cavendish type torsion balance experiments:} {\it Left Panel}: The scheme of the historic Cavendish experiment. The attraction between the big source mass and the smaller test masses is measured precisely by detecting the slow torsion angle change. Picture taken from~\cite{cavendish1798experiments}.  {\it Right Panel}. The version of a modern torsion pendulum gravity sensor by Kapner \textit{et al}. This is an upgraded version of the so-called {\it missing mass} torsion balance instrument. This experiment has been use to test extensions of Newtons square-law gravity by Yukawa type extensions as predicted by models for dark energy.  Reprinted figure with permission from~\cite{Kapner:2007}. Copyright 2007 by the American Physical Society.}
\label{cavendish}
\end{center}
\end{figure}

{\it D. Torsion pendulums.---}While the first experiment has been performed by Cavendish in 1798~\cite{cavendish1798experiments}, the same basic principle is still used in the best experiments to measure classical gravity.  A set of two test masses attached to a torsion pendulum is positioned close to a set of larger masses and the gravitational attraction is monitored by measuring precisely the torsion angle, while the test mass assembly is moving slowly towards the bigger masses, see \fref{cavendish} (left). 

Modern versions of the same idea of a torsion balance are amongst the most sensitive gravity sensors, to test the inverse-square Newton law---even below the dark energy length scale~\cite{Kapner:2007}, see \fref{cavendish} (right). The very diverse optomechanical devices are related and will be discussed in the next section.

{\it E. Optomechanics experiments.---}Optomechanical devices span a wide range of sizes and masses. Crucially the optomechanical coupling allows for both ultra-precise detection of the mechanical motion and preparation of sensitive classical and non-classical states of the mechanics~\cite{Aspelmeyer:2014}. Optomechanical devices have reached the quantum domain which promises for ultra-precise gravity sensing and even the exploration of the large mass quantum domain, see \fref{massrange}. 

\begin{figure}[t!]
\begin{center}
\includegraphics[width=0.4\linewidth]{{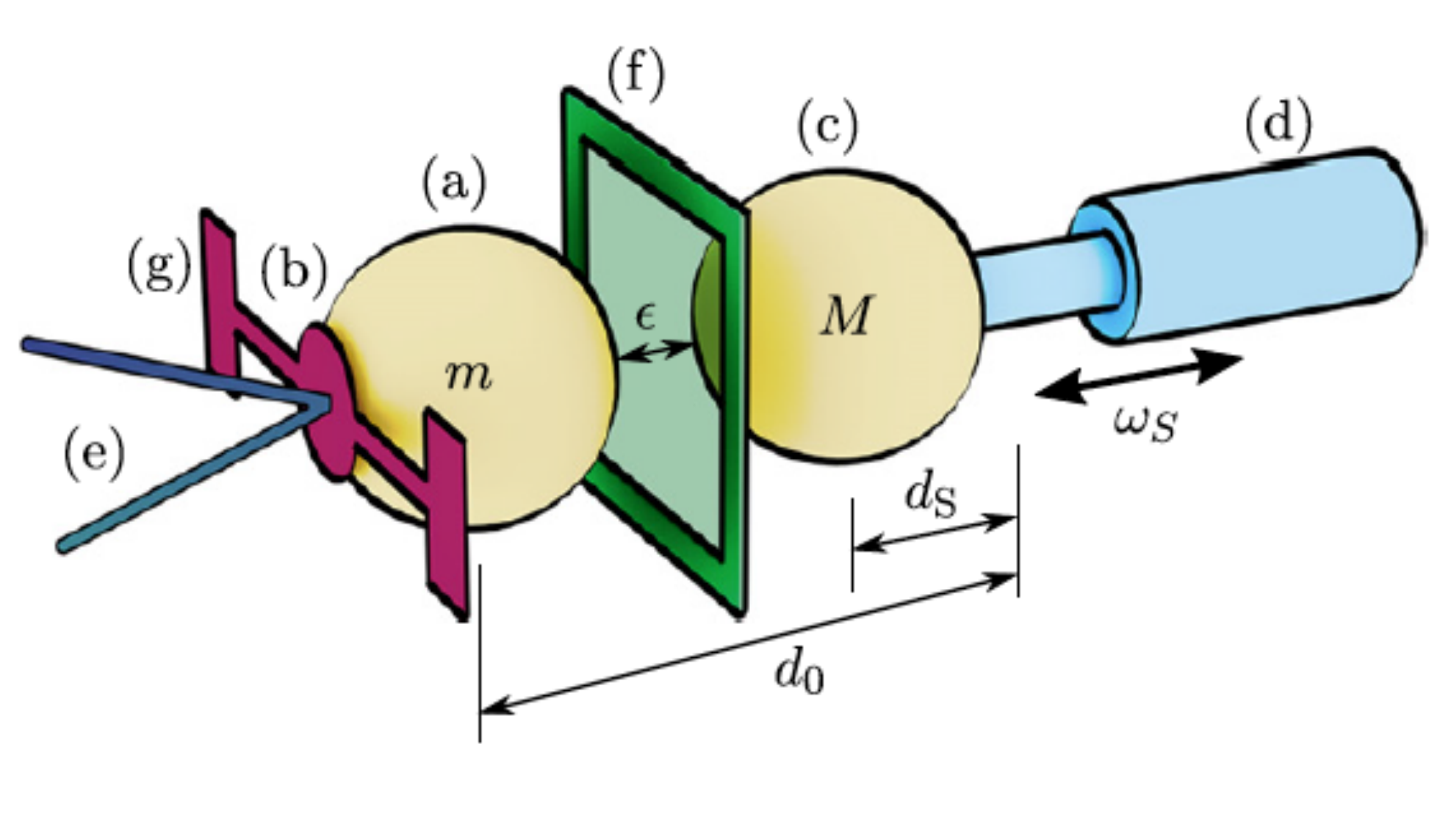}}
\hspace{0.1\linewidth}
\includegraphics[width=0.4\linewidth]{{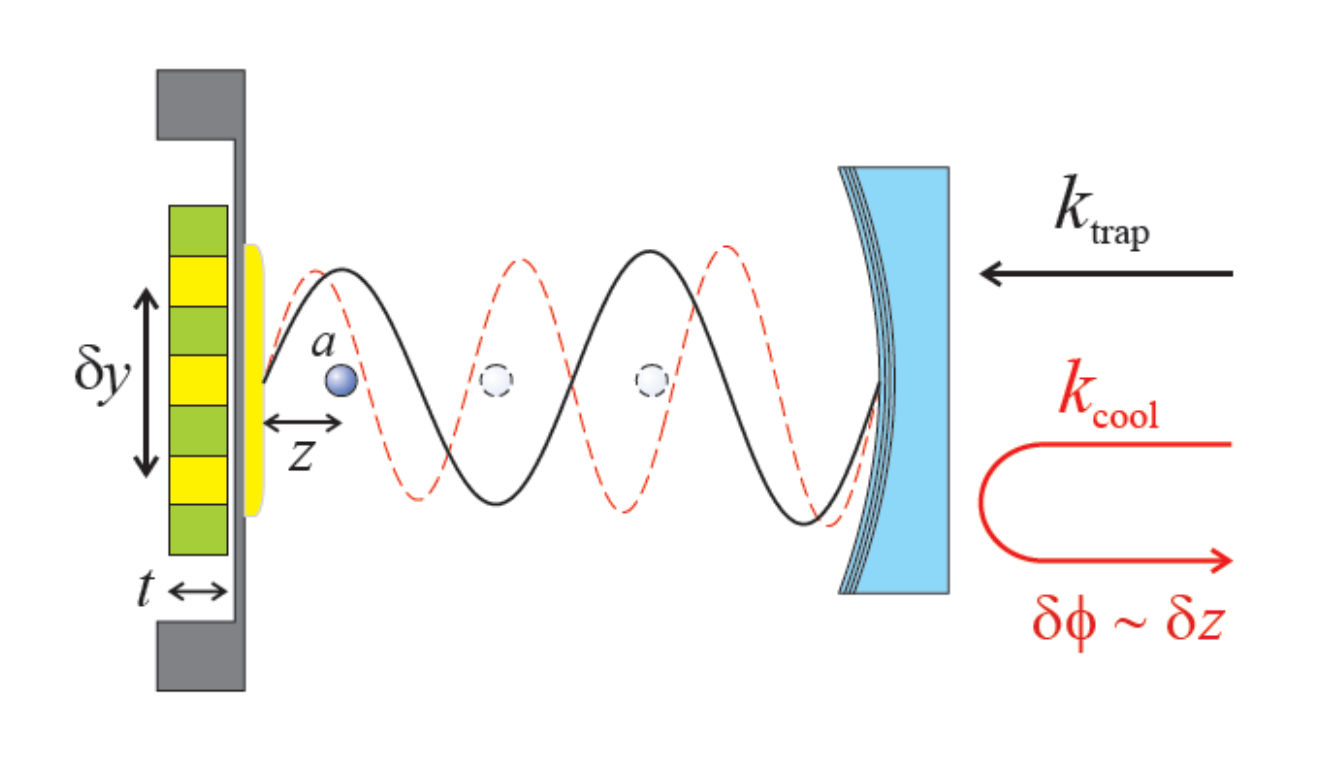}}
\caption{{\bf  Optomechanical experiments to test gravity:} {\it Left Panel}: A test-mass m (a) is loaded on a micromechanical device (b). A source mass M (c) is located at a COM distance $d_0$ from the test mass and is modulated
through a drive motor (d) with maximum amplitude $d_S$. The displacement of the test mass cantilever is read out optically (e). Other, non-gravitational forces are further suppressed by a shielding membrane (f). (g) labels the mounting support structure. Picture and caption taken from~\cite{schmole2016micromechanical}.  {\it Right Panel}. Levitated optomechanics proposal and experiment to test gravity. Proposed experimental geometry. A sub-wavelength dielectric microsphere of radius a is trapped with light in an optical cavity. The sphere is positioned at an anti-node occurring at distance $z$ from a gold-coated SiN membrane. Light of a second wavelength $\lambda_{cool}=2\lambda_{trap}/3$ is used to simultaneously cool and measure the center-of-mass motion of the sphere. The sphere displacement $\delta z$ results in a phase shift $\delta\phi$ in the cooling light reflected from the cavity. For the short-range gravity measurement, a source mass of thickness $t$ with varying density sections is positioned on a movable element behind the mirror surface that oscillates harmonically with an amplitude $\delta y$. The source mass is coated with a thin layer of gold to provide an equipotential. Reprinted figure and caption with permission from~\cite{geraci2010short}. Copyright 2010 by the American Physical Society.}
\label{optomech}
\end{center}
\end{figure}

First optomechanical devices including torsion balance type sensors have been realized and approached the standard quantum limit in sensing~\cite{kim2016approaching}. A first proposal has been suggested to use optomechanical devices to perform gravity tests with mg masses~\cite{schmole2016micromechanical}, which would push down the low mass limit for gravity experiments (see \fref{optomech}, left). Here we refer to the mass of the test mass itself, instead of the interaction with a big external mass as for atom interferometry. Clamped cantilever type devices coupled to optical~\cite{norte2016mechanical} and microwave cavities~\cite{teufel2011sideband} have been realized as well as levitated optomechanical devices~\cite{Chang:2010, romero2010toward, barker2010doppler, neukirch2015nano, yin2013optomechanics, kiesel2013cavity, gieseler2012subkelvin, millen2015cavity, asenbaum2013cavity, rashid2016experimental}. The latter are very young, being seemingly a hybrid between the nanofabricated optomechanical structures and the atomic and molecular matter-wave systems, have a great potential for gravity sensing~\cite{geraci2010short} as their mechanical quality factor ($Q$) is unprecedentedly high, as for example achieved for the driven rotation of a levitated nanorod: $Q=10^{11}$~\cite{Kuhn:2017} (an illustration of one of the proposals is shown in \fref{optomech}, right). Beside optical levitation, also magnetic levitation of superconductors is under development~\cite{Pino:2016, Prat-Camps:2017}. All this is certainly of high potential, but still has to be shown to work as an ultra-precise gravity sensor in the quantum domain.  

A clear challenge for all the nanometer and micrometer scale gravity tests---such as the optomechanics based experiments---is the competition with other interactions in the close proximity of surfaces, such as for example van der Waals and Casimir--Polder interactions. The latter can easily overwhelm the effects of gravity, while clever experimental tricks have been proposed to keep such dispersion force effects constant while the gravity effect is modulated~\cite{geraci2010short}. Needless to say, all sorts of noise and heating effects generated by the trapping field, the environment, collisions with background gas particles, and so on, will set bounds on the detectable magnitude of the desired gravity effect.

{\it F. Tests of general relativity by different types of experiments.---}Beside the experiments to measure classical Newtonian gravity there are also attempts to detect effects as predicted by classical general relativity. At present, most experiments are on the proposal level or are attempted based on technologies with the highest sensitivities as general relativity effects are usually smaller than the Newtonian effects. No deviations from predictions of general relativity have been reported. Torsion pendulums, cold atom interferometers, and atomic clocks are typical (proposed and used) devices for such tests. This is a very broad and active field and we can here only mention some experiments performed. Examples include: i) Tests of the equivalence principle~\cite{schlamminger2008test, vessot1979test, lammerzahl2001gyros, hohensee2011equivalence}. A very creative idea for an experiment in this context involved mimicking of  gravitation by acceleration of nanomechanical elements~\cite{katz2015mesoscopic}. ii) Tests of the universality of free fall, where two test masses of different mass are dropped at the same time and their free fall times are compared. Such experiments have been performed with two different atomic species on Earth~\cite{schlippert2014quantum} and have been proposed for space~\cite{aguilera2014ste}. iii) Tests of Lorentz invariance, a topic, with a dedicated review article~\cite{mattingly2005modern}. iv) The proposals for using split path interferometers (for light or for matter) to detect general relativistic time dilation effects. We will come back to this topic in \sref{Gdeco} on gravitational decoherence. v) And last but not least, the general relativistic effect of {\it gravitational red shift} which has attracted a lot of discussion in the atom interferometry community~\cite{Mueller:2010, Wolf:2010, Mueller:2010a, wolf2011does, hohensee2012comment, Giulini:2012b, lan2013clock}.

There are many more experiments to test general relativity such as the Shapiro experiment~\cite{shapiro1968fourth}, which are beyond the table-top of a laboratory and therefore we will not discuss them here.\footnote{Note: During the preparation of the this review a first paper has been published by Asenbaum \etal (Nature, 2017) on an atom interferometer experiment to show a tidal effect of space time.}


\subsection{Proposals for experimental tests of the Schr\"{o}dinger--Newton equation} \label{SNtest}
The \sne\ assumes classical gravity to affect the wave function directly. In a way, the wave function acts as a mass (and not only a probability) distribution and the self-gravity of that spatial mass distribution affects the evolution of the spatial wave function. The \sne\ is presented in \sref{SNtheo}. So far, there has been no experimental test (direct or indirect) of predicted \schr--Newton effects, while there are a number of proposals for such tests. We briefly review proposals for both direct and indirect experimental tests. We note that if the evolution of a matter-wave can be studied experimentally, it will as well provide a direct test of the quantum superposition principle~\cite{Bassi:2013}.

The regime of the \sne\ is sometimes called the semi-classical regime as gravity remains classical---even Newtonian---but this should not be confused with the regime of quantum field theory on curved spacetime, which is also sometimes called semi-classical. In the latter case, quantum matter fields are described in locally flat Minkowski frames on an overall curved (general relativistic) spacetime,
which leads to the prediction of, e.\,g., the Unruh effect and Hawking radiation, and is therefore the mathematical foundation of black hole thermodynamics~\cite{Wald:1994}. The curvature of spacetime is,
however, given as an external feature in this latter case.
The former case of the \sne, to the contrary, assumes that spacetime must be treated as fundamentally classical, and describes the curvature resulting from the quantum matter itself, i.\,e.~self-gravity.

It is obvious that semi-classical gravity, in connection with the Copenhagen view on quantum mechanics,
allows for faster-than-light signaling, and is incompatible with the more radical extension of that view, the many worlds interpretation. The \sne, therefore, can only seem reasonable when it is discussed in combination with the question of wave function collapse~\cite{Bahrami:2014}.
That obvious conflict between the two approaches holds the potential for an experimental test where two alternative views on the world predict a different outcome of the same experiment. The experiment will be of the type discussed in \sref{Gsup}. There has been already  an attempt to perform an experiment~\cite{Page:1981} which was supposed to test semi-classical gravity in the sense of the \sne.
It is, however, a purely classical experiment together with incomplete theoretical arguments~\cite{Mattingly:2005} which failed the condition to have, for the same degree of freedom, both sufficiently large mass, in order to see a gravity effect, and to be in a quantum mechanical state. It can therefore not be regarded as a valid experiment to test semi-classical gravity. (See also \sref{SNtheo} for a discussion of formal problems of the \sne.)

Formally, the \sne\ is very similar to the physics of many-body systems in the Hartree approximation, such as the nonlinear Gross--Pitaevskii equation~\cite{Gross:1961,Pitaevskii:1961}. However, it is worth noting that the \schr--Newton nonlinearity is delocalized over the range of the affected physical system, contrary to the Gross--Pitaevskii equation. The nonlinearity of the equation makes analytical solutions of the \sne\ rather difficult to obtain, although some analytical properties are known~\cite{Tod:1999}.

\subsubsection{Proposed direct tests of \sne: wave function expansion}
\begin{figure}[t!]
\begin{center}
\includegraphics[width=0.7\linewidth]{{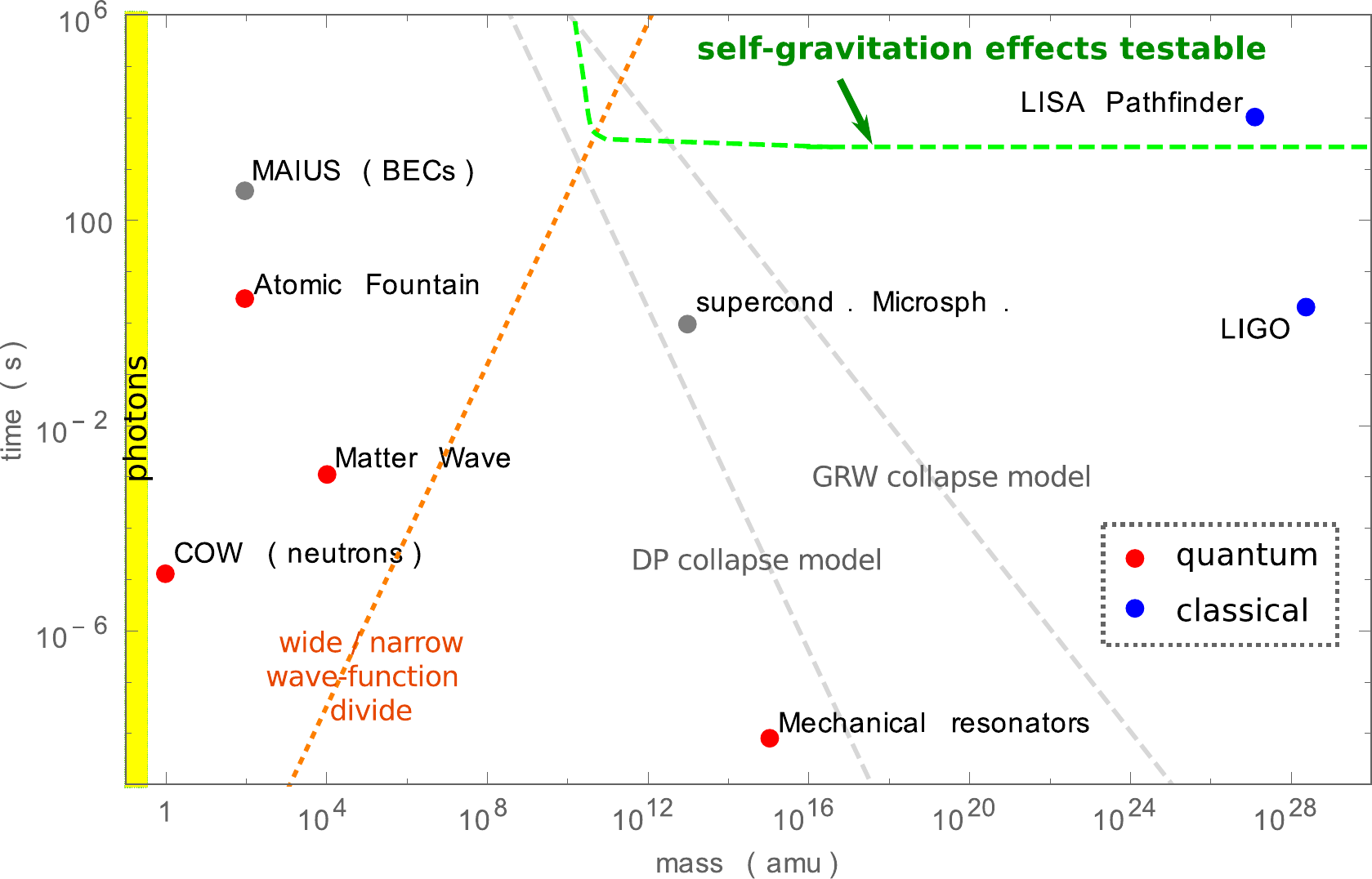}}
\caption{{\bf Direct Test of Schr\"{o}dinger--Newton wave function evolution:} The mass-time plot to illustrate the parameter range which needs to be reached for direct \schr--Newton wave function evolution experiments. This clearly needs to be done without external gravity and other forces/interactions and therefore an experiment in space appears a likely option. The red area shows the parameter range for a proposed space mission to test the \schr--Newton effect.}
\label{SNspace}
\end{center}
\end{figure}

The direct test of the \sne\ is by studying the free expansion of the wave function of sufficiently massive objects. Then a contraction of the wave function according to the \schr--Newton self-gravity effect should have a consequence on that expansion, competing with its natural Schr\"{o}dinger's dynamics spread. Clearly, because of the weakness of the gravitation interaction, the mass has to be sufficiently large while the object has to remain in a quantum mechanical state which can be described by a center-of-mass quantum wave function, meaning the spatial extent of the wave function should be detectable for the full duration of the evolution. See \fref{SNspace} for the mass-time parameter space required to observe the predicted \schr--Newton effect directly, which has been studied extensively~\cite{Moroz:1998, Carlip:2008, Giulini:2011, Giulini:2012, Bahrami:2014, Colin:2016, Grossardt:2016b}, while analytical solutions of the \sne\ are difficult and even numerical simulations are non-trivial.

One possible experimental scenario would be a molecule interferometry experiment~\cite{Gerlich:2007}, as such matter-wave experiments probe spatial superposition states of large molecules. The \schr--Newton contraction effect could also be observed for a free expansion of a singular wave function originated from a point in space, without the necessity of interferometric methods. The key is that the mass of the evolving quantum object has to be comparably large, much larger than the mass achieved in present molecule interferometry experiments. Cold atoms and even BECs of atoms (see \sref{ExpCG}\,C), which benefit from the multitude of coherent manipulation, control, and cooling schemes do not seem to have large enough mass in order to show the \schr--Newton inhibition effect on the free expansion.
Clearly, in order to measure the SN effect one needs both a large mass of the object and access to the coherent quantum evolution of the wave function.
The high mass and the long expansion times to be studied challenge the experimental realization. 

Should direct tests of the \schr--Newton equation be done in space? At this point, there seems to be no other way to allow the wave function expansion for long enough, typically some hundred seconds, see \fref{SNspace}. Proposals to levitate massive particles (optically or magnetically) and, therefore, to compensate for the drop in Earth's gravity have not been realized and are more problematic for \schr--Newton tests. The levitated tests rely on proposed techniques to accelerate the wave function expansion artificially by optical or magnetic field gradients~\cite{Pino:2016, scala2013matter}. That acceleration would have the potential to wash out completely the fragile \schr--Newton effect. 
\begin{figure}[t!]
\begin{center}
\includegraphics[width=0.5\linewidth]{{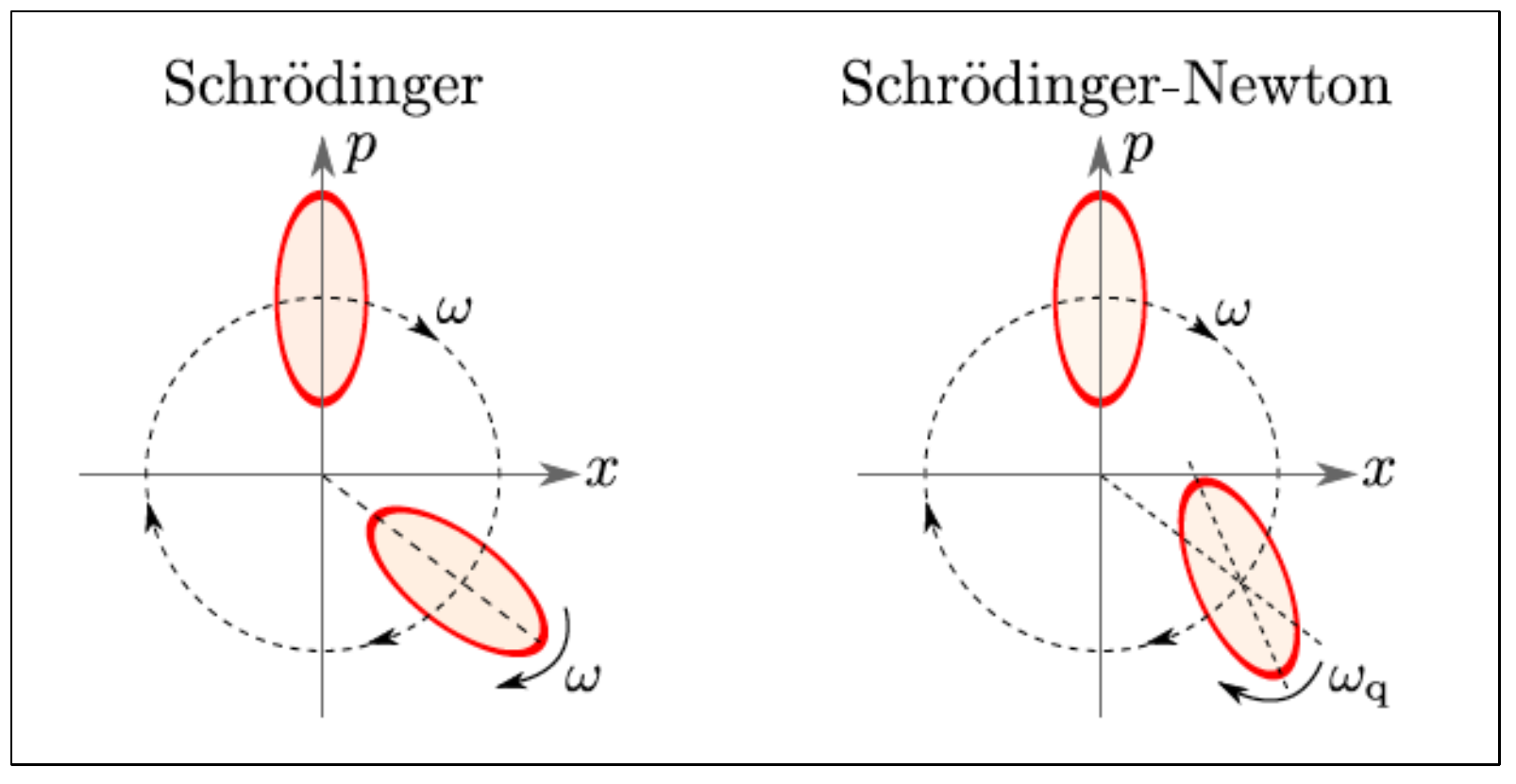}}
\hspace{0.05\linewidth}
\includegraphics[width=0.35\linewidth]{{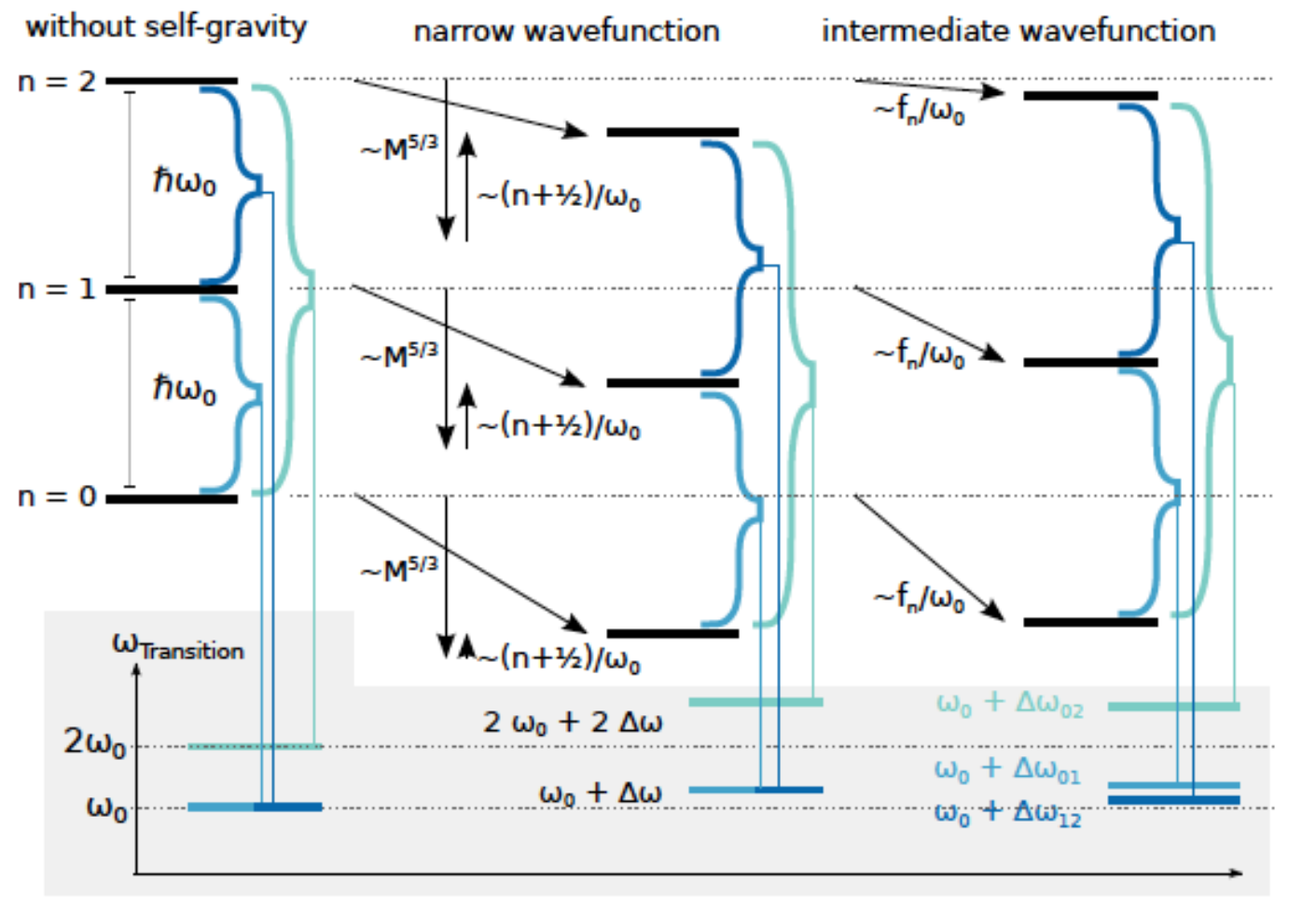}}
\caption{{\bf Indirect Tests of the \sne:} {\it Left Panel:} Phase space plot of mechanical squeezed state with extra rotation of state distribution according to the \schr--Newton effect. Left side: according to standard quantum mechanics, both the vector $(\langle x\rangle, \langle p\rangle)$ and the uncertainty ellipse of a Gaussian state for the center of mass of a macroscopic object rotate clockwise in phase space, at the same frequency $\omega=\omega_{CM}$. Right side: according to the center-of-mass \sne, $(\langle x\rangle, \langle p\rangle)$ still rotates at $\omega_{CM}$, but the uncertainty ellipse rotates at $\omega_q=(\omega^2_{CM}+\omega^2_{SN})^2$.  Reprinted figure with permission from~\cite{Yang:2013}. Copyright 2013 by the American Physical Society. 
{\it Right Panel:} Schematic overview of the effect of the \sne on the spectrum. The top part shows the first three energy eigenvalues and their shift due to the first order perturbative expansion of the Schr\"{o}dinger--Newton potential. The bottom part shows the resulting spectrum of transition frequencies. In the narrow wave function regime (middle part), all energy levels are shifted down by an n-independent value minus an $n$-proportional contribution that scales with the inverse trap frequency. In the intermediate regime, where the wave function width becomes comparable to the localization length scale of the nuclei, this $n$-proportionality does no longer hold, leading to a removal of the degeneracy in the spectrum. Picture and caption taken from~\cite{Grossardt:2016}.}
\label{SNopto}
\end{center}
\end{figure}

\subsubsection{Proposed indirect tests of \sne}
Indirect \schr--Newton effects have been predicted for optomechanical systems which are comparably massive and on the verge to be quantum, see \fref{massrange}. Such effects are very small, can be overwhelmed by noise effects in the experiments, but can be done on the table-top. Therefore, these tests represent a serious experimental challenge, while proposed to be possible with available technology. Two optomechanics experimental cases, as well as the study of the \schr--Newton dynamics in nonlinear optics analogs are mentioned:

{\it A. \schr--Newton rotation of squeezed states.---}The mechanical motion of an optomechanical system, clamped or levitated, is squeezed. Quantum squeezing of clamped optomechanics has been realized experimentally already~\cite{wollman2015quantum, pirkkalainen2015squeezing}, while a classical analog has been demonstrated for a levitated system~\cite{rashid2016experimental}. An optical homodyne detection of both field quadratures of the mechanical state is plotted and shows the cigar-shaped state, see in \fref{SNopto} (left). The \sne\ predicts an extra rotation of the squeezed phase-space distribution~\cite{Yang:2013}, i.\,e.~the internal oscillation of the squeezed state width
$\erw{(x-\erw{x})^2}$ is not in phase with the external oscillation of position $\erw{x}$.

{\it B. \schr--Newton energy shifts of mechanical harmonic oscillator.---}~A further theoretical study~\cite{Grossardt:2016} predicts \schr--Newton related shifts of the energy levels of the stationary states of the quantum harmonic oscillator describing the optomechanical system, see for an illustration of the multiple energy shift effects \fref{SNopto} (right). There, different effects for the so-called wide and narrow wave function regimes are predicted for the situations that the spatial extent of the center-of-mass wave-function is larger (wide wave function regime) or smaller (narrow wave function regime) than the physical size of the massive object. A detailed experimental scenario has been worked out and awaits its realization in an actual laboratory.

{\it C. Nonlinear optics simulation of the \sne.---}Specific delocalized nonlinearities in optical systems, typically just a piece of glass with a large refractive index, show a very similar type of dynamics for the propagation of light through that system if compared to \schr--Newton dynamics. The analogue holds at least in (1+2) spacetime dimensions. The analogue provides an interesting option to study the dynamics of the \sne\ in a parameter regime complementary to numeric simulations. Some experiments have been already performed~\cite{bekenstein2015optical, roger2016optical} to study cosmological settings of the \sne\ such as exotic Boson stars. The main question remains, what can we ultimately learn from optics analogue experiments. Although these experiments do not genuinely test gravity, the study of the formally analogue dynamics, which is hard to calculate and even simulate otherwise, can lead to a better understanding of gravity, as well. See \sref{QGtest}~C for more tests of gravity based on analogues.


\subsection{Proposed tests of Quantized Gravity}\label{QGtest}
Here we look into predicted experimental tests for approaches that assume that gravity is quantum on the fundamental level, string theory and loop quantum gravity being the two most prevalent candidate theories.  Arguments why gravity should be fundamentally quantum include a thermodynamic argument~\cite{padmanabhan2011lessons}, and we will revisit in our brief section on the holographic principle. Again we focus on tests which are predicted for the low energy regime of table-top experiments.

The consequences of quantum gravity theories for experiments are very tiny effects and no test has been successfully performed yet. Also none of the candidate theories is regarded as a fully consistent theory of quantum gravity at their present stage of theoretical development.
String theory is essentially an extension of quantum field theoretical principles from point particles to extended objects, strings, which uses flat Minkowski spacetime as its starting point~\cite{green1987superstring}, and introduces gravity by predicting a new particle, the graviton.
A direct test in particle accelerator type experiments would take extremely high energies---far beyond the scales achievable in nonrelativistic table-top experiments. A potentially testable prediction on the table-top is related to the higher dimensional background spacetime used in string theory. This predicts a deviation of Newton's inverse square law for sufficiently large gravitational interaction strengths, i.\,e.~in the close proximity of two masses. It is similar to fifth force tests by measuring the gravitational interaction at close proximity of masses where also effects of dark energy candidates~\cite{burrage2015probing}, e.\,g.~the Chameleon field are expected. Again, the most precise gravity sensors (torsion balances, atom interferometers, optomechanics) compete in these experiments, while no evidence for a deviation from the inverse square law has been measured~\cite{Kapner:2007, burrage2015probing, hamilton2015atom, hsu2016cooling}. Loop quantum gravity follows a more conservative approach, starting from a re-formulation of general relativity, and attempting to directly (canonically) quantize it, thereby introducing operators for geometrical features such as length, area and volume~\cite{rovelli1998loop, ashtekar2004background}.

There are other quantum gravity predicted experimental consequences to be testable on the table-top for quantized gravity and we will mention some here.

{\it A. Minimal length scale tests.---}The basic assumption is that spacetime has a minimum length scale at which point it becomes discrete and therefore quantum, as explained in \sref{sec:min-length}. While it has been speculated that the Planck length could provide such a minimal length scale, there is no consensus that this is actually the case. There is already an excellent review of minimum length scale models and their possible experimental tests~\cite{plato2016gravitational}, so we do not repeat a review of those topics here. We only mention some aspects, which seem most interesting to us---mostly related to proposed experimental tests.

Some of the proposed experiments to measure physics at the Planck length are based on the related modification of quantum mechanical relations, such as the canonical uncertainty relation $\Delta x \Delta p \geq\hbar/2$~\cite{garay1995quantum}, which can be measured precisely in quantum optomechanical experiments~\cite{Pikovski:2012}. The proposal is to measure $\Delta x$ and $\Delta p$ of the mechanical harmonic oscillator by quantum optics techniques to the ultimate precision, in order to detect a deviation as predicted by Planck-scale physics, such as doubly special relativity~\cite{amelino2002doubly, magueijo2003generalized}, string theory~\cite{gross1988string, amati1987superstring, witten1996reflections}, the principle of relative locality~\cite{amelino2011principle}, and the deformation of the canonical commutator~\cite{kempf1995hilbert, das2008universality}. The predicted effect is however tiny and optomechanics experiments have not reached the required sensitivity to observe any deviation. 

No deviation from the standard quantum mechanical relations has been detected yet, while experiments at the sub-millikelvin cooled mechanical motion of the gravitational wave detector AURIGA have been used to set upper bounds on Planck scale modifications~\cite{marin2013gravitational}. However, AURIGA weights several tons and, arguably, it is not a table-top experiment anymore. Such upper bounds have been lowered by an optomechanical experiment using masses of around the Planck mass, but in a purely classical motional state~\cite{bawaj2015probing}.

{\it B. Bekenstein's idea for a table-top experiment to measure discreteness of spacetime.---}This proposal is related to the minimum length scale ideas as it aims to resolve experimentally the discrete structure of spacetime as in principle predicted by the quantized version of general relativity. Bekenstein's idea was to hunt for discrete steps in the precisely detected motion of a macroscopic glass block. The discrete steps are thought to originate from ripples of spacetime at the Planck (or any other minimum) length. While not surprising, the interesting line of thought is that a mechanism is needed to amplify the (proposed) tiny rippled structure of spacetime in order to make it detectable. In Bekenstein's proposal the amplification is realized by nonlinear optics effects. This is an approach which remains to appear fruitful to follow further. 

Two major problems have been identified with this idea: 1) There is no definite size of those ripples of spacetime predicted by any quantum gravity theory or model. 2) The optical amplification principle is speculative and has not been demonstrated even for larger modulations of the motion of a macroscopic glass block. More clearly, the motion of the glass block has to be detected relative to the propagation of a single photon in a nonlinear optical medium (glass), while it is not clear if the motion of the photon itself is not also affected by the same quantum gravity effect which affects the motion of the macroscopic glass block.

However we remain positive about the initial idea as the proposal has certainly inspired experimentalist to think about possible tests of quantum gravity on the table-top.

{\it C. The idea of the holographic principle.---}The idea of the holographic principle provides an interesting connection between gravity and thermodynamics. It is, that all physics (via the statistical interpretation of entropy) within an enclosed volume is encoded on its surface,  was proposed first in the so-called brick wall model~\cite{hooft1985quantum} and soon found adaptation for string theory in the so-called AdS/CFT correspondence~\cite{maldacena1999large}. AdS/CFT is very popular in string theory, as it provides an approach to circumvent non-trivial calculations of a higher dimension ($n+1$) theory (in anti deSitter spacetime, hence AdS) to the corresponding lower dimensional ($n$) conformal field theory (CFT). This approach ultimately leads to the lively discussed idea of the so-called firewall~\cite{bousso1999covariant, hayden2007black}.

The link to thermodynamics comes as physical degrees of freedom are considered, as they might appear on the event horizon of a black hole~\cite{bekenstein1973black}.  The mathematical framework of thermodynamics, which is the very same as quantum field theory on curved spacetime as discussed in \sref{SNtest}, allows for quantum effects in that gravity scenario, such as Hawking radiation~\cite{hawking1975particle}. In turn thermodynamic properties such as temperature and entropy can be associated with the extreme gravity situation of a black hole, and is therefore going beyond general relativity. While the holographic principle provides an exciting link between quantum mechanics and gravity, it remains very hard to test by any experiment. 

If true, this principle means that the total number of degrees of freedom in the universe is finite, according to the Bekenstein bound~\cite{bekenstein1974generalized}, in contradiction to infinite Hilbert-space quantum mechanics.
The fascinating question is if the principle is universal and therefore could in principle also be tested in a less general relativistic domain - such as by quantum table-top experiments. There is also an interesting and potentially fruitful link to information theory, through the link to von Neumann entropy. Interestingly, there is a link to quantum optics to so-called area laws, which use the same informational link of entropy between the area and the enclosed volume~\cite{verstraete2006criticality, eisert2010colloquium}. This may open the door for experimental tests of the holographic principle by scenarios involving quantum entanglement in the laboratory.

Experiments with optical interferometers have been proposed to be used to test Planck scale quantum geometry models, some of which are mentioned in \sref{QGtest}~A, including those with holographic bounds which induces noise in the interferometer measurements~\cite{kwon2016interferometric}.

Again, as black holes and especially quantum physics on the black hole horizon are notoriously hard to reach by experiments, analogue experiments have been proposed to test formally equivalent physics on artificial event horizons, which can be realized in the laboratory~\cite{philbin2008fiber, leonhardt2002laboratory, schutzhold2007quantum, novello2002artificial, volovik2003universe}. Physical systems used to generate event horizons and to detect quantum physics there are nonlinear optics~\cite{belgiorno2010hawking} and flowing water~\cite{weinfurtner2011measurement} amongst others. While analogue experiments clearly give insight into the dynamics of formal equivalent mathematics, it remains unclear, per definition, whether the same physics applies to the case involving actual gravity. Gravity analogue experiments certainly help to ask more precise questions about the case with gravity.


\subsection{Gravitational decoherence effects}\label{Gdeco}
Tests of gravitational decoherence are based on the the straight-forward approach to generate a spatial superposition state (or any other non-classical state) of a massive particle and test if such a state decoheres according to (classical or quantum) gravity. Clearly, the experimental challenge is the preparation of such a state of sufficient mass. Typical experiments involve matter-wave interferometers and quantum optomechanics. The state of the art of experiments have been already covered in \sref{ExpCG}. While the largest mass is given again by molecule interferometry, some of the effects (such as time dilation) are more promising to be tested in smaller mass systems such as cold atom interferometry, as those can be prepared in larger size superposition states to pick up a larger dephasing or decoherence effect. Although on first sight it appears that only massive systems can be used for the test, it becomes clear that general relativity effects also exist for photons~\cite{zych2012general}. 

{\it A. Gravitational decoherence affecting superpositions.---}In matter-wave interferometry experiments, the proposed effects from general relativistic time dilation~\cite{Zych:2011, Pikovski:2015}, discussed in \sref{sec:gd-time} may be tested. Time dilation leads to a dephasing effect in a matter-wave interferometer for the propagation of the wave function along the two different arms---ultimately resulting in a reduction of the visibility of the interference pattern. The effect has been predicted to scale with the number of all internal degrees of freedom, which are involved in the energy-momentum tensor on the right hand side of Einstein's equations (and which, therefore, affect the spacetime curvature and result in a gravitational force). 

Atom interferometry tests of the time dilation effect appear most promising at the moment. They profit from the high control of the center-of-mass motion of cold atoms, e.\,g.~in a 10~\meter\ fountain and with sensitivity on the verge of $10^{-19}$, while the theoretical details of the effect are still debated. As a universal decoherence effect to explain the evident macroscopic quantum to classical transition, it is clear that time dilation decoherence, should it exist, is weaker by many orders of magnitude than know environmental effects such as decoherence due to collisions by an even very diluted background gas~\cite{Carlesso:2016a}, which leaves the usefulness of the general relativity effect in question.

To be more precise, each (internal) degree of freedom of the particle is regarded as a clock running at a typical frequency, but depending via general relativistic time dilation on the local gravitational environment. Then each single clock, if separated between the two different paths of an interferometer, will be sensitive to the relative duration of time and, therefore, dephase. This experiment has been realized as a proof of principle experiment with atomic chips~\cite{margalit2015self}, where the much larger spatial separation in other atomic interferometers~\cite{kovachy2015quantum} will help to improve the sensitivity to observe the predicted effect, discussed in \sref{sec:gd-time}.

{\it B. Gravitational effects in dynamical reduction models.---}Dynamical reduction or collapse models have been formulated to explain the quantum to classical transition on a fundamental level and in complement to decoherence models~\cite{Bassi:2013}. While the physical reason for the collapse to occur is explained by the existence of a universal classical and random noise field, the physical origin of that field is still debated.
The Di\'osi--Penrose model and Adler's model suggest that gravity is the cause of the collapse. This has been reviewed in \sref{sec:oxfc}.

The best way to test those models is by large mass matter-wave interferometry, where the mass has to be beyond the presently reached limit of molecule interferometry by many orders of magnitude. This means that testing such models requires the preparation of large masses in non-classical states and optomechanical or magnetomechanical systems look most promising for the test~\cite{bose1997preparation, bose1999scheme, vanner2011pulsed}. Proposed experiments along those lines involve~\cite{marshall2003towards, Pino:2016}.

Also indirect, non-interferometric experiments can be performed. \Fref{Fig_parameters} shows the state of the art with respect to Adler's model. Future experiments, proposed in order to close the remaining gap in the parameter plot, involve those to generate large and massive quantum superpositions~\cite{romero2011large, asadian2014probing, Pino:2016, bateman2013near, wan2016free}. Such experiments are currently under development in the laboratories.
A similar analysis for the Di\'osi--Penrose model (and K{\'a}rolyh{\'a}zy's model) is still missing.

{\it C. Gravitational wave induced decoherence.---}Gravitational waves, which are a predicted consequence of general relativity~\cite{Einstein:1916,Ehlers:1962}, have attracted much attention recently as a first experimental proof of their existence has been made by the LIGO experiment~\cite{Abbott:2016} in 2016, which is promising that gravitational waves may soon serve as new tool for astronomy. The LIGO experiment is a sophisticated version of a Michelson interferometer with kilometer long arms. It is, therefore, a large-scale physics experiment and we will not discuss it here in much more details. A first step has been taken to implement a space based version of a Michelson type gravitational wave interferometer with the very successful LISA pathfinder mission~\cite{armano2016sub}, which is of course also way beyond the scale of the table-top experiments discussed here.

There are further proposals to detect the effect of gravitational waves in more table-top like experiments. However, it will become clear that also the scale of these seemingly table-top ideas goes easily beyond the laboratory scale. As we have discussed in \sref{ExpCG}~C, in atom interferometers there are sensitive spatial superposition states of massive particles which have been proposed to be used as a probe for gravitational waves. One can distinguish two ways to detect gravitational waves by atom interferometry, depending on whether gravitational waves are assumed to {\it 'decohere' or 'not to decohere'} matter-waves: The effect described by Lamine \etal~\cite{Lamine:2006} is {\it decoherence} of the matter-wave by collision with a gravitational wave, while in Dimopoulos \etal~\cite{dimopoulos2009gravitational} the matter-wave undergoes a {\it dephasing}, which is precisely measured by a differential measurement between two widely separated atom interferometers using a common laser. 

For the latter case and with technology such as the 10~\meter\ Stanford atomic fountain on ground with strain sensitivity of up to $10^{-19}/\sqrt{\hertz}$ in the frequency band of 1--10~\hertz, the gravitational wave measurement would be complementary to LIGO. A recent proposal is about a technique to use atom interferometers for a single baseline detector for gravitational waves with improved strain sensitivity on ground~\cite{graham2013new}. Similar ideas based on atom interferometry have been proposed for a space based experiment in the LISA-type arrangement and the achievable strain sensitivities are predicted to be $10^{-20}/\sqrt{\hertz}$ for the same gravitational wave frequency spectrum as LISA. A further low orbit scenario with a 30~\kilo\meter\ baseline has been proposed~\cite{hogan2011atomic}, aiming for strain sensitivity of less than $10^{-18}/\sqrt{\hertz}$ in the gravitational wave frequency band of 50~\milli\hertz\ to 10~\hertz, which would be complementary to LIGO-type instruments. This idea is the basis for a proposed space mission.

For the case of decoherence due to scattering with gravitational waves, where the gravitational wave hits the matter-wave, there is the HYPER proposal by Reynaud and co-workers~\cite{Reynaud:2004}. They evaluate the significance of the decoherence process associated with the stochastic background of gravitational waves, and they show it has a tiny effect on HYPER-like atomic interferometers. In the work of Lamine \etal~\cite{Lamine:2006} they work out that gravitational waves would ultimately limit matter-wave interferometers by a decoherence effect, while the effects are very tiny and much smaller than all other decoherence effects discussed in this review. They estimate that the decoherence effect on a matter-wave would be visible for a molecule interferometer type of experiment~\cite{arndt2014testing} of path separation on the order of 1~\meter, for $10^{-18}$~\kilogram\ molecules traveling at a speed of 1~\kilo\meter\per\second---which is clearly way beyond existing experiments and technology in the laboratories, while not un-thinkable for a space based experiment. The interferometric path separation of 1~\meter\ remains the biggest experimental challenge. 

Other ideas to probe gravitational waves on more table-top like experiments are to look for unique heating and noise effects in optical lattice atomic clocks~\cite{kolkowitz2016gravitational}, or with superfluid $^4$He droplets~\cite{Singh:2016}. Furthermore, levitated optomechanics has been proposed to be used to probe gravitational waves~\cite{arvanitaki2013detecting}, where they also work out how the effect to detect gravitational waves would scale with mass. Such an instrument could detect gravitational waves in the 50~\kilo\hertz--330~\kilo\hertz\ frequency range with a $10^3$ times improved sensitivity compared to LIGO assuming a 150~\nano\meter\ radius levitated microdisk in a 100~\meter\ optical cavity. Clearly the 100~\meter\ cavity line is beyond a strict table-top scale.

Long-term competitors of LIGO type optical interferometers to detect gravitational waves are the Schenberg instruments or parametric transducers~\cite{aguiar2008schenberg} in Leiden and Sao Paulo which have the expected strain sensitivity of $10^{-22}/\sqrt{\hertz}$ in the narrow frequency band of 3~\kilo\hertz\ to 3.4~\kilo\hertz. There, the gravitational wave scattering effect on the precisely controlled motion of a magnetically trapped macroscopic perfect sphere is used to detect gravitational waves. 

{\it D. Competing effects for matter-wave interferometry.---}In order to be able to see such gravity effects and how they collapse or decohere the wave function in matter-wave based experiments, all competing environmental decoherence processes have to be suppressed, which is the major experimental challenge in order to perform the experiments. Dominating decoherence effects are due to collisions with background gas, collisional decoherence~\cite{hornberger2003collisional}, and the effects due to exchange of thermal radiation between the quantum system and the environment~\cite{hackermuller2004decoherence, romero2011quantum, bateman2013near}. Magnetic levitation of superconducting microparticles, by definition, avoids all effects related to internal temperature radiation as the experiment is cryogenic and on top of that all noises related to lasers are removed as well~\cite{Pino:2016}. This represents a huge advantage compared to optomechanics test. Furthermore, vibrations set serious constraints to all mechanics based tests of wave function collapse and gravity.

{\it E. The case for space.---}Ultimately a test of gravity decoherence and gravity induced collapse of the wave function would benefit from large masses of the particles in superposition states as well as long lifetimes of those superposition states in order to observe the extremely weak effects. The space proposal on macroscopic quantum resonators (MAQRO)~\cite{Kaltenbaek:2016} would be able to fulfill all those conditions. A community has started to work towards such a test in space and to propose a related mission. There are also many attempts to perform atom interferometry in space, which have been already mentioned and cited in \sref{ExpCG}~C on atom interferometry and \sref{Gdeco}~C on gravitational wave decoherence.


\subsection{The gravity of a quantum state}\label{Gsup}

A somehow different approach to the question of the interplay between quantum mechanics and gravity is given here. The setting goes back to Feynman~\cite{Feynman:1995} and amounts to the question: what is the gravitational field generated by a massive quantum superposition? Is it the superposition of the two gravitational fields generated by the two terms of the superposition, as predicted by quantum gravity? (Feynman thought this was the most logic answer.) Or is it the sum of the two gravitational fields, as predicted by the \sne\ and perhaps by any consistent theory, which keeps gravity fundamentally classical? Or does the superposition decay because of gravity, as assumed by Penrose and developed in gravity-related collapse models?

This question challenges directly our understanding of gravity. But it goes further than this. It is directly related to another fundamental question: what is the mass, in a quantum universe? Is it a parameter entering the \schr\ equation, more or less like the electric charge is? Or does it have a different origin, as general relativity suggests? Is the mass delocalized along with the wave function, or is it always localized in space?

For years these questions were purely speculative, because a detectable gravitational field requires a large enough mass, which can be hardly put in a spatial quantum superposition. This is why an experiment along these lines has never been done. But recent technological developments in quantum optomechanics are changing things. Though not in a truly quantum regime yet, appreciably large masses are controlled and manipulated with extreme accuracy, and very sensitive force and position measurements can be done. Now a feasible experimental proposal can be thought of, although the details still need to be sorted out~\cite{Bahrami:2015}.

A related proposal is to use the nonlinearity of gravity interaction to generate quantum entanglement between two distant massive systems in order to prove the fundamental non-classicality of gravity~\cite{Kafri:2015}.
 
It is worth mentioning another interesting effect---retarded gravity---which has been speculated.  Using the language of a classical gravity, the claim is that the center of the gravitational field does not coincide with the center of mass originating the field. As a consequence, there is a retarded gravity effect, which can be observed in specific experimental conditions~\cite{diosi2014equation, Yang:2015}. Experiments, most likely torsion balances or optomechanical devices, could aim to resolve it.

\section{Conclusions}

After about half a century of intense research in unifying quantum theory and general relativity, only recently a significant part of the scientific community started reformulating the problem from scratch, in a perhaps less ambitious but equally interesting way: to challenge the interplay between the two theories, both theoretically and experimentally, rather than trying to combine them in a single framework.

The field of gravitational decoherence is relatively young, and this explains the variety of approaches and  the lack of unity, as it emerges  from the present review. In this summary we will present some of the most relevant open problems, at least in our view, and possible future directions of research.

From the theoretical point of view, two lines of research have been initiated: to understand the effect of gravitational noise on quantum systems, at the nonrelativistic level, which is more easily accessible from the experimental  point of view; and to speculate whether gravity might modify quantum theory, as a resolution of the quantum measurement problem. The first strategy was reviewed in \sref{sec:decohgrav}, the second in \sref{sec:oxfc}. 

As far as the first line is concerned, a rather large variety of models exists which take different approaches and, therefore, are hard to compare and put on a common ground. Most models are based on, or at least motivated by, assumptions about quantum gravity, such as the existence of a minimal length scale or a bath of thermalized gravitons. Interestingly, much less work has been focused on effects of classical spacetime fluctuations, and a comparison of the different approaches discussed in \sref{sec:decohgrav} would be desirable. A comparison of the effects of a classical spacetime to quantum gravity seems necessary also, in order to identify effects that can truly be considered as quantum gravity fingerprints.
Despite being the most straightforward one from its theoretical origin, the time dilation induced decoherence effect~\cite{Pikovski:2015}, which should be considered as an effect of non-inertial reference frames rather than a genuinely gravitational decoherence, only raised interest very recently.
There is an ongoing debate about the interpretation and testability of this effect.

Coming to the second research line,
the fact that the program of quantizing gravity turned out not to be as straightforward as with the other known forces, and is yet unresolved, together with the conceptual problems of quantum theory, has lead part of the scientific community to explore the possibility that quantum theory needs first to be reconsidered and that gravity might play a role in this; Penrose~\cite{Penrose:2014} refers to this as ``gravitization of quantum mechanics'', as opposed to the ``quantization of gravity''. As we have seen, the formulation of the problem is simple: quantum mechanics does not explain the absence of macroscopic superpositions. On the other hand, gravity is the stronger, the larger the system. Hence one might speculate that gravity plays a role in preventing macroscopic superpositions. 

We have reviewed four options in \sref{sec:oxfc}: the Di\'osi--Penrose model; Adler's idea on the existence of an irreducibly complex, rapidly fluctuating component of the metric; K{\'a}rolyh{\'a}zy's model; and the Schr\"odinger-Newton equation. They first three models predict the existence of a stochastic gravitational background. The first two also predict  a non-standard (in particular, nonlinear) coupling between matter and the background, which induces the collapse of the wave function in space, in agreement with the Born rule. The third one instead appears more like a standard decoherence model of a system in a noisy gravitational environment. The fourth model predicts a self-gravitational attraction of different parts of the wave function, opposing its natural spread. 

These models do not resolve the quantum-gravity problem, and have very much the flavor of provisional models. So in this sense they cannot compete with much better developed theories such as string theory. However they are interesting and relevant, as they challenge the, so far incomplete, mainstream view according to which the world is quantum, gravity included. They are also interesting because they all predict effects,  which can be tested with   (a reasonable improvement of)  current technology, as discussed in \sref{sec:experiments}.

From the theoretical point of view, the obvious work which needs to be done, is to consolidate these models, ideally deriving them from what can be rightfully regarded as a new theory of matter and gravity. From a phenomenological point of view, it is crucial to conceive new experimental scenarios, where their effects can be tested, possibly in cheap table-top settings. And then experiments need to be done. Such experiments would probe the quantum-gravity interplay, one of the most fascinating and mysterious areas of current research.

From the experimental point of view, precise experiments have been performed to detect classical Newtonian gravity ($G$ and $g$), which, however, are still less precise than the tests of all other fundamental constants. Such experiments include atom interferometers where the test mass is in a non-classical superposition state. Different values for $G$ are found from different types of experiments~\cite{rosi2014precision}. The exact knowledge of the precise distribution of atoms in the typically big source mass has been identified as the limiting factor to increase precision further.  It can be speculated that more precise tests of both $G$ and $g$ can be performed when both the test and the source mass are comparable in size, sufficiently massive to observe gravity effects, but well maintained within the quantum regime. 

While matter-wave interferometer quantum experiments have been performed in the low mass regime, see \fref{massrange}, the higher mass range, all the way up to milligram masses---the working regime of torsion balances---is almost unexplored by any experiment and definitely not by any quantum experiment. Optomechanical devices are one candidate to bridge this enormous mass gap, while being in a quantum mechanical state and very massive at the same time. Especially levitated mechanical systems hold promise to test new physics in that new mass range.
A variety of theoretical proposals and ideas for the interplay between quantum mechanics and gravity will become testable in the same mass range. Such ideas include the here discussed gravitationally induced decoherence and collapse models, the \sne, the gravity of a quantum state, and quantum gravity effects.

\ack
We wish to thank colleagues for enlightening discussions on the topics discussed in this
review---some of them quite some time ago:
Stephen L. Adler,
Peter Aichelburg,
Markus Arndt,
Markus Aspelmeyer,
Peter Barker,
James Bateman,
Robert Beig,
Jacob Bekenstein,
Sougato Bose,
Matteo Carlesso,
Lajos Di\'osi,
Sandro Donadi,
Detlef D\"urr,
Daniele Faccio,
Giulio Gasbarri,
Domenico Giulini,
Saikat Gosh,
Klaus Hornberger,
Bei-Lok Hu,
Rainer Kaltenbaek,
Myungshik Kim,
Renate Loll,
Stefan Nimmrichter,
Mauro Paternostro,
Igor Pikovski,
Jess Riedel,
Tejinder P. Singh,
Andrea Smirne,
Marko Toro{\v{s}},
Michael Vanner,
Andrea Vinante.
AB acknowledges financial support from the University of Trieste (FRA\,2016) and INFN,
AG acknowledged funding by the German Research Foundation (DFG), and
HU acknowledges funding by the Leverhulme Trust (RPG-2016-046).
The authors further thank ICTS Bangalore for hospitality during the program ``Fundamental Problems in Quantum Physics'', during which several of the ideas presented here were discussed.

\section*{References}
\providecommand{\newblock}{}

\end{document}